\documentclass{IEEEtran}

\usepackage{graphicx}
\usepackage{caption}
\usepackage{subcaption}
\usepackage[svgnames, rgb]{xcolor}  
\usepackage{tikz}
\usepackage{url}
\usepackage{comment}
\usepackage{color, colortbl}
\usepackage{array}

\usepackage{cite}
\usepackage{multirow}
\usepackage{authblk}

\graphicspath{{figures/}{images/}}

\definecolor{Yellow}{rgb}{1,1,0}
\newcolumntype{P}[1]{>{\centering\arraybackslash}m{#1}}
\def\checkmark{\tikz\fill[scale=0.4](0,.35) -- (.25,0) -- (1,.7) -- (.25,.15) -- cycle;} 

\newcommand{\ar}[1]{\textcolor{black}{#1}}

\newcommand{\arnau}[1]{\textcolor{black}{#1}}
\newcommand{\arnautwo}[1]{\textcolor{black}{#1}}

\usepackage{color}

\begin{document} 

\title {A Review of AI-enabled Routing Protocols for UAV Networks: Trends, Challenges, and Future Outlook}

\author[1]{Arnau Rovira-Sugranes}
\author[2]{Abolfazl Razi}
\author[3]{Fatemeh Afghah}
\author[4]{Jacob Chakareski}
\affil[1]{School of Informatics, Computing and Cyber Systems, Northern Arizona University, Flagstaff, USA}
\affil[2]{School of Computing, Clemson University, Clemson, USA}
\affil[3]{Department of Electrical Engineering, Clemson University, Clemson, USA}
\affil[4]{College of Computing, New Jersey Institute of Technology, Newark, USA}
\renewcommand\Authands{ and }


\maketitle

\begin{abstract}
Unmanned Aerial Vehicles (UAVs), as a recently emerging technology, enabled a new breed of unprecedented applications in different domains. This technology's ongoing trend is departing from large remotely-controlled drones to networks of small autonomous drones to collectively complete intricate tasks time and cost-effectively. An important challenge is developing efficient sensing, communication, and control algorithms that can accommodate the requirements of highly dynamic UAV networks with heterogeneous mobility levels. Recently, the use of Artificial Intelligence (AI) in learning-based networking has gained momentum to harness the learning power of cognizant nodes to make more intelligent networking decisions \arnautwo{by integrating computational intelligence into UAV networks.} An important example of this trend is developing learning-powered routing protocols, where machine learning methods are used to model and predict topology evolution, channel status, traffic mobility, and environmental factors for enhanced routing.

This paper reviews AI-enabled routing protocols designed primarily for aerial networks, \arnautwo{including topology-predictive and self-adaptive learning-based routing algorithms,} with an emphasis on accommodating highly-dynamic network topology. \arnautwo{To this end, we justify the importance and adaptation of AI into UAV network communications. We also address, with an AI emphasis, the closely related topics of mobility and networking models for UAV networks, simulation tools and public datasets, and relations to UAV swarming, which serve to choose the right algorithm for each scenario.} We conclude by presenting future trends, and the remaining challenges in AI-based UAV networking, for different aspects of routing, connectivity, topology control, security and privacy, energy efficiency, and spectrum sharing \footnote{This material is based upon the work supported by the National Science Foundation under Grant No. 2008784.}.

\end{abstract}

\section{Introduction} 
UAVs are an emerging technology that has opened its way into many fields and is expected to continue impacting the future of human life in the coming years. UAVs have already been utilized in many applications to provide fast, low-cost, on-demand, and precise monitoring and actuation services while minimizing human intervention and life-threatening risks.  
This covers many applications including transportation \cite{transportation}, traffic control \cite{traffic-monitoring}, surveillance \cite{surveillance}, border patrolling \cite{border-patrolling}, search and rescue \cite{search-rescue}, disaster management \cite{erdelj2017help}, wireless network connectivity \cite{connectivity,ChakareskiNMXAR:19}, smart agriculture and forestry \cite{forestry}, and remote immersion via mobile virtual reality \cite{chakareski2017drone,chakareski2019uav,KhanCG:20}. Drones are also widely used in the military domain. For instance, LOCUS is a project by the US navy to utilize a swarm of autonomous drones to perform coordinated military attacks \cite{LOCUST}.
In addition to these commercialized use cases, many new applications are under design and implementation in academia and industry. 
For instance, surveying and mapping \cite{serveyingandmapping}, volcano monitoring \cite{volcanomonitoring}, UAV control by the brain \cite{UAVcontrolbybrain}, early warning of severe weather \cite{casa}, plant protection \cite{plantprotection}, airborne wind energy harvesting systems \cite{windenergy}, robotic herding of a flock of birds \cite{FlockofBirds}, Amazon Prime Air \cite{swarm3}, and UPS drone delivery service \cite{swarm4} are only a few examples of many projects in their infancy steps.

Compared to piloted aircraft, satellite-based imaging, and ground-based sensing and actuation platforms, UAVs offer several advantages, including a small size, low operational and maintenance cost, less human intervention requirements, less operational hazard, autonomous control, more controlled imaging with adjustable zoom and angle of view, and higher maneuverability levels\cite{Gupta_survey}. Therefore, the UAV market has experienced continued growth in the past decades, from an estimated \$19.3 billion in 2019 to a projected \$45.8 billion market by 2025, which represents a Compound Annual Growth Rate (CAGR) of 15.5\% from 2019 to 2025 \cite{uavmarket}.

\begin{figure*}[t]
 	\centering
	\includegraphics[width=1.75\columnwidth]{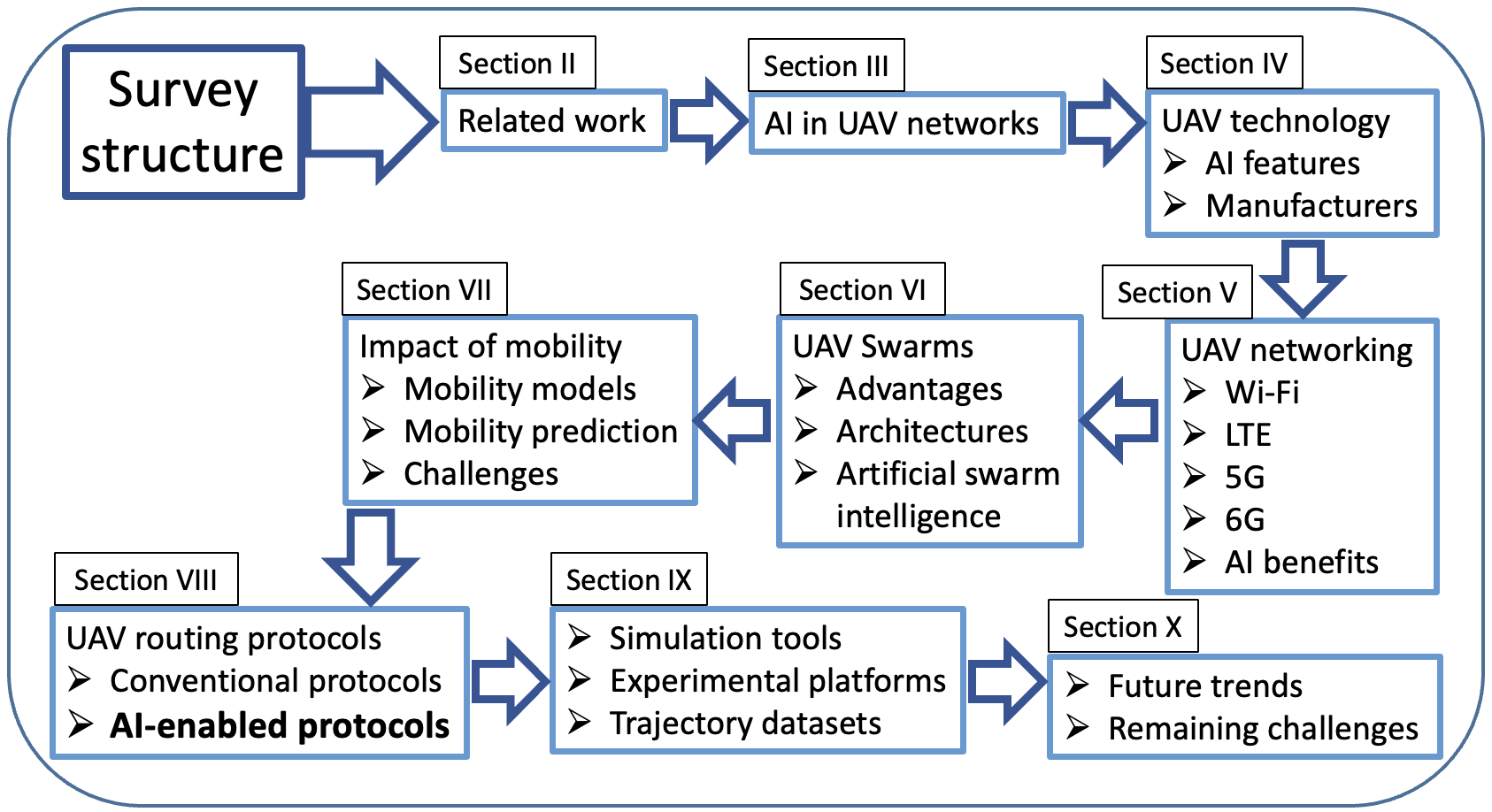}
	\caption{\arnautwo{The organization of this survey paper.}}
    \label{surveystruct}
\end{figure*}

Despite the following advantages in using drone technology, there still exist numerous challenges, and technical issues for implementing networking and control protocols for UAV-based infrastructures \cite{Gupta_survey}. For instance, the limited payload of UAVs translates into constraints in power consumption, communication range, and computational limits that in turn may cause difficulties for networking, robust control, information acquisition and processing, autonomy, and task coordination. Another issue is the extreme dynamicity of UAV networks due to their high speed, heterogeneous maneuverability levels, and obstacle-sparse flight zones, compared to ground vehicle networks with more predictable motion trajectories dictated by road patterns. Therefore, communication, control, path-planning, and information acquisition protocols that are primarily designed for ground platforms deem inefficient for UAV networks. For instance, connectivity of vehicular networks along roads necessitates optimal positioning of nodes or regulating their speed in a one-dimensional subspace that is not directly applicable to UAV networks. Likewise, autonomous driving and collision avoidance for drones have different constraints and requirements, compared to similar tasks in self-driving cars.

The goal of this survey paper is two-fold. First, we review UAV networks' features, including UAV technology, networking protocols, and swarms, with a focus on characterizing the impact of mobility on network topology, connectivity, and networking performance metrics. In this respect, we highlight the importance of having an accurate mobility prediction system for more efficient networking. Second, we review routing protocols designed for UAV networks, emphasizing AI-enabled routing protocols, which present better outcomes for high mobility networks \cite{futureAI1, futureAI2,futureAI3}. \arnautwo{Furthermore, we addressed the closely related topics of mobility models for UAV networks, simulation tools and public datasets, and relations to UAV swarming, which serve to choose the right algorithm for each scenario, as an additional contribution of this paper.}

\subsection{\arnau{Motivation}} \label{sec:motivation}

Routing protocols developed for Vehicular Ad-Hoc Networks (VANETs) and other ground networks are not well-positioned to accommodate the requirements of UAV networks. As stated earlier, drones enjoy a higher mobility degree of freedom, compared to ground vehicles. This leads to a more vivid and fast-changing topology in comparison to VANETs \cite{VANET}. Also, the lower node density of Flying Ad-hoc Networks (FANETs), compared to VANETs, raise connectivity issues due to the drones' limited communication ranges \cite{issuereview}. In terms of channel modeling, fading, and diversity phenomena, FANETs benefit from more accessible Line-of-Sight (LoS) links between UAVs and the use of smart directional antennas for collective beamforming, and similar techniques can offer higher gains \cite{beamforming1, beamforming2}. From a different point of view, it also highlights the necessity of developing more accurate localization and tracking technology for aerial networks. 

Furthermore, conventional routing protocols developed for VANETs merely rely on the node's prior information or current perception of the network topology, and do not perform well in maintaining connectivity. 
They either impose a large overhead for constantly updating the global network information or require a time-consuming route setup phase.
Also, UAV networks are structure-free and not consistent with centralized routing protocols. 
Appropriate routing protocols for UAV networks should have properties like low complexity, low overhead, and preferably without the need for global knowledge and lengthy route setup stages, as discussed in \cite{greedySECON}. The new generation of self-adaptive learning-based and topology predictive routing protocols learn the state of the network by experiencing and predicting dominant trends and constantly adapting to both minor or abrupt changes. This approach leads to a higher packet delivery ratio and energy efficiency.
In this paradigm, decisions are made based on the anticipated network topology, and not solely based on the current state. These key requirements promote AI-enabled routing protocols to achieve superior performance \cite{futureAI1, futureAI2,futureAI3}. \arnautwo{Therefore, using AI methods to predict motion patterns of freely flying UAVs in a 3-dimensional space is an integral part of AI-based UAV protocols, while routing protocols for a 2-dimensional network of cars along the highway (as in VANETs) may not necessarily need this computationally-expensive component.} This survey paper is devoted to highlighting recent developments in \ar{the AI-based routing protocols} and analyzing their benefits and drawbacks when used in realistic situations.

\begin{table*}[ht]
\caption{Most recent survey papers for routing protocols in Wireless Sensor Networks.}
\label{surveys_routing_table}
\begin{tabular}{|p{0.5cm}|p{0.7cm}|p{6cm}|p{6cm}|p{3.2cm}|}
\hline
\rowcolor{gray}
\textbf{Year} & \textbf{Survey} & \textbf{Content included} & \textbf{Drawback} & \textbf{Application domain} \\
\hline
\multirow{4}{*}{2020} & \cite{survey1} & SDN-based routing, monitoring, cellular, satellite, security, placement, evaluation tools and future challenges. & Does not include routing protocols beyond SDN-based methods. & SDN-based UAV networks \\ \cline{2-5}
               & \cite{survey5} & UAV classification, application, mobility models, routing protocols classification, challenges and open issues. &  The review of predictive methods is not complete. It also excludes self-adaptive learning-based routing protocols. & UAV networks \\ \cline{2-5}
                & \cite{issuereview} & UAV communication networks issues, characteristics, design, applications, routing protocols, quality of service, and future open research areas. & Does not consider learning-based routing protocols.  & UAV networks \\ \cline{2-5}
                & \cite{survey6} & Network architecture and design, routing protocols including performance analysis and QoS metrics, and opening issues. &  Routing classification only includes a few methods. & UAV networks \\ \cline{2-5}
                & \cite{survey7} & UAV-UGV coordination, data gathering, monitoring, cellular communications, disaster management, computing and UAV-assisted routing. & Only considers routing methods for UAV-assisted networks, excluding routing protocols for UAV-to-UAV communications. & UAV-assisted networks \\ \cline{2-5}
                & \cite{survey4} & Routing protocols for UAV-aided vehicular ad hoc networks with open research issues and challenges. & Only considers routing protocols for UAV-aided vehicular networks, excluding routing for UAV-to-UAV communications. & UAV-aided vehicle networks \\
\hline
2019 & \cite{survey13} & UAV design, architecture, routing protocols, open issues and research challenges. & AI-enabled routing protocols and their impact in dynamic networks are missing.  & UAV networks\\ \cline{2-5}
                 & \cite{survey8} & Architecture, mobility models, routing techniques and protocols with a comparative study. Future challenges are also included. & Self-adaptive learning-based methods are not considered. & UAV networks\\ \cline{2-5}
                 & \cite{survey14} & UAV routing schemes, including objectives, challenges, routing metrics, characteristics, and performance measures, along with open issues. & It only briefly mentions adaptive routing protocols, missing most of the AI-enabled routing protocols. & UAV networks\\
\hline
2018 & \cite{survey3} & Routing protocols comparison and open research issues. & Does not include AI-enabled routing protocols, among others. & UAV networks\\
\hline
\multirow{4}{*}{2017} & \cite{survey2} & Single-layer and cross-layer routing, challenges and open research directions. & Most of the routing protocols are suitable for vehicular networks, but not defined for UAV Networks. & Vehicular ad-hoc networks \\ \cline{2-5}
                 & \cite{survey12} & UAV architectures, projects, characteristics, applications and routing protocols, with emphasis in UAV security challenges. & Does not consider AI-enabled routing protocols and future trends. & UAV networks \\ \cline{2-5}
                 & \cite{survey9} & Position-based routing protocols with a detailed description and comparative study. Also, mobility models and UAV applications are described. & Includes position-based routing protocols only, which is just one type of UAV routing. & UAV networks \\
\hline
\rowcolor{lightgray}
\multicolumn{2}{|c|}{Our survey} & UAV technology, UAV networking, UAV swarm formation, mobility models, UAV routing protocols, tools and public datasets to simulate real UAV network environments, future trends and remaining issues for UAV networking. & NA & UAV networks \\
 \hline
\end{tabular}
\end{table*}

The rest of this paper is organized as presented in Figure \ref{surveystruct}. \arnau{In Section \ref{sec:relatedwork}, we investigate recently published survey papers to highlight the new content and additional aspects covered by our paper. Next, we emphasize the role of AI methods in improving the performance of UAV networking in Section \ref{sec:AI}.} In Section \ref{sec:uavtech}, a review of UAV technologies used in military, industrial and commercial applications is provided. Section~\ref{sec:networking} presents commonly used networking protocols, and UAV swarm formation methods are presented in Section \ref{sec:swarms}. In Section \ref{sec:impactmobility}, commonly-used UAV mobility models, and their impact on key networking characteristics, including connectivity, channel models, network topology, and routing efficiency, is investigated. 
Fundamentals of conventional and AI-enabled routing protocols, along with their stability under dynamic conditions, are provided in Section \ref{sec:routing}. Tools, public datasets, and remote experimentation infrastructures for testing routing protocols are reviewed in Section \ref{sec:tools}. Future trends and remaining issues are discussed in Section \ref{sec:futuredirections}, followed by concluding remarks in Section~\ref{sec:conclusion}.

\section{\ar{Related work}}  \label{sec:relatedwork}


There are a few recent review papers that survey routing protocols for ground and aerial networks. Other related papers that review UAV networks survey a broader set of aspects. Table \ref{surveys_routing_table} summarizes key topics covered in these surveys, along with key topics missing in each paper. To our knowledge, no paper provides a comprehensive and up-to-date review of AI-based routing protocols for aerial networks, which is our central focus.

Here, we closely discuss the most notable papers from the last three years only, since newer papers usually tend to improve previous reviews and cover the most recent developments. In addition to recent studies, we also consider two fairly older papers for their remarkably unique content. 
One paper is \cite{survey10}, which provides a thorough review of routing protocols in inter-vehicle communication systems. This paper covers broadcast-based routing, multicast, and geocast-based routing, as well as unicast-based routing protocols, which is perhaps the most complete review for routing protocols developed for vehicular networks.
Another review is a seminal paper by Gupta et al. \cite{Gupta_survey} published in 2015, which offers a broad outlook and comprehensive review of important issues in UAV networks. Also, it reviews the concept of routing in networks subject to severe delays and disruptions, which is unique among the published papers.

A survey paper by Awang et al. in 2017 \cite{survey2} provides a review of routing protocols for vehicular ad hoc networks describing existing single-layer and cross-layer routing algorithms. It offers a fluent review of routing protocol for VANETs along with a clear description of the advantages and disadvantages of each method. However, most of the methods mentioned in this paper are designed for VANETs, and not suitable for FANETs with substantially different constraints and requirements. 

A review of routing protocols and security challenges in UAV networks is provided in \cite{survey12}. This paper reviews different routing protocols developed for dynamic networks. Nevertheless, it does not include an important and emerging trend of AI-enabled routing protocols.

Another paper \cite{survey9} provides a comprehensive review of position-based routing protocols for UAV networks. It nicely classifies routing protocols with a detailed description of each category. Also, the routing algorithms are compared based on various criteria and performance metrics. However, only position-based routing protocols are mentioned, excluding many other types of routing protocols developed for UAV networks.


In 2018, a paper titled ``Routing protocols for Unmanned Aerial Vehicles" \cite{survey3} compared the routing protocols from different standpoints. This paper sheds light on which methods are suitable for UAV Networks under different network conditions and application-oriented requirements. However, the provided review is not comprehensive, and many efficient routing protocols such as hierarchical, probabilistic and AI-enabled methods are not discussed.

In 2019, a survey paper offered a complete review of UAV network design, architectures, routing protocols, open issues, and research challenges \cite{survey13}. Deterministic, stochastic, and social-network-based routing protocols are discussed, along with a qualitative comparison of their major features, characteristics, and performance. However, AI-enabled routing protocols and their role in accommodating dynamic networks are missing.

In \cite{survey8}, a comparative review of major existing routing protocols developed for FANETs, along with a careful analysis of their performance under different design constraints and planning strategies, is provided. However, the important class of self-adaptive learning-based methods is not discussed.
Similarly, \cite{survey14} offers a comprehensive review of routing schemes in FANETs, including objectives, challenges, routing metrics, characteristics, performance measures, and open issues. However, most of the AI-enabled routing protocols are not discussed.

The following are a few survey papers published in 2020.
A recently published paper \cite{survey1} reviews Software-Defined Network (SDN) and Network Function Virtualization (NFV) for UAV-assisted monitoring, cellular, and satellite communications systems. 
More specifically, this paper reviewed SDN-based routing. However, it only considers Air-to-Ground scenarios.

A review of mobility models and routing algorithms for FANETs is provided in \cite{survey5} and \cite{issuereview}, with the inclusion of UAV communication networks issues. Nonetheless, they exclude an important class of learning-based routing algorithms. A comparative analysis of emerging routing protocols for UAV networks under different conditions is provided in \cite{survey6}. However, it includes only the position-based methods and ignores different implementations of proactive, reactive, and AI-enabled routing protocols.

Two recent works \cite{survey7, survey4} offer an exhaustive review of communication protocols, applications, and security issues of UAV-assisted ground and vehicular networks. However, their center of attention is UAV-assisted routing, which excludes many routing protocols developed for the more general class of UAV-to-UAV communication in FANETs. 
In summary, almost all of the previously published surveys do not pay enough attention to the emerging class of AI-enabled routing protocols, which can be considered the most appropriate class of routing protocols for extremely dynamic UAV Networks.

\section{\arnau{Artificial  Intelligence in UAV networks}} \label{sec:AI}

\begin{figure}[h]
 	\centering
	\includegraphics[width=1\columnwidth]{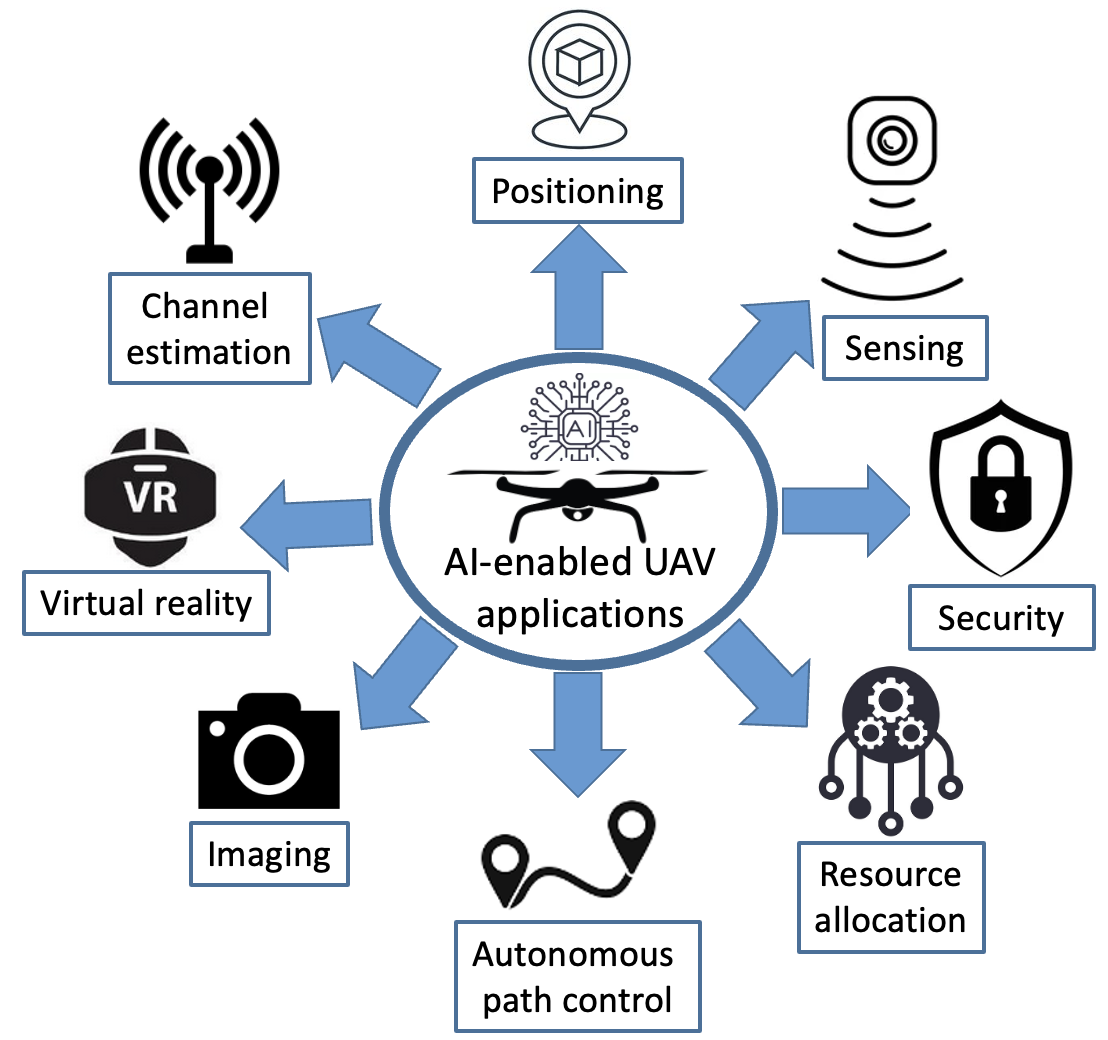}
	\caption{\arnau{AI-enabled UAV applications.}}
    \label{aiuavapp}
\end{figure}

\arnau{AI-based solutions help to solve complex problems related to UAV networking and operation by integrating computational intelligence into different aspects of UAV networks. The key idea is to incorporate AI algorithms into networking and control protocols to assist UAVs in perceiving the networks' and environments' overall conditions based on their limited observations. AI can also empower UAVs to process and interpret common patterns and anticipate future states and events when making decisions.}

\arnau{The benefits of using AI in UAVs are countless. For example, AI-based decision-making with real-time data allows continuous feedback in inaccessible areas to keep functions alive \cite{AIbenefit}. Also, training nodes by experience, in most cases, would result in more accurate results than taking actions blindly. Gathering information can facilitate resource management for energy optimization and trajectory design to avoid obstacles. However, these benefits come at a higher computational cost. In contrast to sophisticated scenarios, the value added by the AI methods in more straightforward scenarios, especially with no or limited training dataset, is negligible. Therefore, each scenario should be carefully investigated to analyze the benefits and drawbacks of using AI methods.}

\arnau{In \cite{AIapp_survey}, applications of AI methods to UAV-enabled wireless networking are listed. This paper summarizes different learning approaches including supervised and unsupervised learning, reinforcement learning and federated learning. Some areas where researchers introduced AI-based solutions include positioning and detection \cite{futureAI5, positioning1, positioning2}, channel estimation \cite{channelest1, channelest2}, virtual reality applications \cite{virtual1, virtual2, chakareski2019uav}, imaging \cite{imaging1}, autonomous path control \cite{autonomous1, autonomous2}, scheduling and resource allocation \cite{scheduling1}, security \cite{security1} and sensing \cite{sensing1}, as shown in Figure \ref{aiuavapp}.}
\arnau{Our paper focuses on the applications of AI in routing protocols while mentioning how AI is embedded into UAV technology, networking protocols and swarms, in general.}

\begin{figure}[h]
	\centering
	\includegraphics[width=1\columnwidth]{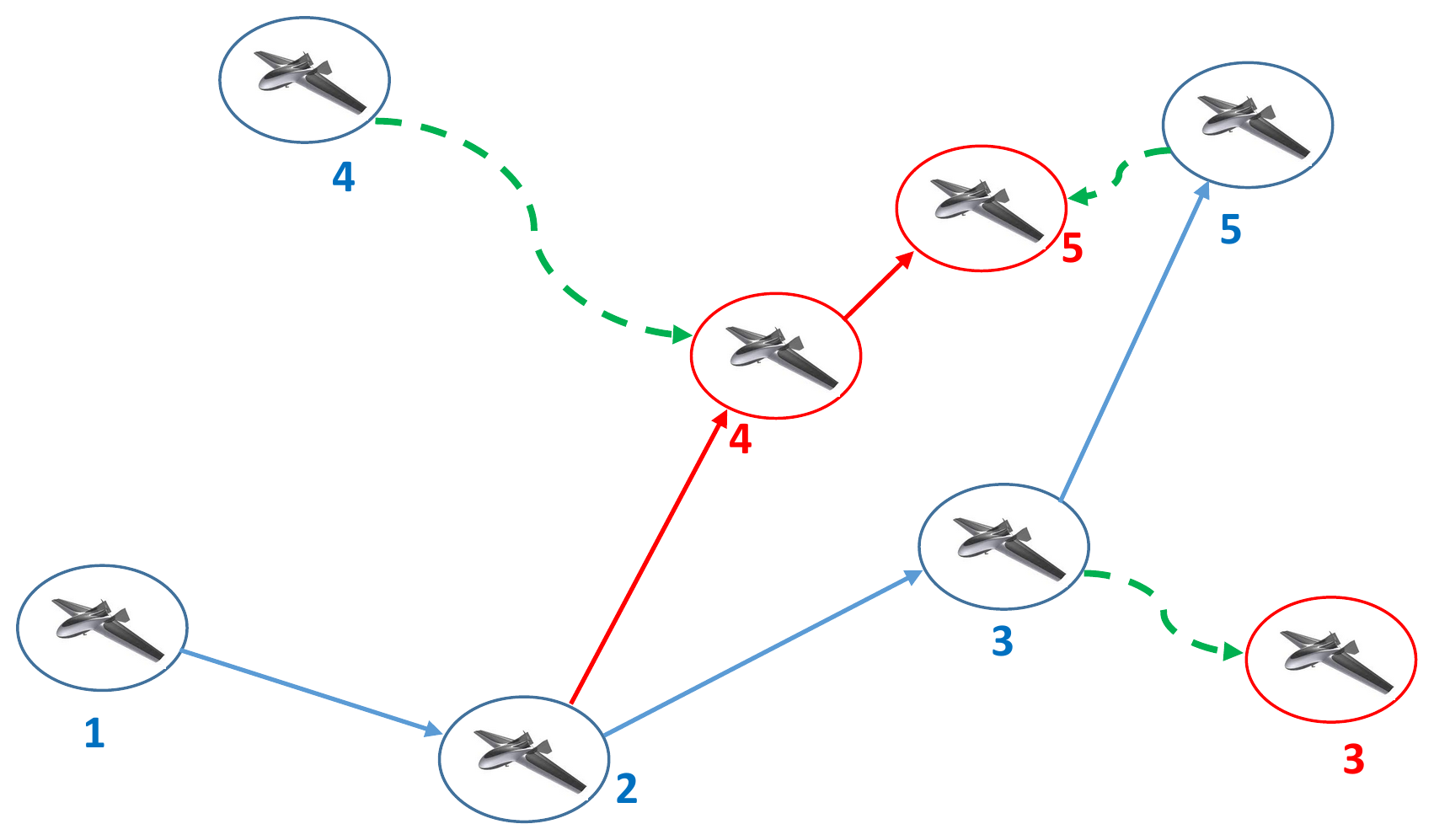}
	\caption{\footnotesize \arnau{Scenario showing the importance of predictive routing \cite{arnau2}}.}
     \label{predictive_routing}
  \end{figure}

\arnau{Figure \ref{predictive_routing} presents a scenario on the importance of including AI techniques for optimal routing. Node 1 intends to send a packet to node 5 through the optimal path. The edge metrics represent an arbitrary performance metric such as distance, energy consumption, delay, bitrate, or a combination of these metrics. 
Consider a network of freely moving UAVs that transmit their info through queued and delay-tolerant communication. The network topology can change substantially during the transmission session while the data packets are waiting in the intermediate nodes' transmission buffer. 
Therefore, the optimal path, if found by the source node based on the initial network topology using a typical shortest path algorithm, may not remain optimal throughout the transmission. 
In Figure \ref{predictive_routing}, the blue and red circles, respectively, present the original and the updated positions of the nodes (after motions shown by dashed green arrows). A conventional algorithm would determine (1-2-3-5) as the optimal path from source node $1$ to destination $5$ (represented by blue arrows) based on the original positions (blue circles). In contrast, a \textit{predictive routing} selects the path shown by red color (1-2-4-5), taking into account the predicted network topology change (i.e., the position of the nodes when met by the data packets), while the packet is waiting in the transmission buffer of node $2$. }

\section{UAV technology: military, industrial and commercial drones} \label{sec:uavtech}

Like many other technologies, the use of drones initiated in military domain, and soon afterwards, made its way into industrial and commercial application. One of the key motivating factors was using drones in risky environments, and inaccessible areas with harsh conditions to minimize human risk.

\begin{figure}[h]
 	\centering
	\includegraphics[width=1\columnwidth]{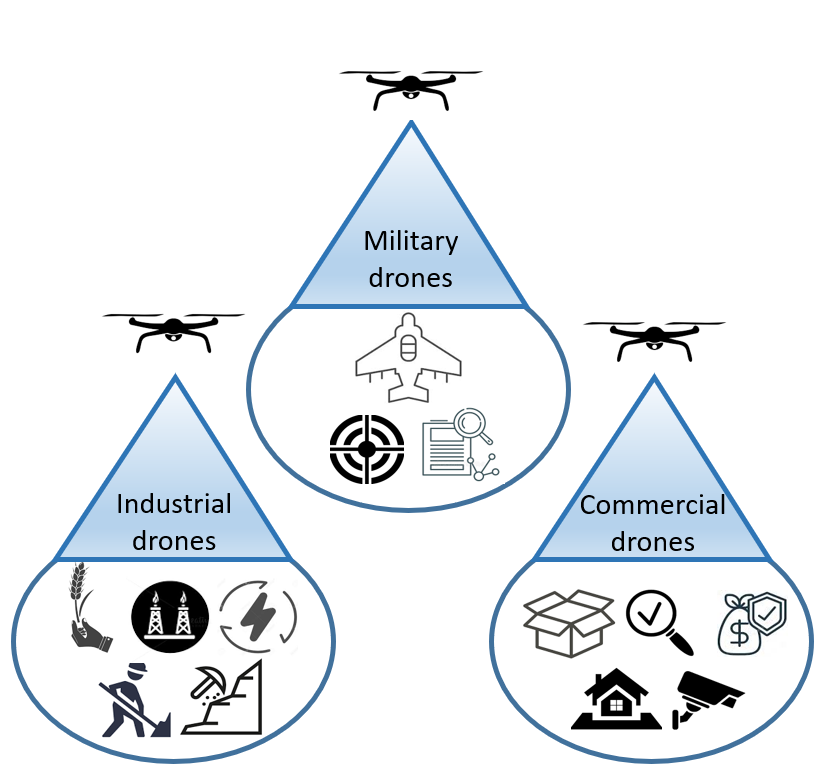}
	\caption{Military, industrial and commercial drones.}
    \label{uavtypes}
\end{figure}

Military-grade drones typically utilized advanced features (e.g., stealth), custom-built sensors, equipment, and weaponry, appropriate for military and reconnaissance missions in hostile environments \cite{militarydrones1}. 
For instance, hyperspectral and LiDAR sensors, AI-based object recognition, quantum cryptography, and multi-spectral targeting systems with infrared sensors are usually utilized in more advanced technology, compared to commercial products \cite{militarydrones3, militarywilson}.
For their proven performance and success in reducing casualties, the military sector invests heavily in the research and development of military drones \cite{militarydrones4}. 
UAVs in the military can be categorized based on their weight, range, speed, and specific capabilities. Based on their weight, we can differentiate class I, class II and class III drones. Class I refers to micro, mini, or small drones that weigh less than 150 kgs. Class II includes tactical drones between 150 and 600 kgs. Lastly, class III are strategic drones weighing more than 600 kgs \cite{militarydrones1}.

Drones in industrial settings are used in a broad range of applications such as smart agriculture, forestry, mining, construction, weather and climate control, power plants, structural monitoring (buildings, dams, and bridges) and the energy sector (oil and gas refineries). In addition to high definition cameras, industrial drones are typically equipped with different types of sensors, including but not limited to LiDAR, GPS, range finders for collision avoidance, PIN diodes for motion detection, and pressure gauges \cite{industrialdrone}. Environmental and climate sensors such as temperature, humidity, air pollution, and chemical sensors are also embodied in industrial drones when necessary.


More basic drones are also used by ordinary people for regular tasks, ranging from hobby and entertainment to more complicated monitoring tasks. The applications of commercial drones are countless and include shipping and delivery, inspection, real state, security, insurance, life habitat monitoring, border patrolling, structural monitoring, entertainment, sports monitoring, fire monitoring, flood prediction, smart agriculture, forest monitoring, volcano monitoring, fishery, weather report, etc. Some deliverables from drones include 3D maps, orthomosaic, and actionable reports \cite{commercialdrone}. Commercial drones must meet the regulations set by the Federal Aviation Administration (FAA) for safe operation. For instance, commercial drones should have a maximum weight of 55 pounds and should operate at or below 400 feet above the ground when in uncontrolled (Class G) airspace. Otherwise, specific authorizations should be obtained for flying in controlled airspace (Class B, C, D, and E) \cite{faa}.

Different types of UAV technology, including military, industrial, and commercial drones, are displayed in Figure \ref{uavtypes}.

\subsection{\arnau{AI features}}

\arnau{UAV technology has evolved in recent years. Modern UAVs are equipped with onboard computation boards powered by embedded circuitry, Graphical/Tensor Processing Units (GPU/TPU), and FPGA boards that allow running light-weight deep learning (DL) algorithms for AI applications \cite{AIchip3}. These AI chips, along with different sensors, allow UAVs to realize some levels of intelligence to improve performance in various applications, including those presented in Figure \ref{aiuavapp}. An alternative way is using UAVs with high payload capacity (e.g., xFold rigs Dragon X12 U11 drone with a payload size of 100 lbs) that can carry onboard huge computation servers. 
The motivation for embedding AI features into UAV technology comes from the importance of realizing low latency and fast processing of data for real-time applications. Therefore, using deep learning with AI-enabled chips by the UAVs can offer superior performance than streaming the raw data and running the applications on the ground-based processing centers \cite{AIchip}. This approach also substantially reduces the communication cost and satisfies the low-power requirements of UAVs compared to aggregating the information by the UAVs and running the applications on the cloud. Online processing also is desirable for real-time applications. However, this is not a general recipe for all applications. There exist some scenarios (e.g., low-cost single-hop communication for a single drone with relatively huge computation load), where raw video streaming and offloading the computation load to the cloud servers is advantageous \cite{AIchip1}. 
Regardless of the computation load distribution between the drone and the ground servers, the benefits of AI algorithms are achievable.}

\arnau{AI hardware in modern UAVs consists of computing, storage, and networking parts \cite{AIchip2}. Computing has been developing rapidly in recent years. However, storage and networking aspects still need more research to satisfy UAVs' diverse requirements. Particularly, there is a need for long-term storage and networking protocols for linking equipment to servers. Regarding the AI chip design, various technologies are available and under development, including GPU, TPU, reconfigurable Neural Processing Unit (NPU), neuromorphic chip architectures, and analog memory-based technologies. Based on the application-specific requirements and constraints, we can incorporate one or some of these designs. Lastly, we must acknowledge that producing AI hardware is a complex process \cite{AIchip4}. Consequently, many of the tech leaders, such as Apple, Google, Microsoft, Intel, IBM, Nvidia, etc., are competing to design and build the most innovative AI technology, and we expect to witness more breakthrough developments in the coming years \cite{AIchip5}.}

\subsection{Drone Manufacturers}

Drone market is expected to grow to \$63.6 billion by 2025, with 2.4 million global shipments by 2023, increasing at a 66.8\% Compound Annual Growth Rate (CAGR) \cite{dronemanufacturers}. 
A large portion of the commercial drone market share in the US (about 77\%) belongs to the DJI company. Intel, Vuneec, Parrot, GoPro, 3DR, HolyStone, Autel, SenseFly, and Kespry are among the top 10 drone market shares in the US \cite{dronemanufacturers1}.
Also, many companies use drones to provide 3rd party aerial solutions for different applications. For instance, PrecisionHawk is a North Carolina-based company that offers smart agriculture solutions \cite{precisionhawk}). Fortem Technologies offers AI-enabled interceptor drones that can hunt intruding drones \cite{dronetech}.
The number of these companies grows larger than 100, and a list of such companies can be found in \cite{uavcompanies}.
Commercial drones can offer service in many domains including emergency response, disaster relief, disease control, fighting crime, etc. \cite{dronetech1}.

\section{UAV networking protocols} \label{sec:networking}

\begin{figure}[h]
 	\centering
	\includegraphics[width=0.8\columnwidth]{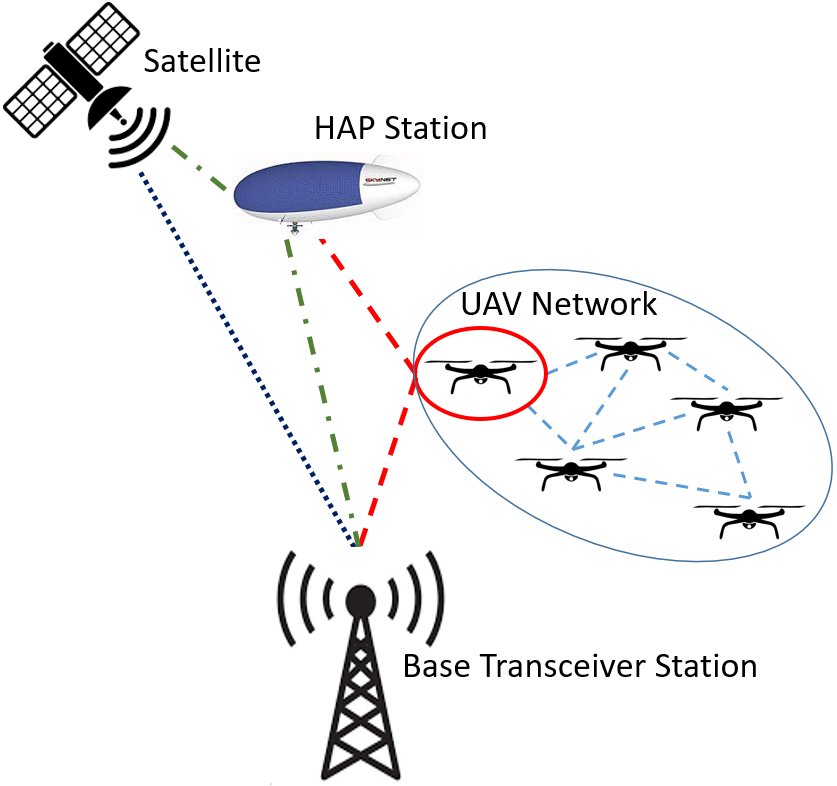}
	\caption{UAV Networking from different standpoints.}
    \label{fig:uav-links}
\end{figure}

\begin{figure*}[h]
 	\centering
	\includegraphics[width=1.5\columnwidth]{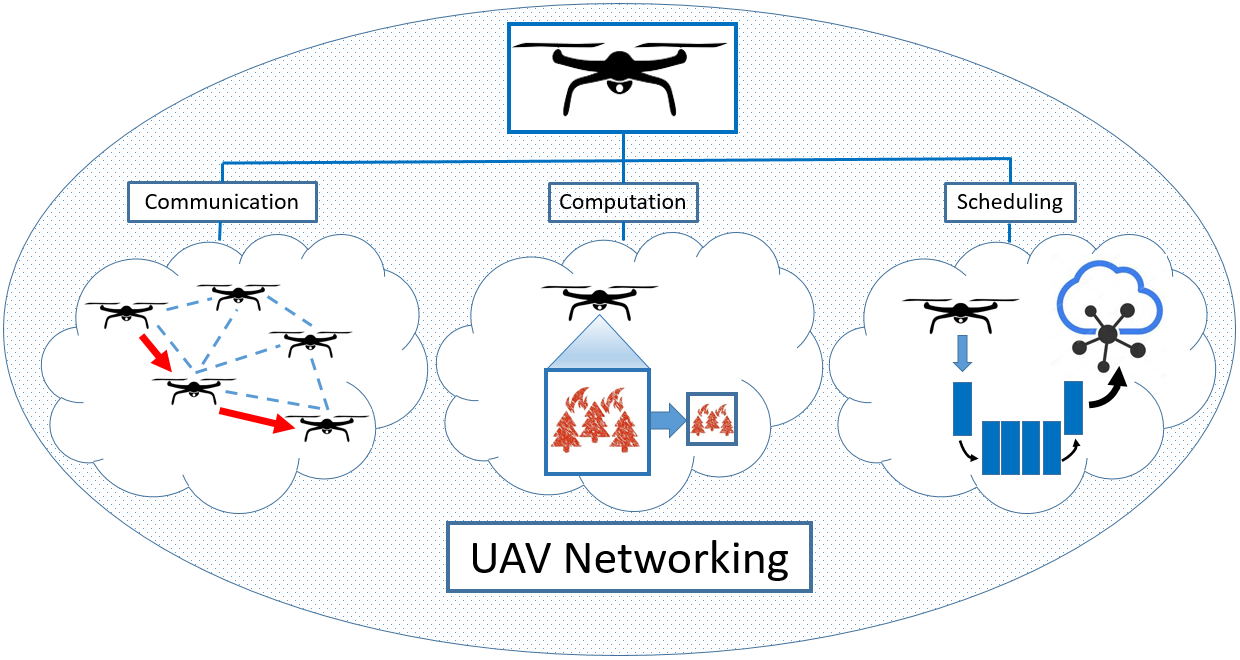}	
	\caption{UAV networking: different aspects.}
    \label{fig:uav-networking}
\end{figure*}

Different communication protocols can be used to transfer data among drones, to/from satellite and aerial control units through Air to Air (A2A) communications, and from Air to Ground (A2G) (Figure \ref{fig:uav-links}). Ground-based stations include standalone control units, larger servers, Internet gateways, and edge computers. 
As shown in Figure \ref{fig:uav-networking}, networking can be studied from different perspectives, especially from the communication, computation, and scheduling requirements and constraints. 
Most of the aerial monitoring platforms utilize ground-based or web-based servers for bulk processing, where drones collect and offload raw information into processing units. However, alternative methods such as on-the-fly processing using embedded light-weight GPUs/TPUs, Mobile Edge Computing (MEC), and fog computing for accelerated and near real-time processing are gaining more momentum recent years \cite{edgecomputing, fogcomputing}.

In most cases, the operation area is vast, far beyond the Line of Sight (LoS) communication range of a single UAV, hence using networked UAV platforms is unavoidable to ensure connectivity. 
One of the key design questions is choosing the best wireless technology (e.g., WiFi, LTE bands) with enough capacity and an acceptable Quality of Service (QoS) both from the technical and feasibility, and economic points of view. Some efforts have been made to create a nationwide and high-speed broadband wireless network for public safety communications. For example, FirstNet offers a solution to deploy, operate, maintain, and improve the first high-speed, nationwide wireless broadband network dedicated to public safety that could apply to UAV networks in the near future \cite{firstnet}.
The potential broadband wireless technologies include WiFi, 4G Long Term Evolution (LTE), 5G (with the 3rd Generation Partnership Project (3GPP) standard on 5G communication for drones), satellite communications, and dedicated public safety systems such as TETRA and APCO25.
Also, 
Long Range Wide Area Network (LoRaWAN), 
which enables long-range communications at low powers, when high throughput is not necessary 
\cite{lora, LORA1, LORA2}.

The following is a review of the most commonly used protocols in UAV networks, emphasizing their capability in handling the dynamicity of the network topology.

\subsection{Wi-Fi}
Most commercial UAVs use Wi-Fi (IEEE 802.11 series) for their communications, especially to a ground station (e.g., command and control commands in the uplink and video streaming in the downlink), as a low-cost, scalable, and affordable solution. Wi-Fi-based UAV networks can also be used for wireless backhauling \cite{backhauling}. Also, inter-drone communications can be powered by Wi-Fi provided that one node is defined as the Access Point (AP) to implement a local WLAN. This node may or may not provide access to the Internet. 
Apparently, one drawback of Wi-Fi is handling mobility and hand-off between the base stations, limiting the operation range of drones within direct access to the AP to a few miles. Although the throughput of Wi-Fi (theoretically between 54 Mbps for 802.11a to as high as 2.4 Gbps for 802.11ax \cite{intelwifi}) is relatively lower than LTE and 5G, it is sufficient for most applications, including real-time high-resolution video streaming. In scenarios that long-range connectivity is required, Wi-Fi loses the game to licensed wireless systems when such networks are available. Some drones develop their proprietary communication protocol on top of Wi-Fi. For instance, the XFold spy x8 KDE U3 drone by Xfold Rigs \cite{xfoldrig} comes with a Futaba Commercial \textcolor{blue}{14-channel.}



\subsection{LTE}
LTE systems offer airborne connectivity beyond the LoS communications for drones. They improve the throughput and network connectivity due to the hard and soft hand-over mechanism \cite{skylimit}. Recent years have witnessed a surge of activities in using terrestrial LTE networks to provide connectivity to UAVs. A collaborative project has been initiated by FAA and the National Aeronautics and Space Administration (NASA) in the U.S. since January 2017 to build a system using LTE technology. To better understand the potential of LTE for small UAVs, the 3GPP has formed a study group to investigate enhanced LTE support for aerial vehicles since March 2017 \cite{lte}.
The most notable drawback of using LTE and other cellular systems is the need to register drone transmitters with a service provider, which increases the operation cost and restricts the operation of drones to areas covered by the service provider. This is why using cellular systems is not as popular as Wi-Fi for drones.

Another key issue is that the LTE propagation plannings typically aim to serve the ground users; hence, the propagation maps are not optimized for aerial nodes. Therefore, LTE radio planning requires substantial revisions to serve UAV networks, especially when they scale up to large networks at higher altitudes and with a high-varying topology.



\subsection{5G}

Similar to LTE, 5G is also considered for drone communications when higher bitrates beyond 2.4 Gbps are required. It also enables the concept of \textit{Internet of Things} (IoT) for drones, where a drone serves as a \textit{thing} \cite{lagkas2018uav}.
UAV-assisted communications have several promising advantages such as facilitating on-demand deployment, high flexibility in network reconfiguration, and enabling long-range LoS communication links. 
In some scenarios, drones serve as 5G radio stations (also called AP) to extend the coverage of 5G networks for ground users, especially for sensitive applications, such as public safety and post-disaster management \cite{sekander2018multi,naqvi2018drone, selim2018post}. 




\subsection{\arnau{6G}}


\arnau{The demand for higher throughput and bigger numbers of devices never stops, and 6G is on the way to serve these requirements. 6G, the next generation for wireless communication, is expected to provide intelligent, secure, reliable, and limitless connectivity at rates 100 times faster than 5G \cite{6G1}. Similar to 5G, we expect that the UAV networks' diverse requirements such as low latency, reliability, and energy efficiency will be better served by 6G networks, and aerial nodes will be an integral part of 6G networks. Also, network intelligence is envisioned to be a key feature of 6G, which can assist in many connectivity-related applications \cite{6G2}, for example, in a blockchain-based solution for UAV communications \cite{6G3}. Some of the challenges brought by the futuristic concept of \textit{connected sky} include high mobility, interference, and connection to down-tilted antennas. It is expected that aerial nodes, when integrated into terrestrial nodes, will be instrumental in covering such issues and enhancing the 6G user experience.}


\subsection{\arnau{AI benefits in UAV networking}}


\arnau{Artificial intelligence for UAV networking can help with the reliability, connectivity, and security of wireless communication by offering data-driven solutions for key challenges of interference management, mobility management, and handover, cyber-physical attacks, and authentication \cite{AInetworking1}. For instance, \cite{AInetworking2} uses AI to predict transmission success and failures, to anticipate and avoid networking issues.}

Among the aforementioned communication protocols, the most appropriate one should be selected based on the application-specific constraints and requirements. Several research efforts have been devoted to implementing new networking algorithms on top of these protocols, most of which have not yet been commercialized. For instance, the idea of spectrum sharing and spectrum leasing for drones is proposed to extend the connectivity of drones for high-speed and temporary service when wireless coverage is accessible, which integrates WiFi and Cellular access \cite{shamsoshoara2020autonomous,shamsoshoara2019distributed}.
Also, beamforming can extend the communication range further and reduce the interference \cite{jalali2017beam}. 
Indeed, UAVs provide a realistic scenario for distributed and cooperative beamforming, since transmit antennas can be spatially separated among UAVs \cite{muralidharan2017energy}. The authors of this paper have proposed AI-enabled routing \cite{ArnauWiSEE}, compression \cite{arnaucompression}, and task coordination \cite{arnauscheduling} protocols to minimize the unnecessary information exchange among the UAVs and prolong their mission time.


\section{UAV swarms} \label{sec:swarms}

The concept of using UAV swarms is gaining more momentum in recent years. The idea is to use a sheer number of drones, in most cases miniaturized and limited-capability drones, to collectively perform a complicated mission with no or minimal operator intervention. This approach mitigates the drawbacks of using a single drone, such as limited allowable payload and limited sensing and actuation capabilities. A general architecture for task order in swarm environments is presented in \cite{swarm1}. 


\subsection{Swarming advantages}

The main advantages of using UAV swarms include shortening the task completion time, extending the coverage area, and reducing the operation cost. In the military domain, UAV swarms also increase the tactical mission's success rate by eliminating the reliance of the mission on a single drone's functionality, which can be subject to cyber-attacks and hijacking by an adversary. Using UAV swarms can also increase the unpredictability of the mission and overwhelm the enemy's defense system with a large number of potential targets in an interactive battleground. Further, UAV swarms are used to collectively find and fight enemy targets \cite{swarmmilitary}. These ideas were behind the US navy's LOCUST project to design UAV swarm attacks \cite{LOCUST}. Also, swarm systems can be equipped with anti-jamming systems to more efficiently block cyber attacks \cite{swarmantijamming}.
The use of UAV swarms is not limited to military applications. They enable search and rescue missions over big areas \cite{swarmsearch}. Some agriculture duties such as watering or identifying sick plants are time-consuming, and using UAV swarms with minimal operator intervention would increase the efficiency of precision agriculture \cite{swarm2}. 
More applications for UAV swarms can be found in \cite{swarmsurvey}.

    \begin{figure*}[t]
    \centering
    \subfloat[Infrastructure-based]{%
      \includegraphics[width=4.5cm]{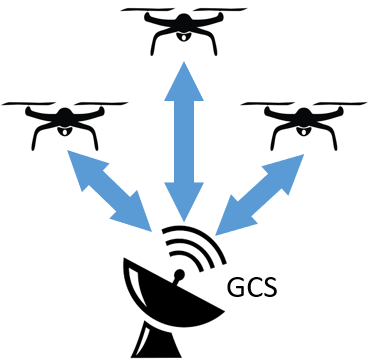}%
      \label{fig:uavswarms1}%
    }\qquad
    \subfloat[Ad-hoc structure-free]{%
      \includegraphics[width=4.5cm]{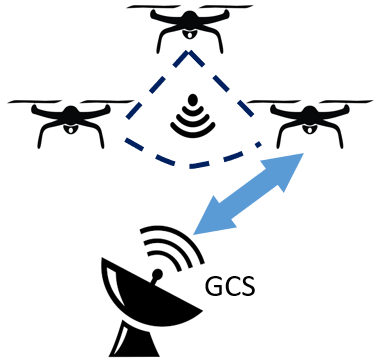}%
      \label{fig:uavswarms2}%
    }\qquad
    \subfloat[Hybrid architecture.]{%
      \includegraphics[width=4.5cm]{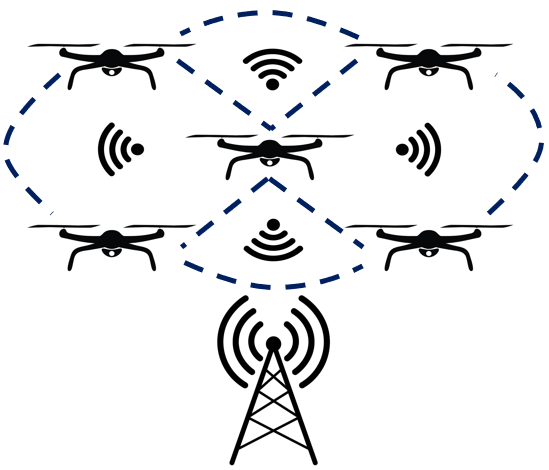}%
      \label{fig:uavswarms3}%
    }
    \caption{Different architectures for UAV swarms.}
    \label{fig:uavswarms}
    \end{figure*}

\subsection{Swarm architectures}


\arnau{UAV swarm refers to scenarios when a sheer number of drones, most of which with similar sensing, networking, and actuation capabilities, collectively perform a designated monitoring or operation tasks \cite{swarmarchitectures}. For example, a swarm of drones can carry out a coordinated military attack \cite{LOCUST}.} 
UAV swarms function in the following different ways from the control and coordination point of view.

\subsubsection{\arnau{Infrastructure-based swarm architecture}} 
In this architecture, a central node, mostly a \arnau{Ground Control Station (GCS)}, collects and processes real-time information from all swarm members; and sends back control commands (e.g., navigation, sensor actuation, sampling rate, camera control, etc.) to the UAVs, as shown in Figure \ref{fig:uavswarms1}. For instance, pre-planned paths can be revised during the mission to avoid collisions when facing unexpected obstacles based on the live video streaming by member UAVs \cite{Bekmezci}. The key advantages of this architecture include (i) feasibility of mission for low-capability UAVs by offloading the computation load to a more capable GCS, (ii) global optimality of the resulting decisions, and (iii) no need for complex networking algorithms for inter-UAV communications \cite{swarm5}. Also, realizing asymmetric security protocols such as Public Private Infrastructure (PKI), which requires Central Authority (CA) to allocate digital signatures is feasible \cite{vranics2019electronic}. However, it suffers from the well-known drawbacks of central systems such as (i) sensitivity to the GSC malfunction, hijack, or cyber threats, and (ii) the restriction of the mission area to the accessible range of the GCS.


\subsubsection{\arnau{Ad-hoc structure-free architecture}} 
This architecture allows direct communication between UAVs with no need for APs or routers, and can utilize distributed decision making (Figure \ref{fig:uavswarms2}). This method eliminates the sensitivity of the mission to the GCS function and relaxes the constraints on the mission area. The cost paid is the need for more capable UAVs for local decision making and a routing protocol to accommodate dynamic network topology \cite{swarm6,swarm7}.
    
\subsubsection{\arnau{Hybrid swarm architecture}} 
This architecture makes use of a cellular network to connect UAVs while using distributed decision-making without the need for a central controller (Figure \ref{fig:uavswarms3}). This architecture leverages the strengths of network-based and structure-free architectures by enabling long-range missions with reliable networking among drones while not relying on a central controller's function \cite{swarm2}. This architecture takes advantage of the high mobility and networking efficiency of M2M communication in LTE, and 5G wireless systems \cite{swarm8, swarm9, swarm10, swarm11}.

\subsection{\arnau{Artificial swarm intelligence}}

\arnau{Artificial swarm intelligence, also known as swarm AI, is a technology that combines real-time inputs and uses algorithms to optimize the overall performance of the swarm \cite{swarmAI1}.
The initial idea comes from the swarm intelligence found in natural systems, including ant colonies, bird flocking, animal herding, etc.}


\arnau{For UAV networks, it represents an aerial system that uses AI and data-driven methods to control the drones to achieve the designated goal \cite{swarmAI2}. In military drones, AI can be used to enhance UAV swarm operation by increasing the range, accuracy, mass, coordination, intelligence, and speed, with a potential impact on security and strategic stability \cite{swarmAI3}. Other examples include AI-based flight control for autonomous drones for real-time positioning without a centralized controller \cite{swarmAI4}, as well as flight control for detection, localization, and tracking tasks while relying only on local spatial, temporal, and electromagnetic information \cite{swarmAI5}. In the upcoming years, we expect to see how UAV swarm AI can revolutionize many existing systems and create a new breed of applications. A key concern is the negative impacts that may be brought by the excessive power of autonomous UAVs that can jeopardize people's privacy and security. This issue is more critical when drones' control units are hijacked by cyber attacks. Therefore, a hot research area is developing security and privacy-preserving protocols for next-generation UAV networks.}

\section{Impact of mobility on communications}  \label{sec:impactmobility}

In this section, we investigate the impact of node mobility on data transmissions. We first review popular UAV mobility models as well as the techniques used to predict UAV mobility and network topology. We also review the challenges that mobility brings to connectivity control and optimal routing.

\subsection{UAV mobility models}
\label{mobility_models}

Mobility models are used to describe, model, and emulate UAV motion trajectories. Generative modes typically incorporate location, speed, and direction changes as model parameters. The following is the list of commonly-used mobility models that facilitate the analysis of UAV networks.


\subsubsection{\arnau{Random WayPoint (RWP) \cite{camp2002survey,broch1998performance}}} 
This segment-wise model includes linear and independent motions with constant speed and direction between a set of points called waypoints. Also, UAVs decide on their next action based on some fixed probabilities, and their motion does not depend on neighbor nodes. According to \cite{WANG2010399}, RWP has two important variants, Random Walk Model (RWM) and Random Direction Model (RDM) \cite{yoon2003sound}. This model can be used for both rotary and fixed-wing drones, and is the most appropriate for missions with pre-path planning.

\subsubsection{\arnau{Levy process} \cite{gonzalez2008understanding}} 
This mobility model is similar to the random walk mobility model, with a distinction that the steps-lengths are not constant, rather random values that follow a power-law distribution. This mobility model is more appropriate for rotary drones.

\subsubsection{\arnau{Gauss Markov Mobility Model (GMMM) \cite{kumari2015survey}}} 
This model is used to simulate the movement of UAVs in swarms, which incorporates controlled randomness to the speed and direction equations. It prevents sudden stops and sharp turns within the simulation region, to realize smooth and more realistic trajectories, especially for fixed-wing UAVs, and targeted missions.

\subsubsection{\arnau{Semi random circular movement \cite{bekmezci2014connected,WANG2010399}}}
Used to simulate the curved movements of UAVs when they hover at a constant altitude (e.g., for collecting imagery). It has a uniform node spatial distribution and outperforms random waypoints (in terms of connectivity and scanning coverage over 2D disk). The nodes move around co-centered circles to cover destinations located on the circle perimeters. Uniform distributions are used for directional and angular velocities and the node pause at each destination. Once a circle scan is finished, the node switches to the next circle.

\subsubsection{\arnau{Mission Plan-Based mobility model (MPB) \cite{lee1999demand}}}
In this mobility model, UAVs are aware of trajectory information and move according to a predetermined path to the mission area, where potential information is available. Start and end points are randomly assigned, but the velocity and the flight time are given.
    
\subsubsection{\arnau{PaPaRaZzi Mobility model (PPRZM) \cite{6842277}}}
It is a stochastic mobility model that combines various models with a Markov state diagram. Each UAV chooses a movement type from a set of predefined motion patterns with different parameters. Each motion state's parameters are initialized randomly according to a given distribution and remain unchanged until the transition to another state with different motion parameters. A set of common motion patterns includes "Stay-at" "Way-Point", "Eight", "Scan", and "Oval", which are used in the original implementation of PPRZM. Results show that this model outperforms the RWP model in terms of geometric and network performance, since it brings the flexibility of switching between different modes.

Most of the aforementioned models consider independent motion trajectories for each drone. However, in some scenarios (e.g., UAV swarms), the motion of network members can be highly correlated. Several models are proposed to address this concept and realize correlated trajectories.

\subsubsection{\arnau{Distributed pheromone repel model \cite{kuiper2006mobility}}} 
This model uses a pheromone map to guide UAVs in reconnaissance scenarios. Each UAV maintains its own pheromone map, and scans the area of the corresponding map. The UAVs share information every few seconds to create a global pheromone map. UAVs turn right, left, or go straight with probabilities based on the pheromone smell. UAVs prefer areas with a low pheromone smell, so new areas are scanned.  This model results in good scanning properties, but does not consider the network connectivity between UAVs that serve different areas. 
    
\subsubsection{\arnau{Hybrid Mobility model with Pheromones (H3MP) \cite{hybrid_mobility}}}
This model is suitable for search and rescue applications. This hybrid mobility model combines Markov chains and pheromones to adapt to dynamic environments. Markov chains guide UAVs to promising areas, and pheromones guide information sharing that allows mobility management through UAVs. Results show the superiority of this method in detecting and tracking targets, compared to other pheromone-based methods.
    
\subsubsection{\arnau{UAV fleet mobility model \cite{connectivity}}} 
This mobility model incorporates the remaining energy level, the area coverage, and network connectivity into the mobility decision criterion. After receiving information from its neighbors, each UAV determines its next movement based on these criteria. The direction and the speed of the UAVs are calculated using weighted vectors considering neighbor UAVs. Results show that this method outperforms random motion methods in terms of coverage and connectivity.

These models clearly are more appropriate to design motion paths for specific missions, and less appropriate for modeling general UAVs networks.
Overall, a proper mobility model should be adopted for each application based on the drone types, the mission requirements, and the utilized path planning method.

\subsection{Mobility prediction}

Network nodes use radar-based and visual target tracking to perceive the network topology, at least in their close neighborhood. The purpose of mobility prediction is to go one step beyond and anticipate the future locations of the objects that form the network. These methods can be categorized into the following two mainstream trends, data-driven, and model-based methods.

\subsubsection{\arnau{Data-driven}}
This approach includes data mining and fuzzy methods, where frequent motion patterns are exploited by analyzing relatively large datasets. These methods indirectly capture the influence of the natural and human-made textures, user behavioral habits on the spatial and temporal variations of node mobility. 
For instance, \textit{TAPASCologne} is a project to collect and publish datasets of vehicle motion patterns in the city of Cologne, in Germany with application to cellular network design \cite{TAPASCologne1,TAPASCologne2}. Similar data-driven methods are proposed to model the motion patterns of pedestrians \cite{pedestrian-motion-modeling1999image,ossama2011extended,bennewitz2005learning}, vehicles \cite{vehicle-motion-modeling2009mobility,vehiclemotion2006learning}, animals \cite{lee2015unifying} and other mobile users. In a similar line of research, node mobilities are not exploited directly, rather traffic distribution trends are extracted \cite{larsen2007route,tan2016short}. Although these methods are useful in the network planning phase, they cannot be used for real-time networking decisions by the network nodes. 
    
\subsubsection{\arnau{Model-based}}
In this approach, the smoothness of the motion paths is used to predict the future locations of mobile objects,  typically in an online fashion. These methods include piecewise segment methods \cite{choi2006learning}, Hidden Markov Models (HMM)~\cite{bennewitz2005learning}, Levy flight process \cite{gonzalez2008understanding}, Bayesian methods \cite{aoude2011mobile}, Manifold learning \cite{lee2012identification}, Kalman filtering \cite{grewal2011kalman}, fuzzy zone-based method \cite{aoude2011mobile} and mixture Gaussian models \cite{aoude2011mobile}. Each method relies on a model that is appropriate for a different class of mobile objects, including pedestrians, indoor mobile users, vehicles, etc. Indeed, these methods rely on an underlying model, which is customized for a specific object with a different mobility model such as random walk \cite{lee2009slaw}, Random Waypoint (RWP) \cite{ming2015interference}, HMM~\cite{bennewitz2005learning}, Gaussian Morkov Mobility Model (GMMM) \cite{lu2005predictiveMobility}, Brownian motion \cite{groenevelt2006relaying}, Linear model via Durbin's curve \cite{owen2015implementing}, and mixture models \cite{wiest2012probabilistic}. Consequently, they are not applicable to a network of heterogeneous nodes and fail in balancing between the randomness and predictability of node's mobility.

\subsection{Mobility-related networking challenges }

High mobility nodes, especially when not properly predicted, pose critical challenges to the communication performance in terms of connectivity and routing optimality. Mobility of nodes translates to the network dynamicity that can disrupt the information exchange by losing connectivity and undermining the routing efficiency. In extreme cases, the network can breakdown into isolated islands. 
Different approaches can be taken to overcome the loss of connectivity in UAV networks. One main approach is \textit{topology control} to avoid connectivity issues, which can be seen as jointly optimizing the networking and control aspects. For instance, the idea of the dynamic placement of new nodes in locations to cover connectivity holes is introduced in \cite{chandrashekar2004providing}. ML methods can assist with achieving this goal by modeling and predicting network topology, traffic mobility, spectrum availability, and channel states. An online learning procedure is used in \cite{connectivity0} to adjust the UAVs to their radio transmission parameters, based on the perceived topology while revising their flight paths. The authors of \cite{connectivityAnt} propose a Chaotic Ant Colony Optimization approach (CACOC) to maximize the coverage area while preserving the connectivity. 
Another method to improve network connectivity is ECORA \cite{connectECORA}. This method uses geographic protocols considering positioning prediction and link expiration time by excluding links with approaching expiration time from the path selection algorithm. 
These methods usually aim to enhance network connectivity by controlling the network topology. Recently, the idea of using predictive and self-adaptive learning-based routing protocols gained a lot of attention to use ML methods to enhance networking efficiency. One approach is predicting network topology changes and incorporating the predicted network topology into the networking decisions  \cite{ArnauWiSEE,alsamhi2019convergence}. These methods are more appropriate for separating the networking layer from mission-based path-planning algorithms.

\section{Routing protocols for UAV networks}  \label{sec:routing}



This section reviews routing protocols, emphasizing the role of ML methods to accommodate the requirements of UAV networks.

\begin{figure*}
\centering
  \includegraphics[width=1.5\columnwidth]{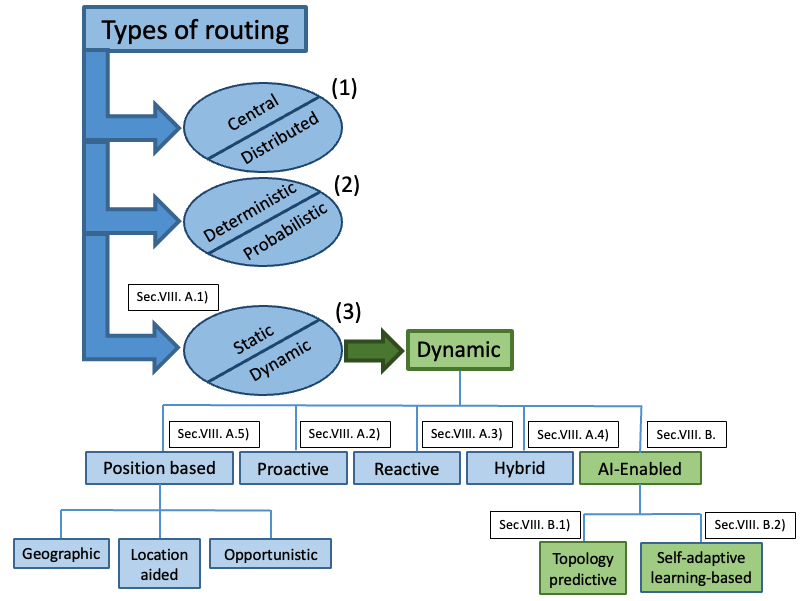}
\caption{\arnau{Routing classification based on different criteria and dynamic types of routing protocols.}}
  \label{typesrouting}
\end{figure*}

Routing protocols can be categorized in many different senses, as shown in Figure \ref{typesrouting}. 
Regardless of their reaction to topology changes, routing algorithms can be divided into centralized and distributed algorithms (Figure \ref{typesrouting}, layer 1). In centralized algorithms, the global network topology is known for a central node, which can share with network members. Therefore, the source node can execute a local algorithm to find the optimal end-to-end path. In distributed algorithms, the nodes are not aware of the entire network topology, and they have partial knowledge about their local subnetwork topology. In these methods, either sequential methods are used to break down finding the optimal end-to-end path into piece-wise sub-problems; or methods like "hello messages" are used to discover the network topology.

Routing algorithms can also be classified into deterministic and probabilistic algorithms based on the optimization approach or equivalently the resulting route's randomness (Figure \ref{typesrouting}, layer 2). Deterministic routing refers to a non-stochastic decision-making policy, where the resulting routes are fully and explicitly determined under given assumptions, network conditions, and decision rules. 
On the other hand, in probabilistic algorithms, the resulting routes are probabilistic; hence the actual paths are selected in run-time by the nodes based on a set of rules and probabilities.

From a different standpoint, we can classify routing protocols into static and dynamic routing protocols (Figure \ref{typesrouting}, layer 3). In static routing protocols, the route is established based on the initial network topology without considering the changes which occur during the transmission. These algorithms are appropriate for static networks and low-volume transmissions. On the other hand, in dynamic routing protocols, the resulting end-to-end path can change over time to accommodate node mobility. Therefore, they are more suitable for UAV networks and will be discussed in more detail in this paper. 
Dynamic routing algorithms have different variants based on how paths are determined in response to network topology changes. Main categories include proactive, reactive, hybrid, position-based, topology predictive, and self-adaptive learning-based routing methods.

To investigate the performance of different routing protocols in UAV networks, we first present the key characteristics of these routing methods in Table \ref{types_of_routing_table}. Next, we review different implementations of each category and investigate their use for highly dynamic UAV networks.

\begin{table*}[t]
\caption{Types of routing.}
\label{types_of_routing_table}
\centering
{\footnotesize
\begin{tabular}{|p{3.5cm} | p{1.4cm} | p{1.8cm} | p{1.95cm} | p{1.25cm}  | p{1.35cm} | p{2.25cm}| p{1.2cm}|} 
\hline
\rowcolor{gray}
\textbf{Routing protocol} & \textbf{Central / distributed} & \textbf{Deterministic / Probabilistic} & \textbf{Scalable} & \textbf{Mobility} & \textbf{Global info} & \textbf{Discovery message} & \arnau{\textbf{AI aspect}}\\
\hline
Static & Central & Deterministic & Yes & Static & Yes & No & \arnau{No}\\
\hline
Position-based & Distributed & Deterministic & Yes & Low speed & No & No & \arnau{No}\\
\hline
Proactive & Distributed & Deterministic & Bigger overhead & Low speed & Yes & Hello message & \arnau{No}\\
\hline
Reactive & Distributed & Deterministic & Longer RREQ & Low speed & No & RREQ \& RREP & \arnau{No}\\
\hline
Hybrid & Distributed & Deterministic & Complexity & Low speed & No & Hello message & \arnau{No}\\
\hline
Hierarchical & Distributed & Deterministic & Yes, into clusters & Low speed & Yes & Hello message & \arnau{No}\\
\hline
Probabilistic & Distributed & Probabilistic &  Adds complexity & Low speed & Yes & No & \arnau{No}\\ 
\hline
\rowcolor{lightgray}
\textbf{Topology predictive} & Distributed & Deterministic & Yes & Dynamic & No & Depends & \arnau{Yes}\\
\hline
\rowcolor{lightgray}
\textbf{Self-adaptive learning-based} & Distributed & Deterministic & Yes & Dynamic & No & Depends & \arnau{Yes}\\
\hline
\end{tabular}
}
\end{table*}

\subsection{Conventional routing protocols} 

In this section, we briefly review routing protocols that were mainly introduced for low-speed ad hoc networks. These routing protocols do not adapt to high mobility and abrupt changes we find in UAV networks. Therefore, most of them are not applicable for high-speed UAV applications.

\subsubsection{Static routing protocols}
Static protocols are mainly designed for networks with static or slow-varying topology, meaning that the optimal end-to-end path for any source-destination pair does not change over time. Static routing algorithms consider the initial network topology when finding the best path. 
Generally, in this approach, the global network topology is known to a central node (which can also be shared with network members). Therefore, the optimal paths for all source-destination nodes are calculated and programmed in terms of routing tables. In other words, each intermediate node receives a packet, passes the packet to the next node determined by the routing tables based on the destination address. 
\arnau{These algorithms are suitable for structured networks, but some modified versions are proposed for UAV networks. Examples of such algorithms include shortest path algorithm (Dijkstra's and Bellman-Ford's algorithms) \cite{dijkstra}, shortest-path-aided back-pressure \cite{backpressure}, Multi-Level Hierarchical Routing (MLHR) \cite{routingreview2014}, Load Carry and Deliver Routing (LCDR) \cite{routingreview2014}, and Data-Centric Routing \cite{routingreview2014}.}

\subsubsection{Proactive routing protocols}

Proactive routing protocols can be implemented as table-driven methods, where optimal paths are found for all source-destination pairs based on the global network topology. The routing tables are filled accordingly at all nodes to guide the packets link-by-link to their final destinations. 
In applications like UAV networks, routing tables should be updated if optimal paths are changed due to varying network topology or link budgets.
The main advantage of this approach is that the route can easily be calculated and established. In case of a link breakdown, new links can be re-established quickly. However, it may impose a large overhead for time-consuming topology exploration and path discovery, such as using hello packets to learn the network topology. \arnau{Some important implementations of proactive routing algorithms designed for UAV networks include Optimized Link State Routing (OLSR) \cite{OLSR}, Directional Optimized Link State Routing (DOLSR) \cite{DOLSR}, Multidimensional Perception and Energy Awareness OLSR (MPEAOLSR) \cite{MPEAOLSR}, Dynamic Dual Reinforcement Learning Routing (DDRLR) \cite{CoutinhoMMRSCK:15}, Destination Sequenced Distance Vector (DSDV) \cite{DSDV}, BABEL \cite{babel}, Cluster head Gateway Switch Routing (CGSR) \cite{GCSR}, Wireless Routing Protocol (WRP) \cite{WRP}, Topology Broadcast based on Reverse Path Forwarding (TBRPF) \cite{TBRPF} and Better Approach To Mobile Ad hoc Network (BATMAN) \cite{batman}}.

\subsubsection{Reactive routing protocols}

Reactive routing protocols create on-demand routing information. It means that the route discovery process is executed only when a transmission session has to be established. The main benefit of this approach is its reduced overhead, especially in the low-traffic regime. On the other hand, in case of a route failure, the re-establishment of a new route can take a long time. \arnau{The following are some important implementations of reactive routing protocols: Dynamic Source Routing (DSR) \cite{DSR}, Ad hoc On Demand Distance Vector (AODV) \cite{AODV}, Dynamic Topology-Multipath AODV (DT-MAODV) \cite{DTAODV}, Associativity-Based Routing (ABR) \cite{ABR}, Signal Stability-based Adaptive routing (SSA) \cite{SSA}, Message Priority and Fast Routing (MPFR) \cite{MPFR}, Dynamic Backup Routes Routing Protocol (DBR2P) \cite{DBR2P}, Dynamic MANET On-demand (DYMO) \cite{DYMO} and Time Slotted On-demand Routing (TSOR) \cite{TSOR}.}

\subsubsection{Hybrid routing protocols}

Hybrid routing protocols combine proactive and reactive routing features. The route is initially determined with a proactive protocol. However, a reactive routing protocol is activated when a substantial network topology is recognized or a previously established route breaks. \arnau{Some important implementations of hybrid routing algorithms designed for UAV networks are Zone Routing Protocol (ZRP) \cite{ZRP} and Temporarily Ordered Routing Algorithm (TORA) \cite{TORA}.}

\subsubsection{Position-based routing protocols}

Position-based routing protocols find the optimal route based on the location information. For example, the next node can be selected based on its distance to the current node or to the destination. The key disadvantage of these methods are their dependence on accurate positioning and tracking systems. \arnau{Some important implementations of position-based routing algorithms proposed for UAV networks include: Greedy Perimeter Stateless Routing  (GPSR) \cite{GPSR}, Greedy-Hull-Greedy (GHG) \cite{GHG}, Greedy-Random-Greedy (GRG) \cite{GRG}, greedy forwarding \cite{greedyfor}, Energy-Balanced Greedy forwarding Routing (EBGR) \cite{EBGR}, Greedy Distributed Spanning Tree Routing (GDSTR) \cite{GDSTR}, Cross-layer Link quality and Geographical-aware beaconless Opportunistic routing (Xlingo) \cite{xlingo}, Adaptive Forwarding Protocol (AFP) \cite{AFP}, Reactive-Greedy-Reactive (RGR) \cite{RGR}, scoped flooding and delayed route request RGR \cite{imprvRGR}, beaconless opportunistic routing \cite{Beaconless}, Location-Oriented Directional MAC (LODMAC) \cite{LODMAC}, Extremely Opportunistic Routing (ExOR) \cite{exor} and Location-Aided Routing (LAR) \cite{LAR}.}

\subsubsection{Hierarchical routing protocols}
Hierarchical routing protocols consider nodes arranged hierarchically, where the lower layers can form clusters. Each node typically holds information only about its neighbors stored in a table that is updated through hello packets. Each cluster head communicates with the rest of the cluster heads to select the best path. \arnau{Cluster-Based Routing Protocol (CBRP) \cite{cbrp}, Enhanced Cluster head Gateway Switch Routing (ECGSR) \cite{ecgsr} and Fisheye State Routing (FSR) \cite{FSR1, FSR2} are examples of recently-developed algorithms for UAV networks.}

\subsubsection{Probabilistic routing protocols}
Probabilistic routing protocols find multiple routes from source to destination, which can be selected based on probabilistic mechanisms to cope with network congestion and security. \arnau{Some examples of these algorithms used in UAV networks are random walk routing \cite{randomwalk} and MIMO-based random walk routing \cite{mimorandom}.}

\subsection{AI-enabled routing protocols}

In this section, we study the AI-enabled routing protocols, which use the learning power of ML algorithms for optimal route path selection based on a more accurate perception of the network topology, channel status, user behavior, traffic mobility, etc. These algorithms bridge the two networking and AI research areas to implement modern networking, especially for dynamic UAV networks. These algorithms can be viewed as state of the art and are not included in most previous survey papers. The following is a fairly comprehensive list of AI-based algorithms.

\subsubsection{Topology predictive routing protocols}
The main feature of topology-predictive routing protocols is using ML algorithms to predict the node's motion trajectories (as an approximate of the network topology, if the communication range of nodes is known) and incorporate them into the path selection mechanism. 

Here, we review some of the proposed routing protocols that use mobility or trajectory prediction approaches to enhance the performance of routing algorithms for UAV networks.

\begin{table*}[t]
\centering
\caption{Performance comparison table for topology predictive routing protocols.}
\label{performance_table_predictive}
{\footnotesize
\begin{tabular}{| P{2.7cm} | P{0.75cm} | P{0.75cm} | P{1.4cm} | P{1.4cm} | P{9cm} |}
\hline
\centering
\multirow{2}{*}{\textbf{Algorithm}} &\multicolumn{4}{c|}{\textbf{Objective performance}} & \multirow{2}{*}{\textbf{Results}}\\ \cline{2-5}
 & \textbf{Energy} & \textbf{Delay} & \textbf{Throughput} & \textbf{Connectivity} & \\ 
\hline
PPMAC+RLSRP \cite{PPMAC} & & \checkmark & \checkmark & \checkmark & Compared to DMAC, LODMAC, OLSR, GPMOR, RARP, IMAC+OLSR:
\begin{itemize}
    \item Lowest network delay and longest path lifetime
    \item Highest route setup success and data delivery ratio.
    \item Better successful throughout without retransmissions.
\end{itemize} \\ \cline{1-6}
Predictive Dijkstra's \cite{ArnauWiSEE, arnau2} & & \checkmark & & & Compared to conventional Dijkstra's algorithm: \begin{itemize}
                                        \item Up to 25\% decrease in end-to-end delay for 100 nodes.
                                    \end{itemize} \\ \cline{1-6}
Predictive greedy routing \cite{greedySECON} & \checkmark & & \checkmark & & Compared to conventional Dijkstra's and static greedy:\begin{itemize}
                                        \item Higher probability of success (up to 100\%).
                                        \item Energy consumption reduction (up to 30\%). \end{itemize} \\ \cline{1-6}
P-OLSR \cite{rosati1, rosati2} & & & \checkmark & \checkmark & Compared to OLSR and BABEL: \begin{itemize}
                                        \item Cuts down the outage time by at least 85\%.
                                        \item Achieves a more stable goodput. \end{itemize} \\ \cline{1-6}
GPMOR \cite{GPMOR} & & \checkmark & \checkmark & & Compared to GPSR and GLSR: \begin{itemize}
                                        \item Better packet delivery ratio (up to 250\%).
                                        \item Lower average delay of the network (up to 50\%). \end{itemize} \\ \cline{1-6}
MPCA \cite{MPC} & & & \checkmark & \checkmark & Compared with LID, HD and WCA: \begin{itemize}
                                        \item Up to 500\% increase in clusterhead duration.
                                        \item Up to 70\% lower reaffiliation frequency. \end{itemize}\\ \cline{1-6}
RARP \cite{RARP} & & & \checkmark & \checkmark & Compared to conventional AODV: \begin{itemize}
                                        \item Up to 30\% increase in route setup success rate.
                                        \item Higher average path lifetime. \end{itemize}\\ \cline{1-6}
SFMPRGR \cite{imprvRGR2} & & \checkmark & \checkmark & & Compared to AODV, RGR, and MPRGR: \begin{itemize}
                                        \item 10\% increase in packet delivery ratio.
                                        \item 20\% reduction end-to-end delay. \end{itemize}\\ \cline{1-6}
QGeo \cite{QGeo} & & \checkmark & \checkmark & & Compared to GPSR and QGrid: \begin{itemize}
                                        \item Up to 50\% increase in packet delivery ratio
                                        \item 20\% reduction in end-to-end delay. \end{itemize}\\  \cline{1-6}
PARRoT \cite{parrot} & & \checkmark & \checkmark & & Compared to AODV, OLSR, GPSR and B.A.T.M.A.N: \begin{itemize}
                                        \item Higher packet delivery ratio (at least by 45\%).
                                        \item Lower end-to-end latency \end{itemize}\\ 
                                        \cline{1-6}
FLRLR \cite{fuzzy2} & & \checkmark & & \checkmark & Compared to Fuzzy Logic and ACO: \begin{itemize}
                                        \item Up to 20\% lower average number of hops.
                                        \item Around 30\% higher connectivity. \end{itemize}\\ 
\hline
\end{tabular}
}
\end{table*}

\begin{itemize}
    \item \textbf{Learning-based Adaptive Position MAC protocol} \cite{PPMAC}: 
    This routing protocol proposes an adaptive hybrid communication protocol by integrating a novel Position-Prediction-based directional MAC protocol (PPMAC) and a Self-learning Routing Protocol based on Reinforcement Learning (RLSRP). The performance results show that the proposed PPMAC overcomes the directional deafness problem, which happens when the transmitter fails to communicate with the receiver due to having the receiver's antenna oriented in a different direction. Also, RLSRP provides an automatically evolving and more effective routing scheme, appropriate for autonomous FANETs.
    
    \item \textbf{Predictive Dijkstra's} \cite{ArnauWiSEE, arnau2}: This routing protocol assumes that the intermediate nodes' locations when the packet is supposed to meet them are predicted using ML methods. Then, it incorporates this predictive information into the path selection criterion based on Dijkstra's shortest path algorithm. Results show superior performance compared to the standard Dijkstra's algorithm, especially when higher velocities are applied. The achieved performance gain is dependent upon the prediction accuracy. Two important shortcomings of this method is its reliance to accurate trajectory prediction methods, and the need for global location information exchange.  
    
    \item \textbf{Predictive greedy routing} \cite{greedySECON}: Distance-based greedy routing algorithm solves the issues of \cite{ArnauWiSEE} and relies solely on the UAVs' local observations of their surrounding subnetwork. Each node estimates the location of its neighbors (e.g., using model-based object motion trajectory prediction)  and selects the next node that makes maximum progress toward the destination node. 
    This method adapts to highly dynamic network topology. Moreover, it has low complexity and low overhead with no need for an initial route setup. Simulation results show considerable improvement, compared to centralized shortest path routing algorithms.
    
    \item \textbf{Predictive Optimized Link State Routing (P-OLSR)} \cite{rosati1, rosati2}: This routing protocol is an advanced version of OLSR routing protocol. This algorithm exploits GPS information and calculates an Expected Transmission (ETX) count metric to estimate the quality of the link when finding the optimal path. Numerical results show that the P-OLSR outperforms other algorithms such as OLSR and BABEL under dynamic network topology.

     \item \textbf{Geographic Position Mobility Oriented Routing (GPMOR)} \cite{GPMOR}: This routing protocol proposes an efficient and effective geographic-based routing protocol that uses the Gauss-Markov mobility model to predict the movement of UAVs to eliminate the impact of highly dynamic movements. This approach selects the next hop according to the mobility relationship in addition to the Euclidean distance to make more accurate decisions. Experiment results show that this approach provides effective and accurate data forwarding solutions. Moreover, it decreases the impact of intermittent connectivity and achieves a better latency and packet delay rate than other position-based routing protocols.
    
    \item \textbf{Mobility Prediction Clustering Algorithm (MPCA)} \cite{MPC}: This routing algorithm is appropriate for clustered UAV networks. It finds the highest node reliability to select the cluster head. Then, it predicts the network topology using the Trie data structure dictionary prediction and link expiration time mobility model. Also, this approach ensures the stability of cluster formation.
    
    \item \textbf{Robust and Reliable Predictive (RARP)} \cite{RARP}: This routing protocol combines omnidirectional and directional transmission schemes with dynamic angle adjustment. This method features a hybrid use of unicasting and geocasting routing protocols using the location and trajectory information. The intermediate node locations are predicted using 3-D estimation; then, directional transmission is used toward the predicted location, enabling a longer transmission range and tracking topology changes. The authors show that their method reduces path re-establishment and service disruption time and achieves higher successful packet delivery rates. 
    
   \item \textbf{Scoped Flooding and Mobility Prediction-based RGR (SFMPRGR)} \cite{imprvRGR2}:  This algorithm is a modified version of RGR. This method associates with data packets mobility prediction information, including velocity, direction, and timestamp, to compute the distance between the current node and its neighbors. Then, if the next hop is out of range, the approach directly switches to GGF to save dropped data packets, making it a better approach for dynamic networks.
   
   \item \textbf{Q-learning-based Geographic adhoc routing protocol (QGeo)} \cite{QGeo}: This is an ML-based geographic routing scheme to reduce network overhead in high-mobility scenarios. The basic idea is that nodes make geographic routing decisions distributively, utilizing a reinforcement learning method without knowing the entire network topology. It consists of location estimation, a neighbor table, and a Q-learning module. The location estimation module updates the current location information reported by the GPS or other localization methods. Their results show that QGeo provides a higher packet delivery rate and a lower network overhead than previously reported routing protocols.

    \item \textbf{Predictive Ad-hoc Routing fueled by Reinforcement learning and Trajectory Knowledge (PARRoT)} \cite{parrot}: This is another ML-powered routing protocol, which exploits mobility control information for integrating knowledge about the future motion of the mobile agents into the routing process. Each agent estimates its own future position based on the current position and propagates the result to other nodes. This algorithm achieves higher robustness and a significantly lower end-to-end latency compared to similar algorithms previously reported. 
    This algorithm is appropriate for separating networking from path planning layer or when the paths are planned on the fly, since the nodes are assumed to be unaware of their future locations.
    
    \item \textbf{Fuzzy Logic Reinforcement Learning-Based Routing Algorithm (FLRLR)} \cite{fuzzy2}: 
    This algorithm uses fuzzy logic to determine the neighbor nodes in real-time. Then, by using the future reward method of reinforcement learning, this method reduces the average number of hops through continuous training. Simulation results show lower average numbers of hops and high link connectivity, compared to the Ant Colony Optimization (ACO) algorithm.
 
\end{itemize}

\begin{table*}[t]
\centering
\caption{Performance comparison table for self-adaptive learning-based routing protocols.}
\label{performance_table_selflearning}
{\footnotesize
\begin{tabular}{| P{2.7cm} | P{0.75cm} | P{0.75cm} | P{1.4cm} | P{9cm} |}
\hline
\centering
\multirow{2}{*}{\textbf{Algorithm}} &\multicolumn{3}{c|}{\textbf{Objective performance}} & \multirow{2}{*}{\textbf{Results}}\\\cline{2-4}
 & \textbf{Energy} & \textbf{Delay} & \textbf{Throughput} & \\
\hline
Adaptive Q-Routing Full-echo \cite{adap_q_routing} & & \checkmark & & In comparison to Q-Routing and Dual Q-Routing: \begin{itemize}
    \item Smaller average delivery time.
\end{itemize}  \\ \cline{1-5}
AQRERM \cite{aqrerm} & & \checkmark & & Compared to Q-Routing, DRQ-Routing and AQFE: \begin{itemize}
                                        \item Better overshoot and settling time of the learning.
                                        \item Lower average delivery time. \end{itemize} \\ \cline{1-5}
PBQ-Routing \cite{pbQ-Routing} & \checkmark & \checkmark & \checkmark & Compared to Q‐Routing, Epi-R, PRoPHET and HBPR: \begin{itemize}
                                        \item Delivery probability almost gets doubled.
                                        \item Lower energy use and overhead reduced to half.\end{itemize} \\ \cline{1-5}
Q$^2$-Routing \cite{q2routing} & & \checkmark & \checkmark & Compared to AODV and EQ-Routing: \begin{itemize}
                                        \item Lower packet delay.
                                        \item Higher packet success ratio.
                                        \end{itemize} \\ \cline{1-5}
DQ-Routing \cite{dQ-Routing} & & & \checkmark & Compared to Q-Routing: \begin{itemize}
                                        \item Much higher average reward of DQ-routing.
                                        \item More likely to choose best action.
                                        \end{itemize}\\ \cline{1-5}
QNGPSR \cite{QNGPSR} & & \checkmark & \checkmark & Compared to OLSR, AODV and GPSR: \begin{itemize}
                                        \item Up to 65\% lower end-to-end delay.
                                        \item Up to 35\% higher packet delivery ratio.
                                        \end{itemize}\\ \cline{1-5}
ARdeep \cite{ardeep} & & \checkmark & \checkmark & Compared to QGeo and conventional GPSR: \begin{itemize}
                                        \item Up to 30\% higher packet delivery ratio.
                                        \item Up to 30\% lower average end-to-end delay.
                                        \end{itemize}\\ \cline{1-5}
TQNGPSR \cite{TQNGPSR} & & \checkmark & \checkmark & Compared with AODV, OLSR, GPSR, and QNGPSR: \begin{itemize}
                                        \item Outperforms in terms of the packet delivery ratio, end-to-end delay, and throughput.
                                        \end{itemize}\\ \cline{1-5}
QMR \cite{QMR} & \checkmark & \checkmark & \checkmark & Compared to other Q-learning based routing methods: \begin{itemize}
                                        \item Higher packet arrival ratio, lower delay and energy consumption.
                                        \end{itemize}\\ \cline{1-5}
QLFLMOR \cite{fuzzy1} & \checkmark & \checkmark & \checkmark & Compared to conventional fuzzy logic and Q-value-based AODV: \begin{itemize}
                                        \item Lower hop count and energy consumption and longer network lifetime.
                                        \end{itemize}\\ \cline{1-5}
QAGR \cite{fuzzy3} &  & \checkmark & \checkmark & Compared to ARPRL \cite{ARPL}, U2RV \cite{U2RV}, GPSR and AODV: \begin{itemize}
                                        \item Up to 300\% higher in packet delivering ratio.
                                        \item Around 50\% lower end-to-end delay and up to 90\% reduction in average number of hops.
                                        \end{itemize}\\ \cline{1-5}
FESAIQ-Routing \cite{rovirasugranes2021fullyechoed} & \checkmark & & \checkmark & Compared to Q-R, REE-R, PE-R, AFEQ-R and SAHQ-R: \begin{itemize}
                                        \item Reduction in energy consumption between 7\% to 82\%.
                                        \item Increase of up to 264\% in packet delivery ratio.
                                        \end{itemize}\\
\hline
\end{tabular}
}
\end{table*}

\subsubsection{Self-adaptive learning-based routing protocols}

Most learning-based routing protocols use Reinforcement Learning (RL) to make routing decisions by continued and online learning of the environment and their decision consequences on desired performance metrics such as delay, throughput, energy efficiency, and fairness. A key advantage of RL-based algorithms is their abstract formulation which brings independence from topology prediction and channel estimation. This comes from the concept of learning from experience.

\begin{figure}[h]
    \centering
	\includegraphics[width=0.6\columnwidth]{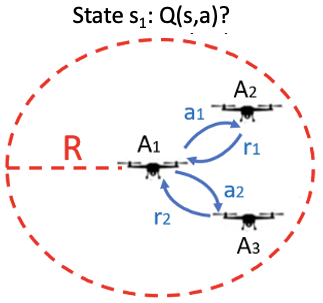}
    \caption{\arnautwo{Illustration of the RL-based routing \cite{rovirasugranes2021fullyechoed}}}
     \label{fig:rl_routing}
\end{figure}

\arnautwo{The concept of RL for optimized routing is shown in Figure \ref{fig:rl_routing}. Initially, the scenario is represented by state $s_1$, where node or agent $A_1$ has two candidate neighbors $A_2$ and $A_3$ to send its packet. Consequently, we must select between action $a_1$ or $a_2$ based on the reward expected for each action $a$ at state $s$, defined as $Q(s,a)$. Once we select the appropriate action, the agent $A_1$ obtains an immediate reward from the environment, $r_1$ or $r_2$, respectively. Next, we start the same process in a new state $s_2$, where decisions are made based on the new environmental conditions and the learned policy in terms of actions-rewards relations. The end goal is to find an optimal policy in which the cumulative reward over time is maximized by assigning optimal actions to each state \cite{reinforcementlearning}.}

RL-based routing was first introduced in \cite{q_routing}, where Q-Routing considered packet forwarding as an application of Q-learning. This method demonstrated superior performance compared to a non-adaptive algorithm based on the pre-computed shortest paths \cite{network_qrouting}.
The essence of Q-Routing is gauging the impact of routing strategies on a desired performance metric by investigating different paths in the \textit{exploration} phase, and using the discovered best paths in the \textit{exploitation} phase.
Exploration imposes an overhead to the system, but is critical for finding newly emerged optimal paths, especially when the network topology undergoes substantial changes. An inherent challenge is to adaptively solve the trade-off between the exploration and exploitation times to accommodate the dynamicity of the network topology.


The following is a summary of learning-based routing protocols mainly based on Q-Routing to show they evolved over time to better serve dynamic UAV networks.

\begin{itemize}
    \item \textbf{Q-Routing} \cite{q_routing}: The first proposed Q-Routing protocol operates based on learning from experience. Each node stores the expected time to the destination through any of its neighbors as Q-values in a Q-table.  Each node selects the next node that minimizes the expected travel time to the destination. Once a packet is received by a node, it sends back the real travel time to the previous node to updates its Q-values for the next round. 
        

    \item \textbf{Predictive Q-Routing (PQ-Routing)} \cite{pred_q_routing}: This method is an extension of the conventional Q-Routing that addresses the exploration-exploitation trade-off and fine-tunes the routing policies for the low network loads. Their approach was based on learning and storing new optimal policies under decreasing load conditions and reusing the learned best experiences by predicting the traffic trend. Their idea was to re-investigate the paths that remain unused for a while due to the congestion-related delays. The probing frequency is an adjustable parameter to be tuned based on the path recovery rate estimate. Their proposed results showed that PQ-Routing outperformed the Q-Routing in terms of both learning speed and adaptability. However, PQ-Routing requires large memory for the recovery rate estimation. Also, it is not accurate in estimating the recovery rate under varying topology change rates (e.g., when nodes start to moving faster). Furthermore, this method only works for delays arising from the queuing congestion and not delays coming from the network topology change. 
    
    \item \textbf{Dual Reinforcement Q-Routing (DRQ-Routing)} \cite{drq_routing}: The key idea of this algorithm is to use forward and backward explorations by the sender and receiver of each communication hop, by appending information to the packets they receive from their neighbors. Simulation results prove that this method learns the optimal policy more than twice faster than the standard Q-Routing. A comparative analysis of learning-based routing algorithms is provided in \cite{comparison}, where the performance of the self-adaptive Q-Routing and dual reinforcement Q-Routing algorithms are compared against the conventional shortest path algorithms. Their results showed that the Q-learning approach performs better than the traditional non-adaptive approach under scenarios with increasing traffic that causes node and link failures. However, Q-Routing does not always guarantee finding the shortest path and does not explore multiple forwarding options in parallel.
    
    \item \textbf{Credence-based Q-Routing (CrQ-Routing) and Probabilistic Credence-based Q-Routing (PCrQ-Routing)} \cite{crq_routing}: These two methods dynamically capture the traffic congestion to improve the learning process to select less congested paths. Both methods adapt to the current network conditions much faster than the conventional Q-Routing.
    
    \item \textbf{Full-echo Q-Routing} \cite{q_routing}: Another technique proposed to accelerate the learning speed of conventional Q-Routing is the \textit{full-echo} approach. In conventional Q-Routing, each node only updates the Q-values for its selection (the best neighbor). In contrast, in the \textit{full-echo} routing, a node gets each neighbor's estimate of the total time to the destination, which helps update the Q-values accordingly for each of the neighbors. 
    
    \item \textbf{Full-echo Q-Routing with Adaptive Learning Rate} \cite{adap_q_routing}: A more recent work added adaptive learning rates to the \textit{full-echo} Q-Routing to improve the exploration performance. Results show that this technique reduces the oscillations of the \textit{full-echo} Q-Routing for high load scenarios.
    
    \item \textbf{Adaptive Q-Routing with Random Echo and Route Memory (AQRERM)} \cite{aqrerm}: An extension of the previous work is AQRERM, which improves the performance of the baseline method in terms of the overshoot and settling time of the learning process, as well as the learning stability.
    
    \item \textbf{Poisson's probability-based Q-Routing (PBQ-Routing)} \cite{pbQ-Routing}: This approach uses forwarding probability and Poisson's probability for decision making and controlling transmission energy for intermittently connected networks. Results show that the delivery probability is almost twice bigger than for Q-Routing, and the overhead ratio is reduced to half.
    
    \item \textbf{Delayed Q-Routing (DQ-Routing)} \cite{dQ-Routing}: This routing protocol uses two sets of value functions to carry out random delayed updates to reduce the overestimation of the value function and improve the rate of convergence. Experiments show that this method also improves the learning rate.
    
    \item \textbf{QoS-aware Q-Routing (Q$^2$-Routing)} \cite{q2routing}: This method includes a variable learning rate based on how big are variations in Q-values and ensures the traffic Quality of Service (QoS). Results show that this method outperforms the well-known ad-hoc routing algorithms in dynamic environments under QoS constraints.
    
    \item \textbf{Q-Network Enhanced Geographic Ad-Hoc Routing Protocol Based on GPSR (QNGPSR)} \cite{QNGPSR}: This routing protocol uses Q-network as an approximator to estimate the quality of different routing paths. Then, it makes forwarding decisions based on the estimated Q-values. Also, the neighbor topology information is introduced to estimate the environment and node states. QNGPSR is trained off-line when the signal propagation model is determined. Therefore, online-learning is not necessary, which reduces the computational load. Results show a higher packet delivery ratio and a lower end-to-end delay compared to the original GPSR.

    \item \textbf{Adaptive and Reliable routing protocol with deep reinforcement learning (ARdeep)} \cite{ardeep}: This is a deep RL-based adaptive and reliable routing protocol that formulates routing decisions with a Markov Decision Process model to characterize the network variations automatically. It considers link status, the packet error ratio, the expected connection time of the link, the remaining energy of nodes, the distance between the node and the destination when making routing decisions to precisely infer the network environment and make more appropriate forwarding decisions. Simulation results show that ARdeep outperforms the existing QGeo and conventional GPSR routing protocols.

    \item \textbf{Traffic-aware Q-Network enhanced routing protocol based on GPSR (TQNGPSR)} \cite{TQNGPSR}: Traffic-aware Q-network enhanced geographic routing protocol based on Greedy Perimeter Stateless Routing (GPSR). This algorithm uses the congestion information of neighbors and evaluates the quality of a wireless link by the Q-network algorithm as a traffic balancing strategy. Then, the protocol makes routing decisions based on the evaluation of each wireless link. Results show an improving performance in terms of packet delivery ratio and end-to-end delay.    

    \item \textbf{Q-learning based Multi-objective optimization Routing protocol (QMR)} \cite{QMR}:
    This novel Q-learning-based multi-objective optimization routing protocol adaptively adjusts the learning parameters based on the dynamicity of the network. The authors proposed a new mechanism to explore some undiscovered potential optimal paths while exploiting the acquired knowledge by re-estimating neighboring relationships to select the more reliable next hop. Results show a higher packet arrival ratio, lower delay, and energy consumption than the preceding Q-learning-based routing methods.

    \item \textbf{Q-Learning-based Fuzzy Logic for Multi-Objective Routing algorithm in Flying Ad Hoc Networks (QLFLMOR)} \cite{fuzzy1}:
    This multi-objective Q-learning-based fuzzy logic algorithm facilitates the selection of the routing paths in terms of the per-link and overall path performances. The optimal routing path to the destination is determined by each UAV using a fuzzy system with link-level and path-level parameters. The link-level parameters include the transmission rate, energy state, and flight status between neighbor UAVs, while the path-level parameters include the hop count and successful packet delivery time. Simulation results show that the proposed method can maintain low hop count and energy consumption and prolong the network lifetime.
    
    \item \textbf{Adaptive UAV-assisted Geographic Routing with Q-Learning (QAGR)} \cite{fuzzy3}:
    This algorithm uses fuzzy-logic and Depth-First-Search (DFS) algorithms to calculate the global path routing. As it is designed for UAV-Assisted networks, the vehicle on the ground maintains a fix-sized Q-table that converges with a well-designed reward function and forwards the routing request to the optimal node by looking up the Q-table filtered according to the global routing path. Results show a good performance in packet delivering and end-to-end delay.
    
    \item \textbf{Fully-Echoed Q-Routing with Simulated Annealing Inference for Flying Adhoc Networks (FESAIQ-Routing)} \cite{rovirasugranes2021fullyechoed}:
    This routing protocol is a full-echo Q-Routing with an adaptive learning rate controlled by Simulated Annealing (SA) optimization, where the \textit{temperature} parameter captures the influence of the nodes' mobility on the Q-value update rates. The soft variation of the exploration rate not only optimizes the exploration rate, but also accommodates abrupt changes in the network dynamicity. Simulation results show better performance than previous state-of-art Q-Routing algorithms. 
    
    
\end{itemize}

A summary of these routing protocols is presented in Tables \ref{performance_table_predictive} and \ref{performance_table_selflearning} with some characteristics for each routing protocol, as well as some comparative results to provide an idea of how these routing protocols perform under certain circumstances.

\section{Tools and public datasets}  \label{sec:tools}

In this section, we review tools and public datasets available for simulating real UAV networking environments. We investigate their features and capabilities, emphasizing their use for testing networking solutions (e.g., routing protocols) for UAV networks under different conditions.

\begin{table*}[t]
\centering
\caption{UAVs simulation tools.}
\label{simulation_tools}
\hskip0cm
{\footnotesize
\begin{tabular}{|p{3cm} | p{1.5cm} | p{1.8cm} | p{1.5cm}  | p{1.5cm} | p{1.5cm} | p{1.5cm} | p{1.5cm}|}
\hline
\textbf{Simulator} & \textbf{Free access} & \textbf{ROS Interface} & \textbf{MOCAP} & \textbf{MAVLink} & \textbf{SITL} & \textbf{Obstacles} & \textbf{Usability}\\
\hline
X-Plane \cite{xplane} & No & No & No & Yes & Yes & Yes & Medium \\
\hline
FlightGear \cite{flightgear} & Yes & No & No & Yes & Yes & Yes & Medium \\
\hline
Gazebo \cite{gazebo} & Yes & Yes & No & Yes & Yes & Yes & Easy \\
\hline
JMavSim \cite{jmavsim} & Yes & Yes & No & Yes & Yes & No & Easy \\
\hline
Microsoft AirSim \cite{airsim} & Yes & No & Yes & Yes & Yes & Yes & Medium \\
\hline
UE4Sim \cite{ue4sim} & Yes & No & Yes & No & No & Yes & Medium \\
\hline
\end{tabular}
}
\end{table*}

\begin{table*}[t]
\centering
\caption{UAVs experimental platforms}
\label{experimental_platforms}
\hskip0cm
{\footnotesize
\begin{tabular}{|p{3cm} | p{2cm} | p{2cm} | p{2cm} | p{7.2cm} |}
\hline
\textbf{Experimental platform} & \textbf{Interface} & \textbf{Free access} & \textbf{MAVLink} & \textbf{Compatible pilot software}\\
\hline
QGround-Control \cite{qgroundcontrol} & Graphical & Yes & Yes & PX4 Autopilot, ArduPilot\\
\hline
Mission Planner \cite{missionplanner} & Graphical & Yes & No & Ardupilot\\
\hline
MAVProxy \cite{mavproxy} & Command & Yes & Yes & Ardupilot\\
\hline
UGCS \cite{ugcs} & Graphical & Yes, but limited & Yes & DJI, Innoflight, Micropilot, Mikrokopter, Microdrones, Parrot\\
\hline
\end{tabular}
}
\end{table*}

\subsection{Simulation tools}

UAV simulation tools emulate virtual environments to model UAV flights in close-to-reality situations.
\arnau{It gives the convenience of evaluating the performance of UAV networks in virtual environments at much lower costs and trouble before testing in real scenarios. The choice of the appropriate simulator depends on both the testing objective and the list of features offered by each simulator}.
Some simulators incorporate the Motion of Capture (MOCAP), which allows simulating UAVs' natural movements \cite{simulation_survey}.
Another tool is MAVLink, a lightweight messaging protocol for communicating with drones to test communication protocols and algorithms. Software In The Loop (SITL) is a hardware-free simulation environment that facilitates simulating real-time UAV operations. It includes a c++ code to directly implement autopilot operation on the user's computer for testing \cite{SITL}. 

The list of tools for simulating UAV networking is large and still growing. A comparative analysis of some popular simulation tools, including X-Plane \cite{xplane}, FlightGear \cite{flightgear} (compatible with MATLAB Simulink), Gazebo \cite{gazebo}, JMavSim \cite{jmavsim}, Microsoft AirSim \cite{airsim}, and UE4Sim \cite{ue4sim}, is presented in Table \ref{simulation_tools}. 
\arnau{The first four simulate the UAV motions solely based on physics laws and do not support MOCAP. On the other hand, Microsoft AirSim and UE4Sim support MOCAP by using Unreal Engine 4 (UE4), an open-source tool that simulates UAV movement using physics along with a high-quality trajectory creation engine. AirSim is considered a platform for both AI research and training \cite{MicrosoftAirSim}. AirSim is empowered with deep learning, computer vision, and reinforcement learning features to generate and utilize training datasets \cite{airsim2}. Also, UE4Sim simulator benefits from a built-in and robust DL-based approach for real-time autonomous driving that does not require manually collected training data. For this reason, Microsoft AirSim and UE4Sim are considered two of the best simulators currently available. By combining AI components, these simulators allow researchers to simulate their algorithms in near-real environments, with the opportunity to develop better algorithms for AI-based networking and control tasks.}

\subsection{Experimental platforms}
Experimental platforms enable testing networking protocols in emulated network setups. Most experimentation platforms can be executed on a standalone computer or High-Performance Computing (HPC) servers. However, larger experimentation platforms typically consist of custom-built hardware with a set of simulation and operation software packages, programming environment, and web-based user interface to enable remote experimentation for the research community.

The most commonly-used experimental platforms are Network Simulator (NS-3) \cite{ns-3}, and OPNET \cite{opnet}. NS-3 is an open-source, free, and discrete-event network simulator for Internet and networking systems, enabling testing different layers of networking protocols, including routing protocols in MAC and Network layers. 
OPNET is an open network simulator used to simulate the function and performance of different networking systems. It is known for its power and versatility to create and simulate different network topologies. OPNET Technologies considers requests for free access for academic use.

Some other simulators are developed specifically for UAV networks.
For instance, ROS-NetSim \cite{ros-netsim} is a Robot Operating System (ROS) package, which acts as an interface between robots (UAVs in this case) and network simulators. ROS-NetSim accurately replicates Perception-Action-Communication (PAC) loops. Moreover, ROS-NetSim is tunable to account for a large range of communication fidelity and  complexity.

Another UAV-specific simulator is UB-ANC University at Buffalo’s Airborne Networking and Communications Testbed (UB-ANC) \cite{ub-anc}, which is an open platform that facilitates rapid testing and repeatable comparative evaluation of airborne networking and communications protocols at different layers of the protocol stack. It enables the flexible deployment of novel communications and networking protocols, with emphasis on modularity and extensibility.

A recently established experimentation center is Aerial Experimentation and Research Platform for Advanced Wireless (AERPAW) in the North Carolina State University (NCSU) \cite{AERPAW1, AERPAW2}. This NSF-funded center integrates drones and 5G wireless technology to provide increased coverage and connectivity, high throughput aerial monitoring, and improved signals and location data. The idea is to allow U.S. researchers to test new ways of increasing wireless speed and capacity in an experimental infrastructure, where nodes are mobile with the ability to transmit and receiving radio/video waves from user devices while moving on demand. This can be convenient under disaster relief circumstances, in which existing cellular networks may be damaged. Moreover, it is expected to make an impact in time-sensitive deliveries, smart agriculture, autonomous driving, and accident control applications \cite{AERPAW3}.

Another remote experimentation Platform is the Open Wireless Data-driven Experimental Research (POWDER) that operates as a highly flexible, remotely accessible, end-to-end software defined platform supporting a broad range of wireless and mobile related research \cite{POWDER}. This NSF-funded center has many features and capabilities including a massive MIMO base station and Software-Defined Radios (SDRs) that advances competitors in scale, realism, diversity, flexibility, and access. More advanced routing protocols which require function virtualization and SDR technology can be tested in this environment before implementing in UAV networks.

We can also find experimental platforms that act as a ground control station operating artificial UAVs. These platforms can be used to test communication protocols developed for UAV networks in simulated environments. Depending on the test scenario and objectives, we can select from a list of available software packages.  These platforms include QGround-Control \cite{qgroundcontrol}, Mission Planner \cite{missionplanner}, MAVProxy \cite{mavproxy} and UGCS \cite{ugcs}, as presented in Table \ref{experimental_platforms}.

Some experimental platforms use a graphical interface for user convenience, and some are command-line-based to provide more flexibility. We analyze the inclusion of MAVLink and the tools'  compatibility with pilot software. We found that PX4 Autopilot, Ardupilot, ROS, DJI Pilot, Innoflight, Micropilot, Mikrokopter, Microdrones, and Parrot are open-source autopilot systems capable of controlling autonomous vehicles, with a variety of aircraft operation scenarios, such as aerial mapping, surveying, and more applications.

\arnau{It is worth mentioning that some of the mentioned experimental platforms offer AI capabilities for networking research. For instance, NS-3 has NS3-GYM \cite{ns3gym}, and NS3-AI \cite{ns3ai} extension modules that enable applying AI to network simulations in NS-3. The key idea is to provide a high-efficiency solution to allow data interaction between NS-3 and other AI frameworks and encourage the use of AI in networking research. Also, AERPAW and POWDER contain AI embedded into hardware for superior detection, tracking, and classification of UAVs \cite{AERPAW4}, as well as for spectrum-maximizing resource allocations.}

\begin{table*}[t]
\centering
\caption{UAVs trajectory datasets}
\label{trajectory_datasets}
\hskip0cm
{\footnotesize
\begin{tabular}{|p{2.5cm} | p{5cm} | p{9cm} |}
\hline
\textbf{Trajectory datasets} & \textbf{Size / Length of flight} & \textbf{Description}\\
\hline
Blackbird \cite{blackbird}  & 10 hours of flight data from 168 flights. & Large-scale, aggressive indoor flight dataset collected using a custom-built quadrotor platform. Over 17 flight trajectories and 5 environments at velocities up to 7.0 m/s. \\
\hline
KAUST \cite{KAUST} & 11.172 Gb including all material.  & Trajectories, proxy meshes and images generated for path planning on real and synthetic scenes. It includes a benchmarking tool allowing new trajectories to output camera images for reconstruction.\\
\hline
Mid-Air \cite{midair} & 420k frames representing 79 minutes of drone flight records extracted out of more than 5 hours of flight. & Multi-purpose synthetic dataset for low altitude drone flights. Data corresponding to flight records of a flying quadcopter recorded in different climate conditions, including a test set for benchmarking.   \\
\hline
UZH-FPV \cite{uzh} & Over 27 flight sequences, with more than 10 km of flight distance. & Visual-inertial odometry dataset from a drone racing quadrotor with fast laps around a racetrack trajectories, as well as free-form indoor and out trajectories around obstacles. \\
\hline
\end{tabular}
}
\end{table*}

\subsection{Trajectory datasets}

Trajectory datasets consist of recorded data from real-time UAV flights that are useful for simulation purposes. The idea is to test new methodologies or protocols using trajectory data that mimic real-world situations. 
The bigger the size of the dataset, the more variability of flight trajectories is incorporated. In Table \ref{trajectory_datasets}, we compare some UAV trajectory and imagery datasets available for all users that can be helpful in simulating unmanned aerial monitoring platforms. These datasets include Blackbird \cite{blackbird}, KAUST \cite{KAUST}, Mid-Air \cite{midair} and UZH-FPV \cite{uzh}. We observe that each dataset offers UAV motions with different characteristics. The most appropriate dataset should be selected based on the specific test requirements and conditions. 
There exist many UAV datasets containing UAV imagery, as well as aircraft trajectory datasets. However, they do not exactly reflect UAV motion patterns, which is the main feature in path planning and routing scenarios. Therefore, there is a need to produce more relevant UAV trajectory datasets.
\arnau{Some methods use real data to train deep learning algorithms for trajectory design and recognition \cite{trajectorydeep1, trajectoryml}. Having access to accurate trajectory datasets will allow researchers to develop better solutions for real-world  scenarios.}

\section{Future trends and remaining challenges}  \label{sec:futuredirections}

Despite the recent advances in developing ML-powered networking protocols for UAV networks, there still are challenges and issues that would be the center of attention for coming years. From the communication perspective, most technical challenges arise from the limited payload, processing power, and structure-free and highly dynamic nature of unmanned aerial platforms \cite{challengescomm, challenges_comm, challenges_comp, challengessurvey, challengessurvey3}. This section summarizes part of these challenges and the future outlook of UAV networking technology from different perspectives, including \arnautwo{AI integration, energy efficiency, security, regulations, etc.}

\subsection{AI Integration}

Using AI to accelerate networking is the dominant research trend, as discussed in this paper. AI is shown to offer superior performance for communication, control, and operation of autonomous UAVs under different networking scenarios \cite{futureAI1, futureAI2,futureAI3}. Moreover, it can  optimize network management and reduce their complexity \cite{futureAI4}. 
We believe that the current networking designs have not yet fully utilized the power of AI-based solutions, and further research is on the way  to integrate AI and networking paradigms by using more advanced ML algorithms \cite{futureAI5}, RL-based decision-making \cite{futureAI6}, and deep learning \cite{futureAI7} for different aspects of networking, including sensing, scheduling, routing, spectrum sharing, path planning, and resource allocation. 

\arnautwo{However, it is worth mentioning that the use of AI-based methods can bring up new challenges, and needs to be studied carefully. For instance, AI methods can help to predict the future locations of the network nodes and potential link losses, and hence improve the power consumption by avoiding the transmission of hopeless packets that will be lost during the transmission. However, the use of AI methods can add to the computation complexity and CPU power use in power-constrained UAVs. Therefore, the right choice of the routing protocol should be based on the application-specific requirements and design constraints. Then, as a future solution, we need to study if the effect of including AI techniques into our model is worth the complexity that will be added, which could be affecting other important aspects such as the network lifetime and power consumption.}

\subsection{Connectivity}

Maintaining connectivity for UAV networks remains an important issue considering the limited communication range of commercial UAVs, which is typically limited to a few miles, while the coverage area in some applications like forest fire monitoring, disaster management, and wildlife monitoring can scale to hundreds of miles \cite{islam2019fire,erdelj2016uav,gonzalez2016unmanned}. Connectivity loss can cause packet drop, frequent link re-establishment, shortened link lifetime, prolonged delays, and ultimately disrupt the mission and compromise the user QoE. Recent methods use ML methods to predict network topology change ahead of time and avoid connectivity loss by learning-based routing \cite{ArnauWiSEE, MPC, imprvRGR2, parrot}. Another approach is including the link remaining time or path lifetime into the routing criteria \cite{survey4}. Other methods try to identify and resolve the connectivity loss by methods such as placement of new UAVs and link re-establishment mechanisms \cite{GPMOR}. 

One of the potential future trends would be integrating online path planning methods with networking algorithms to realize topology control with minimal connectivity issues. This approach is feasible in many real-world scenarios where the coverage area is defined, but there is a high degree of freedom in UAV's motion paths (e.g., search and rescue scenarios, regular forest monitoring, etc.).  
Also, connectivity is dependent on the interference caused by objects and environmental factors. Therefore, using more advanced AI methods to predict the influence of network nodes, environment, and surrounding objects on networking quality, can mitigate connectivity issues. In some applications with sparsely distributed nodes, intermittent connectivity is unavoidable. Also, connectivity can be caused by UAVs that ran out of battery. Developing AI methods to predict and accommodate such conditions is another potential research direction \cite{survey5}.

\subsection{Routing}

In section \ref{sec:routing}, we studied AI-enabled routing protocols, which use ML algorithms to predict network topology directly (e.g., predictive routing methods \cite{ArnauWiSEE, greedySECON}) or indirectly (e.g., RL-based methods \cite{QMR, fuzzy1}) and use it for the route selection process. However, there still exist challenges to be addressed.

Most routing protocols consider a fully connected network, whereas in reality, link breakages exist, causing routing protocols to fail \cite{futurerouting1}.
Also, node mobility in routing protocols is mostly developed for and tested in 2D spaces, whereas UAVs move in 3D spaces. The third research challenge is developing vision-based target tracking methods to predict network topology, noting that future UAVs will be equipped with Graphics/Tensor Processing Units (GPU/TPUs), capable of running deep learning methods for video processing. Developing probabilistic and priority-based routing protocols to prioritize packets with critical and confidential content is another potential research direction. Finally, extending RL-based method to accommodate non-linear motions and implementing light-weight routing protocols for miniaturized UAVs are two important remaining challenges.

\begin{table*}[htbp]
\centering
\caption{Remaining issues and future directions}
\label{remaining_issues}
{\footnotesize
\begin{tabular}{|p{2.25cm} | p{7cm} | p{8.5cm} |}
\hline
\centering
\textbf{Open issues} & \textbf{Problems} & \textbf{Future directions} \\
\hline
\begin{itemize}
\item AI integration
\end{itemize} & \begin{itemize}
\item Current networking designs have not yet fully utilized the power of AI-based solutions.
\item \arnautwo{The use of AI methods can add to the computation complexity and CPU power use in power-constrained UAVs.}
\end{itemize} & \begin{itemize}
\item Use ML, RL and DL techniques for different aspects of networking, including sensing, scheduling, routing, spectrum sharing, path planning, and resource allocation.
\item \arnautwo{Study if AI improves performance and does not impact negatively network lifetime and power consumption aspects.}
\end{itemize}\\
\hline
\begin{itemize}
\item Connectivity
\end{itemize} & \begin{itemize}
    \item Link failures bring limited network lifetime.
    \item Communication links between UAVs are extremely vulnerable.
    \item Connectivity is dependent on the interference caused by objects and environmental factors. 
\end{itemize} & \begin{itemize}
    \item Use ML methods to predict network topology change ahead of time and avoid connectivity loss by using learning-based routing. 
    \item Consider techniques that enhance link lifetime by adding link remaining time or path lifetime into routing criteria.
    \item Resolve the connectivity loss by methods such as placement of new UAVs and link re-establishment mechanisms. 
    \item Integrate online path planning methods with networking algorithms to realize topology control with minimal connectivity issues. 
\end{itemize}\\
\hline
\begin{itemize}
\item Routing
\end{itemize} & \begin{itemize} 
\item Most routing protocols consider a fully connected network, whereas in reality, link breakages happen.
\item Node mobility in routing protocols is mostly developed for and tested in 2D spaces, whereas UAVs move in 3D spaces.
\item Non-linear motions are not considered in current routing protocols.
\item Miniaturized UAVs have limited battery lifetime.
\end{itemize} & \begin{itemize}
    \item Design a routing protocol that considers intermittent connectivity into the routing selection process, as well as 3-dimensional space.
    \item Design a topology-predictive routing to accommodate non-linear motions.
    \item Develop vision-based target tracking methods for GPU-powered UAVs to predict network topology.
    \item Develop a light-weight AI-powered routing appropriate for miniaturized UAVs.
\end{itemize}\\
\hline 
\begin{itemize}
\item Energy efficiency
\end{itemize} & \begin{itemize}
\item UAVs are highly constrained in payload.
\item Battery designs are limited in energy optimization trends.
\item Only a small portion of routing protocols considers energy or power as a routing criteria. 
\end{itemize} & \begin{itemize}
\item Develop energy-efficient networking to prolong mission time and extend coverage area as a top priority.
\item Develop new battery technologies, such as hydrogen fuel cells and enhanced lithium-ion batteries.
\item Use RF transmission for WPT.
\item Consider mmWave communications, energy beamforming and placement optimization of  wireless charging stations.
\item Examine energy-aware routing future directions to incorporate energy level in the decision-making criteria to extend path lifetime.
\item \arnautwo{Use energy-efficient ML and DL algorithms for networking, such as energy-efficient convolutional neural networks.}
\end{itemize}\\
\hline
\begin{itemize}
\item Spectrum management
\end{itemize} & \begin{itemize}
\item Spectrum unavailability can cause the loss of command and control of the aircraft.
\item Spectrum remains vulnerable to unintentional or intentional interference.
\end{itemize} & \begin{itemize}
\item Spectrum sharing and spectrum leasing techniques using advanced AI-based solutions.
\item Optimal communications including sensitivity to interference and adaptive antenna steering can optimize data acquisition objectives.
\end{itemize}\\
\hline
\begin{itemize}
\item Security and user privacy
\end{itemize} & \begin{itemize}
\item UAVs are usually subject to different security attacks.
\item Aerial monitoring systems may exchange imagery with people's private information, which requires higher protection levels. 
\item Conventional PKI-based asymmetric security solutions are not feasible due to the lack of central authority to issue digital signatures.
\item Jamming attacks can also disrupt UAV missions.
\end{itemize} &  \begin{itemize}
\item Use hardware-driven security keys for UAV authentication and enable the non-repudiation feature.
\item ML methods can be used to detect and eliminate jamming, or provide an additional reference for positioning verification.
\item Develop secure routing schemes that alleviate security issues while finding optimal paths. 
\end{itemize}\\
\hline
\begin{itemize}
\item Operational regulations
\end{itemize} & \begin{itemize}
\item Lack of or ambiguity of regulations and standards for UAV operations, characteristics, safety requirements, secrecy and privacy considerations, and allowed airspace.
\end{itemize} & \begin{itemize}
\item Develop certification standards and air traffic requirements for UAV operations that are universal.
\item \arnautwo{Use AI software for regulation compliance to ensure safety of confidential information, risks mitigation and instant response to new regulatory requirements.}
\end{itemize}\\
\hline
\end{tabular}
}
\end{table*}

\subsection{Energy efficiency}

UAVs are highly constrained in payload and battery lifetime. The available energy should be optimally used for navigation, sensing, actuation, transmission, and data processing \cite{futureenergy1}. Therefore, developing energy-efficient networking to prolong mission time and extend coverage area is usually considered a top priority in UAV networks \cite{futureenergy2, futureenergy3}. 

A parallel research direction to solve energy issues of UAVs is developing new battery technologies, such as hydrogen fuel cells \cite{futureenergy5}, and enhanced lithium-ion batteries \cite{6futureenergy}. Also, Radio Frequency (RF) transmission can be used for Wireless Power Transfer (WPT) \cite{7futureenergy}, which provides controllable and sustainable energy supply for UAVs \cite{8futureenergy, 9futureenergy}. Further research is required to improve the use of WPT by reducing the distance between charging stations and UAVs, the random energy arrivals, and the scalable nature of UAVs \cite{10futureenergy, 11futureenergy}. Another research direction is considering mmWave communications for UAV networks, energy beamforming, and the optimized placement of wireless charging stations. Integration with 5G, and 6G wireless systems are also research topics related to energy optimization.
A few routing protocols consider energy as a routing criterion. Developing multi-objective and constrained optimization methods for routing protocols to enhance routing efficiency, while maintaining maximal connectivity and minimizing energy consumption is an important future direction. Further research is required to develop energy-aware routing protocols. Some recent works offer storing data and postponing calculations to the future to reduce power consumption while flying \cite{12futureenergy}. 

\arnautwo{Another potential solution can be using energy-efficient ML and DL algorithms for networking, such as energy-efficient convolutional neural networks \cite{futureenergy_neural}. However, new challenges come into place when we use the mentioned techniques. First, it is misbelieved that reducing the energy consumption of the algorithms does not necessarily lead to a reduction of the overall energy consumption. Second, in some scenarios, measuring the energy consumption becomes redundant since energy and time are correlated, and time is already measured. Third, it might be hard to measure the energy consumption, making it time consuming and impractical \cite{energymetrics}. Despite a few scenarios where these statements are true, we can find that reducing the energy consumption of the algorithms used will impact positively the overall energy consumption and also, measuring energy consumption can offer a unique overview, compared to time consumed. Last, there are some solutions that can model the energy consumption of different algorithms \cite{energyml}, as for example Alphabet's DeepMind.}


\subsection{Spectrum management} 

Enabling high-rate, low-latency, and ultra-reliable wireless communications in UAV networks is a necessity for future applications. Currently, UAVs use different communication protocols, including WiFi, LTE, LoRA, and 5G for A2A and A2G communications. 
In recent years, progress has been made in obtaining additional dedicated radio-frequency spectrum (5030-5091 MHz) for drone operations \cite{challenges_comm}. 
In addition to the usual ways of power management, interference control, spectrum-efficient networks, the use of different spectrum sensing, spectrum sharing, and spectrum leasing is considered to extend the service area of UAVs \cite{shamsoshoara2019distributed,shamsoshoara2020autonomous,spectrum1, spectrum2}, especially in unexpected and harsh conditions such as disaster management.

In a different line of research, some models offer using fiber optic communications \cite{fiberoptic}, laser \cite{laser} and LiFi \cite{lifi} to provide a faster and more efficient way for transmitting large amounts of data over long distances to cover the increasing demand for bandwidth. These methods would alleviate the spectrum scarcity issue. However, more research is expected to solve optical communications' specific issues, including sensitivity to interference and adaptive antenna steering.

\subsection {Security and user privacy}
Developing secure and privacy-preserving networking methods is another key challenge for UAV networks. UAVs are usually subject to different security attacks, including physical hijacking, jamming attacks, cyber-attacks, man-in-the-middle attack, intruding malicious nodes, channel interception, and denial of service, especially when flying over adversary territory \cite{survey12}. Also, aerial monitoring systems may exchange imagery with people's private information, which requires higher protection levels. 

A key challenge in structure-less UAV networks is that using conventional PKI-based asymmetric security solutions is not feasible due to the lack of central authority to issue digital signatures. Therefore, methods based on distributed certificate \cite{certificate}, key pre-distribution algorithms \cite{gligor,pakp}, and blockchain security \cite{jensen2019blockchain} are under investigation. Also, the idea of using hardware-driven security keys for UAV authentication and enabling the non-repudiation feature is recently proposed as a potential future direction \cite{alladi2020secauthuav}. Jamming attacks can also disrupt UAV missions, especially when they rely in GPS positioning. Using alternative localization methods can solve this issue \cite{uavlocalization}. Also, machine learning methods can be used to detect and eliminate jamming, or provide an additional reference for positioning verification \cite{jamming2}. Another emerging research trend is developing secure routing schemes that alleviate security issues while finding optimal paths \cite{futurerouting2}.

\subsection{Operational regulations} 

Another hindrance to the widespread use of drone technology is the lack of or ambiguity of sufficient regulations and standards for UAV operations, characteristics, safety requirements, secrecy and privacy considerations, and allowed airspace. In the US, the FAA is responsible for developing certification standards and air traffic requirements for drones. For instance, flying drones above class G airspace and in autonomous modes require special permits from the FAA that can take a long time. Also, international coordination would help develop global regulations, noting that different territories follow different standards and regulations. For instance, currently, there exist three different regions, including (i) region 1 that covers Europe, Africa, and parts of the Middle East, (ii) region 2 that covers America, and (iii) region 3 that covers Asia and the Pacific, which have different frequency bands for UAV operations. \arnautwo{One more potential solution would be using AI software for regulation compliance, which ensures increased safety of confidential information, risks mitigation and instant response to new regulatory requirements. This way, we could find AI-enabled drones authorized for use in different regions, following the operational regulations at each specific area.}

A summary of the remaining challenges and future trends that we reviewed here is included in Table \ref{remaining_issues}.


\section{Conclusion}  \label{sec:conclusion}

This paper reviewed the recent developments in AI-based networking for UAV systems, focusing on accommodating dynamic but predictable network topology. The first observation is departing from single-drone systems to networked autonomous drones to accomplish complicated tasks at lower cost and time. However, conventional communication protocols lack technological and design characteristics to adapt to the dynamicity of such networks. Recently, the idea of using AI-based networking protocols has gained a lot of attention to use the learning power of machine learning methods to model and predict network topology changes, and directly or indirectly incorporate it into the networking decision-making mechanisms. 

Previous survey papers mainly emphasize conventional distance-based, static, reactive, and proactive routing protocols, missing the important class of AI-enabled routing protocols. This paper \arnautwo{reviews newly developed AI-based routing protocols for UAV networks, highlighting the benefits and costs of each type, along with available testing and implementation tools, relations to mobility models and networking protocols, and connection to UAV swarming.} These methods include the direct use of ML methods for topology prediction, as well as learning-by-experience approaches. The former is more accurate and accommodates non-linear motion paths, and the second approach was appropriate for separating networking and topology control layers. Most recent papers report substantial improvements in terms of connectivity control, successful packet delivery rate, transmission delay, and throughput for AI-based routing protocols, compared to conventional methods.

We also reviewed future trends and the remaining challenges for AI-based networking. We identified the need for (i) advanced AI methods to precisely predict networking conditions and environmental factors, (ii) integrative networking and topology control for extended connectivity, (iii) vision-based tracking methods for GPU-powered UAVs, (iv) topology-predictive routing to accommodate non-linear motions, (v) developing light-weight AI-powered routing appropriate for miniaturized UAVs, (vi) distributed structure-free extension of asymmetric security protocols based on key pre-distribution, hardware-driven keys, blockchain, and distributed certificate methods, (vii) ML methods to recognize and combat jamming attacks, (viii) energy-efficient low-power and low-complexity networking \arnautwo{using ML and DL}, (ix) AI-based spectrum sharing and leasing policies, and (x) universal regulation and guidelines for UAV operation, as top key issues that are worthy of investigation by the research community.

\bibliographystyle{IEEEtran}  
\bibliography{ArnauBIB}

\begin{thebibliography}{100}
\providecommand{\url}[1]{#1}
\csname url@samestyle\endcsname
\providecommand{\newblock}{\relax}
\providecommand{\bibinfo}[2]{#2}
\providecommand{\BIBentrySTDinterwordspacing}{\spaceskip=0pt\relax}
\providecommand{\BIBentryALTinterwordstretchfactor}{4}
\providecommand{\BIBentryALTinterwordspacing}{\spaceskip=\fontdimen2\font plus
\BIBentryALTinterwordstretchfactor\fontdimen3\font minus
  \fontdimen4\font\relax}
\providecommand{\BIBforeignlanguage}[2]{{%
\expandafter\ifx\csname l@#1\endcsname\relax
\typeout{** WARNING: IEEEtran.bst: No hyphenation pattern has been}%
\typeout{** loaded for the language `#1'. Using the pattern for}%
\typeout{** the default language instead.}%
\else
\language=\csname l@#1\endcsname
\fi
#2}}
\providecommand{\BIBdecl}{\relax}
\BIBdecl

\bibitem{transportation}
\BIBentryALTinterwordspacing
E.~N. Barmpounakis, E.~I. Vlahogianni, and J.~C. Golias, ``Unmanned aerial
  aircraft systems for transportation engineering: Current practice and future
  challenges,'' \emph{International Journal of Transportation Science and
  Technology}, vol.~5, no.~3, pp. 111 -- 122, 2016, unmanned Aerial Vehicles
  and Remote Sensing. [Online]. Available:
  \url{http://www.sciencedirect.com/science/article/pii/S2046043016300533}
\BIBentrySTDinterwordspacing

\bibitem{traffic-monitoring}
K.~Kanistras, G.~Martins, M.~J. Rutherford, and K.~P. Valavanis, ``A survey of
  unmanned aerial vehicles (uavs) for traffic monitoring,'' in \emph{2013
  International Conference on Unmanned Aircraft Systems (ICUAS)}, May 2013, pp.
  221--234.

\bibitem{surveillance}
Z.~Zaheer, A.~Usmani, E.~Khan, and M.~A. Qadeer, ``Aerial surveillance system
  using uav,'' in \emph{2016 Thirteenth International Conference on Wireless
  and Optical Communications Networks (WOCN)}, July 2016, pp. 1--7.

\bibitem{border-patrolling}
D.~Bein, W.~Bein, A.~Karki, and B.~B. Madan, ``Optimizing border patrol
  operations using unmanned aerial vehicles,'' in \emph{2015 12th International
  Conference on Information Technology - New Generations}, April 2015, pp.
  479--484.

\bibitem{search-rescue}
S.~Waharte and N.~Trigoni, ``Supporting search and rescue operations with
  uavs,'' in \emph{2010 International Conference on Emerging Security
  Technologies}, Sept 2010, pp. 142--147.

\bibitem{erdelj2017help}
M.~Erdelj, E.~Natalizio, K.~R. Chowdhury, and I.~F. Akyildiz, ``Help from the
  sky: Leveraging uavs for disaster management,'' \emph{IEEE Pervasive
  Computing}, vol.~16, no.~1, pp. 24--32, 2017.

\bibitem{connectivity}
M.~Messous, S.~Senouci, and H.~Sedjelmaci, ``Network connectivity and area
  coverage for uav fleet mobility model with energy constraint,'' in \emph{2016
  IEEE Wireless Communications and Networking Conference}, April 2016, pp.
  1--6.

\bibitem{ChakareskiNMXAR:19}
J.~{Chakareski}, S.~{Naqvi}, N.~{Mastronarde}, J.~{Xu}, F.~{Afghah}, and
  A.~{Razi}, ``An energy efficient framework for uav-assisted millimeter wave
  5g heterogeneous cellular networks,'' \emph{IEEE Transactions on Green
  Communications and Networking}, vol.~3, no.~1, pp. 37--44, 2019.

\bibitem{forestry}
M.~R. Brust and B.~M. Strimbu, ``A networked swarm model for uav deployment in
  the assessment of forest environments,'' in \emph{2015 IEEE Tenth
  International Conference on Intelligent Sensors, Sensor Networks and
  Information Processing (ISSNIP)}, April 2015, pp. 1--6.

\bibitem{chakareski2017drone}
J.~Chakareski, ``Drone networks for virtual human teleportation,'' in
  \emph{Proc. ACM Workshop on Micro Aerial Vehicle Networks, Systems, and
  Applications}, Niagra Falls, NY, USA, June 2017, pp. 21--26.

\bibitem{chakareski2019uav}
J.~{Chakareski}, ``{UAV-IoT} for next generation virtual reality,'' \emph{IEEE
  Transactions on Image Processing}, vol.~28, no.~12, pp. 5977--5990, 2019.

\bibitem{KhanCG:20}
M.~Khan, J.~Chakareski, and S.~Gupta, ``{RF-FSO} dual-path {UAV} network for
  high fidelity multi-viewpoint scalable 360-degree video streaming,'' in
  \emph{Proc. IEEE Int'l Workshop on Multimedia Signal Processing}, Tampere,
  Finland, Sept. 2020, pp. 1--6.

\bibitem{LOCUST}
\BIBentryALTinterwordspacing
``{\sc LOCUST:} autonomous, swarming uavs fly into the future,'' April 2015.
  [Online]. Available:
  \url{http://www.onr.navy.mil/Media-Center/Press-Releases/2015/LOCUST-low-cost-UAV-swarm-ONR.aspx}
\BIBentrySTDinterwordspacing

\bibitem{serveyingandmapping}
S.~Siebert and J.~Teizer, ``Mobile 3d mapping for surveying earthwork using an
  unmanned aerial vehicle (uav),'' 2013.

\bibitem{volcanomonitoring}
\BIBentryALTinterwordspacing
J.~Plaza, ``Using drones to monitor volcano activity and save lives,'' Jun
  2018. [Online]. Available:
  \url{https://www.commercialuavnews.com/public-safety/using-drones-to-monitor-volcano-activity-and-save-lives}
\BIBentrySTDinterwordspacing

\bibitem{UAVcontrolbybrain}
\BIBentryALTinterwordspacing
``Brain-controlled drones are here: What's coming in the next five years?'' Sep
  2017. [Online]. Available:
  \url{https://www.sciencedaily.com/releases/2017/09/170929230433.htm}
\BIBentrySTDinterwordspacing

\bibitem{casa}
\BIBentryALTinterwordspacing
``Electrical and computer engineering,'' Feb 2018. [Online]. Available:
  \url{https://ece.umass.edu/news/casa-adds-drone-detection-its-early-warning-monitoring-severe-weather}
\BIBentrySTDinterwordspacing

\bibitem{plantprotection}
Y.~A. Pederi and H.~S. Cheporniuk, ``Unmanned aerial vehicles and new
  technological methods of monitoring and crop protection in precision
  agriculture,'' in \emph{2015 IEEE International Conference Actual Problems of
  Unmanned Aerial Vehicles Developments (APUAVD)}, Oct 2015, pp. 298--301.

\bibitem{windenergy}
\BIBentryALTinterwordspacing
A.~Cherubini, A.~Papini, R.~Vertechy, and M.~Fontana, ``Airborne wind energy
  systems: A review of the technologies,'' \emph{Renewable and Sustainable
  Energy Reviews}, vol.~51, pp. 1461 -- 1476, 2015. [Online]. Available:
  \url{http://www.sciencedirect.com/science/article/pii/S1364032115007005}
\BIBentrySTDinterwordspacing

\bibitem{FlockofBirds}
A.~A. Paranjape, S.~Chung, K.~Kim, and D.~H. Shim, ``Robotic herding of a flock
  of birds using an unmanned aerial vehicle,'' \emph{IEEE Transactions on
  Robotics}, vol.~34, no.~4, pp. 901--915, Aug 2018.

\bibitem{swarm3}
\BIBentryALTinterwordspacing
``Amazon prime air.'' [Online]. Available:
  \url{https://www.amazon.com/Amazon-Prime-Air/b?ie=UTF8&node=8037720011}
\BIBentrySTDinterwordspacing

\bibitem{swarm4}
\BIBentryALTinterwordspacing
``Ups flight forward™ drone delivery.'' [Online]. Available:
  \url{https://www.ups.com/us/en/services/shipping-services/flight-forward-drones.page}
\BIBentrySTDinterwordspacing

\bibitem{Gupta_survey}
L.~Gupta, R.~Jain, and G.~Vaszkun, ``Survey of important issues in \textsc{UAV}
  communication networks,'' \emph{IEEE Communications Surveys \&Tutorials},
  vol.~18, no.~2, pp. 1123--1152, Secondquarter 2016.

\bibitem{uavmarket}
\BIBentryALTinterwordspacing
``Unmanned aerial vehicle (uav) market.'' [Online]. Available:
  \url{https://www.marketsandmarkets.com/Market-Reports/unmanned-aerial-vehicles-uav-market-662.html}
\BIBentrySTDinterwordspacing

\bibitem{futureAI1}
K.-I. Kim, K.-H. Kim, M.~Imran, P.~Khan, E.~Tovar, and F.~Ali, ``Uav-enabled
  healthcare architecture: Issues and challenges,'' \emph{Future Generation
  Computer Systems}, 08 2019.

\bibitem{futureAI2}
M.~{Chen}, H.~{Wang}, S.~{Mehrotra}, V.~C.~M. {Leung}, and I.~{Humar},
  ``Intelligent networks assisted by cognitive computing and machine
  learning,'' \emph{IEEE Network}, vol.~33, no.~3, pp. 6--8, 2019.

\bibitem{futureAI3}
P.~Bithas, E.~Michailidis, N.~Nomikos, D.~Vouyioukas, and A.~Kanatas, ``A
  survey on machine-learning techniques for uav-based communications,''
  \emph{Sensors}, vol.~19, 11 2019.

\bibitem{VANET}
B.~.S, ``Study of ad hoc networks with reference to manet, vanet, fanet,''
  \emph{International Journal of Advanced Research in Computer Science and
  Software Engineering}, vol.~7, p. 390, 07 2017.

\bibitem{issuereview}
H.~Lashari, H.~M. Ali, and A.~Laghari, ``Uav communication networks issues: A
  review,'' \emph{Archives of Computational Methods in Engineering}, 03 2020.

\bibitem{beamforming1}
T.~{Izydorczyk}, G.~{Berardinelli}, P.~{Mogensen}, M.~M. {Ginard}, J.~{Wigard},
  and I.~Z. {Kovács}, ``Achieving high uav uplink throughput by using
  beamforming on board,'' \emph{IEEE Access}, vol.~8, pp. 82\,528--82\,538,
  2020.

\bibitem{beamforming2}
W.~{Yuan}, C.~{Liu}, F.~{Liu}, S.~{Li}, and D.~W.~K. {Ng}, ``Learning-based
  predictive beamforming for uav communications with jittering,'' \emph{IEEE
  Wireless Communications Letters}, vol.~9, no.~11, pp. 1970--1974, 2020.

\bibitem{greedySECON}
M.~Khaledi, A.~Rovira-Sugranes, F.~Afghah, and A.~Razi, ``On greedy routing in
  dynamic \textsc{UAV} networks,'' in \emph{IEEE International Conference on
  Sensing, Communication and Networking (SECON 2018)}, June 2018.

\bibitem{survey1}
O.~{Sami Oubbati}, M.~{Atiquzzaman}, T.~{Ahamed Ahanger}, and A.~{Ibrahim},
  ``Softwarization of uav networks: A survey of applications and future
  trends,'' \emph{IEEE Access}, vol.~8, pp. 98\,073--98\,125, 2020.

\bibitem{survey5}
D.~{Shumeye Lakew}, U.~{Sa’ad}, N.~{Dao}, W.~{Na}, and S.~{Cho}, ``Routing in
  flying ad hoc networks: A comprehensive survey,'' \emph{IEEE Communications
  Surveys Tutorials}, vol.~22, no.~2, pp. 1071--1120, 2020.

\bibitem{survey6}
Q.~Sang, H.~Wu, L.~Xing, and P.~Xie, ``Review and comparison of emerging
  routing protocols in flying ad hoc networks,'' \emph{Symmetry}, vol.~12, p.
  971, 06 2020.

\bibitem{survey7}
B.~Alzahrani, O.~S. Oubbati, A.~Barnawi, M.~Atiquzzaman, and D.~Alghazzawi,
  ``Uav assistance paradigm: State-of-the-art in applications and challenges,''
  \emph{Journal of Network and Computer Applications}, 05 2020.

\bibitem{survey4}
R.~A. {Nazib} and S.~{Moh}, ``Routing protocols for unmanned aerial
  vehicle-aided vehicular ad hoc networks: A survey,'' \emph{IEEE Access},
  vol.~8, pp. 77\,535--77\,560, 2020.

\bibitem{survey13}
M.~Y. Arafat and S.~Moh, ``Routing protocols for unmanned aerial vehicle
  networks: A survey,'' \emph{IEEE Access}, vol.~7, pp. 99\,694 -- 99\,720, 07
  2019.

\bibitem{survey8}
O.~S. {Oubbati}, M.~{Atiquzzaman}, P.~{Lorenz}, M.~H. {Tareque}, and M.~S.
  {Hossain}, ``Routing in flying ad hoc networks: Survey, constraints, and
  future challenge perspectives,'' \emph{IEEE Access}, vol.~7, pp.
  81\,057--81\,105, 2019.

\bibitem{survey14}
M.~Khan, K.-L. Yau, R.~Md.~Noor, and M.~Imran, ``Routing schemes in fanets: A
  survey,'' \emph{Sensors (Basel, Switzerland)}, vol.~20, 12 2019.

\bibitem{survey3}
J.~{Jiang} and G.~{Han}, ``Routing protocols for unmanned aerial vehicles,''
  \emph{IEEE Communications Magazine}, vol.~56, no.~1, pp. 58--63, 2018.

\bibitem{survey2}
A.~{Awang}, K.~{Husain}, N.~{Kamel}, and S.~{Aïssa}, ``Routing in vehicular
  ad-hoc networks: A survey on single- and cross-layer design techniques, and
  perspectives,'' \emph{IEEE Access}, vol.~5, pp. 9497--9517, 2017.

\bibitem{survey12}
M.~Jean~aime, M.-S. Mahmoud, and N.~Larrieu, ``Survey on uaanet routing
  protocols and network security challenges,'' \emph{Ad-Hoc and Sensor Wireless
  Networks}, vol.~37, 03 2017.

\bibitem{survey9}
O.~S. Oubbati, A.~Lakas, F.~Zhou, M.~G{\"u}nes, and M.~B. Yagoubi, ``A survey
  on position-based routing protocols for flying ad hoc networks {(FANETs)},''
  \emph{Veh. Commun.}, vol.~10, pp. 29--56, 2017.

\bibitem{survey10}
C.~Suthaputchakun and Z.~Sun, ``Routing protocol in intervehicle communication
  systems: A survey,'' \emph{IEEE Communications Magazine}, vol.~49, pp.
  150--156, 12 2011.

\bibitem{AIbenefit}
\BIBentryALTinterwordspacing
G.~D. Maayan, ``How do ai-based drones work?'' Sep 2020. [Online]. Available:
  \url{https://heartbeat.fritz.ai/how-ai-based-drones-work-a94f20e62695}
\BIBentrySTDinterwordspacing

\bibitem{AIapp_survey}
M.-A. Lahmeri, M.~A. Kishk, and M.-S. Alouini, ``Artificial intelligence for
  uav-enabled wireless networks: A survey,'' \emph{IEEE Open Journal of the
  Communications Society}, vol.~2, pp. 1015--1040, 2021.

\bibitem{futureAI5}
Q.~{Zhang}, M.~{Mozaffari}, W.~{Saad}, M.~{Bennis}, and M.~{Debbah}, ``Machine
  learning for predictive on-demand deployment of uavs for wireless
  communications,'' in \emph{2018 IEEE Global Communications Conference
  (GLOBECOM)}, 2018, pp. 1--6.

\bibitem{positioning1}
J.~Chen, U.~Yatnalli, and D.~Gesbert, ``Learning radio maps for uav-aided
  wireless networks: A segmented regression approach,'' in \emph{2017 IEEE
  International Conference on Communications (ICC)}, 2017, pp. 1--6.

\bibitem{positioning2}
J.~Kim, C.~Park, J.~Ahn, Y.~Ko, J.~Park, and J.~C. Gallagher, ``Real-time uav
  sound detection and analysis system,'' in \emph{2017 IEEE Sensors
  Applications Symposium (SAS)}, 2017, pp. 1--5.

\bibitem{channelest1}
Y.~Zhang, J.~Wen, G.~Yang, Z.~He, and X.~Luo, ``Air-to-air path loss prediction
  based on machine learning methods in urban environments,'' \emph{Wireless
  Communications and Mobile Computing}, vol. 2018, pp. 1--9, 06 2018.

\bibitem{channelest2}
W.~Xia, S.~Rangan, M.~Mezzavillla, A.~Lozano, G.~Geraci, V.~Semkin, and
  G.~Loianno, ``Generative neural network channel modeling for millimeter-wave
  uav communication,'' 2020.

\bibitem{virtual1}
M.~Chen, W.~Saad, and C.~Yin, ``Deep learning for 360$^\circ$ content
  transmission in {UAV-Enabled} {Virtual} {Reality},'' in \emph{ICC 2019 - 2019
  IEEE International Conference on Communications (ICC)}, 2019, pp. 1--6.

\bibitem{virtual2}
S.~Wang, J.~Chen, Z.~Zhang, G.~Wang, Y.~Tan, and Y.~Zheng, ``Construction of a
  virtual reality platform for uav deep learning,'' in \emph{2017 Chinese
  Automation Congress (CAC)}, 2017, pp. 3912--3916.

\bibitem{imaging1}
K.~Mukadam, A.~Sinh, and R.~Karani, ``Detection of landing areas for unmanned
  aerial vehicles,'' in \emph{2016 International Conference on Computing
  Communication Control and automation (ICCUBEA)}, 2016, pp. 1--5.

\bibitem{autonomous1}
N.~Imanberdiyev, C.~Fu, E.~Kayacan, and I.-M. Chen, ``Autonomous navigation of
  uav by using real-time model-based reinforcement learning,'' in \emph{2016
  14th International Conference on Control, Automation, Robotics and Vision
  (ICARCV)}, 2016, pp. 1--6.

\bibitem{autonomous2}
X.~Liu, Y.~Liu, and Y.~Chen, ``Reinforcement learning in multiple-uav networks:
  Deployment and movement design,'' 2019.

\bibitem{scheduling1}
\BIBentryALTinterwordspacing
G.~Faraci, A.~Raciti, S.~A. Rizzo, and G.~Schembra, ``Green wireless power
  transfer system for a drone fleet managed by reinforcement learning in smart
  industry,'' \emph{Applied Energy}, vol. 259, p. 114204, 2020. [Online].
  Available:
  \url{https://www.sciencedirect.com/science/article/pii/S0306261919318914}
\BIBentrySTDinterwordspacing

\bibitem{security1}
N.~I. Mowla, N.~H. Tran, I.~Doh, and K.~Chae, ``Federated learning-based
  cognitive detection of jamming attack in flying ad-hoc network,'' \emph{IEEE
  Access}, vol.~8, pp. 4338--4350, 2020.

\bibitem{sensing1}
Y.~Liu, J.~Nie, X.~Li, S.~H. Ahmed, W.~Y.~B. Lim, and C.~Miao, ``Federated
  learning in the sky: Aerial-ground air quality sensing framework with uav
  swarms,'' \emph{IEEE Internet of Things Journal}, pp. 1--1, 2020.

\bibitem{arnau2}
A.~Razi, C.~Wang, F.~Almaraghi, Q.~Huang, Y.~Zhang, H.~Lu, and
  A.~Rovira-Sugranes, ``Predictive routing for wireless networks:
  Robotics-based test and evaluation platform,'' in \emph{2018 IEEE 8th Annual
  Computing and Communication Workshop and Conference (CCWC)}, Jan 2018, pp.
  993--999.

\bibitem{militarydrones1}
\BIBentryALTinterwordspacing
J.~Brown, ``Types of military drones: The best technology available today,''
  Feb 2020. [Online]. Available:
  \url{https://www.mydronelab.com/blog/types-of-military-drones.html}
\BIBentrySTDinterwordspacing

\bibitem{militarydrones3}
\BIBentryALTinterwordspacing
``4 military sensor technologies that drones are transporting to the commercial
  market,'' Jan 2019. [Online]. Available:
  \url{https://militaryethernet.com/4-military-sensor-technologies-drones-transporting-commercial-market/}
\BIBentrySTDinterwordspacing

\bibitem{militarywilson}
\BIBentryALTinterwordspacing
J.~Wilson, ``The future of artificial intelligence and quantum computing,'' Aug
  2020. [Online]. Available:
  \url{https://www.militaryaerospace.com/computers/article/14182330/future-of-artificial-intelligence-and-quantum-computing}
\BIBentrySTDinterwordspacing

\bibitem{militarydrones4}
\BIBentryALTinterwordspacing
B.~Knight, ``A guide to military drones,'' Jun 2017. [Online]. Available:
  \url{https://www.dw.com/en/a-guide-to-military-drones/a-39441185}
\BIBentrySTDinterwordspacing

\bibitem{industrialdrone}
\BIBentryALTinterwordspacing
A.~Sharma, ``Everything you need to know about industrial grade drones,'' Jul
  2020. [Online]. Available:
  \url{https://jungleworks.com/everything-you-need-to-know-about-industrial-grade-drones/}
\BIBentrySTDinterwordspacing

\bibitem{commercialdrone}
\BIBentryALTinterwordspacing
Flyability, ``Commercial drones: Industries that use drones, deliverables, and
  our list of the top models on the market.'' [Online]. Available:
  \url{https://www.flyability.com/commercial-drones}
\BIBentrySTDinterwordspacing

\bibitem{faa}
\BIBentryALTinterwordspacing
Dec 2020. [Online]. Available:
  \url{https://www.faa.gov/uas/recreational_fliers/}
\BIBentrySTDinterwordspacing

\bibitem{AIchip3}
R.~Yin, W.~Li, Z.-q. Wang, and X.-x. Xu, ``The application of artificial
  intelligence technology in uav,'' in \emph{2020 5th International Conference
  on Information Science, Computer Technology and Transportation (ISCTT)},
  2020, pp. 238--241.

\bibitem{AIchip}
\BIBentryALTinterwordspacing
R.~Hof, ``Ai-on-a-chip soon will make phones, drones and more a lot smarter,''
  May 2016. [Online]. Available:
  \url{https://www.forbes.com/sites/roberthof/2016/05/07/ai-on-a-chip-soon-will-make-phones-drones-and-more-a-lot-smarter/?sh=75b2d2ba7eef}
\BIBentrySTDinterwordspacing

\bibitem{AIchip1}
T.~Hwang, ``Computational power and the social impact of artificial
  intelligence,'' 2018.

\bibitem{AIchip2}
\BIBentryALTinterwordspacing
C.~Dilmegani, ``Ai chips in 2021: Guide to cost-efficient ai training \&
  inference,'' Jan 2021. [Online]. Available:
  \url{https://research.aimultiple.com/ai-chip/#what-are-its-components}
\BIBentrySTDinterwordspacing

\bibitem{AIchip4}
\BIBentryALTinterwordspacing
K.~Freund, ``Ai hardware: Harder than it looks,'' \emph{Forbes}, Oct 2019.
  [Online]. Available:
  \url{https://www.forbes.com/sites/moorinsights/2019/10/07/ai-hardware-harder-than-it-looks/?sh=23dc41e5471f}
\BIBentrySTDinterwordspacing

\bibitem{AIchip5}
\BIBentryALTinterwordspacing
Freund, ``Intel shows off its ai chips and chops,'' \emph{Forbes}, Jun 2018.
  [Online]. Available:
  \url{https://www.forbes.com/sites/moorinsights/2018/06/01/intel-shows-off-its-ai-chips-and-chops/?sh=272064ad6643}
\BIBentrySTDinterwordspacing

\bibitem{dronemanufacturers}
\BIBentryALTinterwordspacing
I.~Insider, ``Drone market outlook in 2021: industry growth trends, market
  stats and forecast,'' Feb 2021. [Online]. Available:
  \url{https://www.businessinsider.com/drone-industry-analysis-market-trends-growth-forecasts#:~:text=Insider
  Intelligence defines enterprise drones,annual growth rate (CAGR)}
\BIBentrySTDinterwordspacing

\bibitem{dronemanufacturers1}
\BIBentryALTinterwordspacing
L.~Schroth, ``Drone manufacturer market shares: Dji leads the way,'' Jan 2021.
  [Online]. Available:
  \url{https://droneii.com/drone-manufacturer-market-shares-dji-leads-the-way-in-the-us}
\BIBentrySTDinterwordspacing

\bibitem{precisionhawk}
\BIBentryALTinterwordspacing
``Geospatial data analytics for the enterprise.'' [Online]. Available:
  \url{https://www.precisionhawk.com/}
\BIBentrySTDinterwordspacing

\bibitem{dronetech}
\BIBentryALTinterwordspacing
``Dronehunter: Net gun drone capture: Products,'' Mar 2021. [Online].
  Available:
  \url{https://fortemtech.com/products/dronehunter/?gclid=Cj0KCQjwl9GCBhDvARIsAFunhskjqqROz4pMBmxT_bxGxRK5KMyievKr4lH3kejZtH_gLS1qQt0X68kaAsEfEALw_wcB}
\BIBentrySTDinterwordspacing

\bibitem{uavcompanies}
\BIBentryALTinterwordspacing
``Top 100 drone companies to watch in 2020,'' Jan 2021. [Online]. Available:
  \url{https://uavcoach.com/drone-companies/}
\BIBentrySTDinterwordspacing

\bibitem{dronetech1}
\BIBentryALTinterwordspacing
C.~Insights, ``How drones will impact society: From fighting war to forecasting
  weather, uavs change everything,'' Jun 2020. [Online]. Available:
  \url{https://www.cbinsights.com/research/drone-impact-society-uav/}
\BIBentrySTDinterwordspacing

\bibitem{edgecomputing}
W.~{Shi}, J.~{Cao}, Q.~{Zhang}, Y.~{Li}, and L.~{Xu}, ``Edge computing: Vision
  and challenges,'' \emph{IEEE Internet of Things Journal}, vol.~3, no.~5, pp.
  637--646, 2016.

\bibitem{fogcomputing}
A.~V. {Dastjerdi} and R.~{Buyya}, ``Fog computing: Helping the internet of
  things realize its potential,'' \emph{Computer}, vol.~49, no.~8, pp.
  112--116, 2016.

\bibitem{firstnet}
G.~{Baldini}, S.~{Karanasios}, D.~{Allen}, and F.~{Vergari}, ``Survey of
  wireless communication technologies for public safety,'' \emph{IEEE
  Communications Surveys Tutorials}, vol.~16, no.~2, pp. 619--641, 2014.

\bibitem{lora}
V.~Sharma, I.~You, G.~Pau, M.~Collotta, J.~Deok~Lim, and J.~Kim,
  ``Lorawan-based energy-efficient surveillance by drones for intelligent
  transportation systems,'' \emph{Energies}, vol.~11, 03 2018.

\bibitem{LORA1}
V.~{Delafontaine}, F.~{Schiano}, G.~{Cocco}, A.~{Rusu}, and D.~{Floreano},
  ``Drone-aided localization in lora iot networks,'' in \emph{2020 IEEE
  International Conference on Robotics and Automation (ICRA)}, 2020, pp.
  286--292.

\bibitem{LORA2}
O.~Saraereh, A.~Alsaraira, I.~Khan, and P.~Uthansakul, ``Performance evaluation
  of uav-enabled lora networks for disaster management applications,''
  \emph{Sensors}, vol.~20, pp. 1--18, 04 2020.

\bibitem{backhauling}
S.~{Chandrasekharan}, K.~{Gomez}, A.~{Al-Hourani}, S.~{Kandeepan},
  T.~{Rasheed}, L.~{Goratti}, L.~{Reynaud}, D.~{Grace}, I.~{Bucaille},
  T.~{Wirth}, and S.~{Allsopp}, ``Designing and implementing future aerial
  communication networks,'' \emph{IEEE Communications Magazine}, vol.~54,
  no.~5, pp. 26--34, 2016.

\bibitem{intelwifi}
\BIBentryALTinterwordspacing
``Different wi-fi protocols and data rates.'' [Online]. Available:
  \url{https://www.intel.com/content/www/us/en/support/articles/000005725/network-and-io/wireless.html}
\BIBentrySTDinterwordspacing

\bibitem{xfoldrig}
\BIBentryALTinterwordspacing
``xfold spy.'' [Online]. Available: \url{http://www.xfoldrig.com/xfold-spy/}
\BIBentrySTDinterwordspacing

\bibitem{skylimit}
\BIBentryALTinterwordspacing
X.~Lin, V.~Yajnanarayana, S.~D. Muruganathan, S.~Gao, H.~Asplund, H.~Maattanen,
  M.~B. A, S.~Euler, and Y.~E. Wang, ``The sky is not the limit: {LTE} for
  unmanned aerial vehicles,'' \emph{CoRR}, vol. abs/1707.07534, 2017. [Online].
  Available: \url{http://arxiv.org/abs/1707.07534}
\BIBentrySTDinterwordspacing

\bibitem{lte}
\BIBentryALTinterwordspacing
Y.~Zeng, J.~Lyu, and R.~Zhang, ``Cellular-connected {UAV:} potentials,
  challenges and promising technologies,'' \emph{CoRR}, vol. abs/1804.02217,
  2018. [Online]. Available: \url{http://arxiv.org/abs/1804.02217}
\BIBentrySTDinterwordspacing

\bibitem{lagkas2018uav}
T.~Lagkas, V.~Argyriou, S.~Bibi, and P.~Sarigiannidis, ``Uav iot framework
  views and challenges: towards protecting drones as “things”,''
  \emph{Sensors}, vol.~18, no.~11, p. 4015, 2018.

\bibitem{sekander2018multi}
S.~Sekander, H.~Tabassum, and E.~Hossain, ``Multi-tier drone architecture for
  5g/b5g cellular networks: Challenges, trends, and prospects,'' \emph{IEEE
  Communications Magazine}, vol.~56, no.~3, pp. 96--103, 2018.

\bibitem{naqvi2018drone}
S.~A.~R. Naqvi, S.~A. Hassan, H.~Pervaiz, and Q.~Ni, ``Drone-aided
  communication as a key enabler for 5g and resilient public safety networks,''
  \emph{IEEE Communications Magazine}, vol.~56, no.~1, pp. 36--42, 2018.

\bibitem{selim2018post}
M.~Y. Selim and A.~E. Kamal, ``Post-disaster 4g/5g network rehabilitation using
  drones: Solving battery and backhaul issues,'' in \emph{2018 IEEE Globecom
  Workshops (GC Wkshps)}.\hskip 1em plus 0.5em minus 0.4em\relax IEEE, 2018,
  pp. 1--6.

\bibitem{6G1}
W.~Saad, M.~Bennis, and M.~Chen, ``A vision of 6g wireless systems:
  Applications, trends, technologies, and open research problems,'' \emph{IEEE
  Network}, vol.~34, no.~3, pp. 134--142, 2020.

\bibitem{6G2}
\BIBentryALTinterwordspacing
M.~Mozaffari, X.~Lin, and S.~Hayes, ``Towards 6g with connected sky: Uavs and
  beyond,'' \emph{CoRR}, vol. abs/2103.01143, 2021. [Online]. Available:
  \url{https://arxiv.org/abs/2103.01143}
\BIBentrySTDinterwordspacing

\bibitem{6G3}
S.~Aggarwal, N.~Kumar, and S.~Tanwar, ``Blockchain-envisioned uav communication
  using 6g networks: Open issues, use cases, and future directions,''
  \emph{IEEE Internet of Things Journal}, vol.~8, no.~7, pp. 5416--5441, 2021.

\bibitem{AInetworking1}
U.~Challita, A.~Ferdowsi, M.~Chen, and W.~Saad, ``Machine learning for wireless
  connectivity and security of cellular-connected uavs,'' \emph{IEEE Wireless
  Communications}, vol.~26, no.~1, pp. 28--35, 2019.

\bibitem{AInetworking2}
J.~Park, Y.~Kim, and J.~Seok, ``Prediction of information propagation in a
  drone network by using machine learning,'' in \emph{2016 International
  Conference on Information and Communication Technology Convergence (ICTC)},
  2016, pp. 147--149.

\bibitem{shamsoshoara2020autonomous}
A.~Shamsoshoara, F.~Afghah, A.~Razi, S.~Mousavi, J.~Ashdown, and K.~Turk, ``An
  autonomous spectrum management scheme for unmanned aerial vehicle networks in
  disaster relief operations,'' \emph{IEEE Access}, vol.~8, pp.
  58\,064--58\,079, 2020.

\bibitem{shamsoshoara2019distributed}
A.~Shamsoshoara, M.~Khaledi, F.~Afghah, A.~Razi, and J.~Ashdown, ``Distributed
  cooperative spectrum sharing in uav networks using multi-agent reinforcement
  learning,'' in \emph{2019 16th IEEE Annual Consumer Communications \&
  Networking Conference (CCNC)}.\hskip 1em plus 0.5em minus 0.4em\relax IEEE,
  2019, pp. 1--6.

\bibitem{jalali2017beam}
A.~Jalali and L.~Schiff, ``Beam forming and pointing in a network of unmanned
  aerial vehicles (uavs) for broadband access,'' Jul.~18 2017, uS Patent
  9,712,228.

\bibitem{muralidharan2017energy}
A.~Muralidharan and Y.~Mostofi, ``Energy optimal distributed beamforming using
  unmanned vehicles,'' \emph{IEEE Transactions on Control of Network Systems},
  vol.~5, no.~4, pp. 1529--1540, 2017.

\bibitem{ArnauWiSEE}
A.~{Rovira-Sugranes} and A.~{Razi}, ``Predictive routing for dynamic uav
  networks,'' in \emph{2017 IEEE International Conference on Wireless for Space
  and Extreme Environments (WiSEE)}, Oct 2017, pp. 43--47.

\bibitem{arnaucompression}
A.~{Rovira-Sugranes}, F.~{Afghah}, and A.~{Razi}, ``Optimized compression
  policy for flying ad hoc networks,'' in \emph{2019 16th IEEE Annual Consumer
  Communications Networking Conference (CCNC)}, 2019, pp. 1--2.

\bibitem{arnauscheduling}
A.~{Rovira-Sugranes} and A.~{Razi}, ``Optimizing the age of information for
  blockchain technology with applications to iot sensors,'' \emph{IEEE
  Communications Letters}, vol.~24, no.~1, pp. 183--187, 2020.

\bibitem{swarm1}
M.~C. {Jeffrey}, S.~{Subramanian}, C.~{Yan}, J.~{Emer}, and D.~{Sanchez}, ``A
  scalable architecture for ordered parallelism,'' in \emph{2015 48th Annual
  IEEE/ACM International Symposium on Microarchitecture (MICRO)}, 2015, pp.
  228--241.

\bibitem{swarmmilitary}
Z.~{Xiaoning}, ``Analysis of military application of uav swarm technology,'' in
  \emph{2020 3rd International Conference on Unmanned Systems (ICUS)}, 2020,
  pp. 1200--1204.

\bibitem{swarmantijamming}
J.~{Peng}, Z.~{Zhang}, Q.~{Wu}, and B.~{Zhang}, ``Anti-jamming communications
  in uav swarms: A reinforcement learning approach,'' \emph{IEEE Access},
  vol.~7, pp. 180\,532--180\,543, 2019.

\bibitem{swarmsearch}
L.~{Ruetten}, P.~A. {Regis}, D.~{Feil-Seifer}, and S.~{Sengupta},
  ``Area-optimized uav swarm network for search and rescue operations,'' in
  \emph{2020 10th Annual Computing and Communication Workshop and Conference
  (CCWC)}, 2020, pp. 0613--0618.

\bibitem{swarm2}
M.~{Campion}, P.~{Ranganathan}, and S.~{Faruque}, ``A review and future
  directions of uav swarm communication architectures,'' in \emph{2018 IEEE
  International Conference on Electro/Information Technology (EIT)}, 2018, pp.
  0903--0908.

\bibitem{swarmsurvey}
\BIBentryALTinterwordspacing
A.~Tahir, J.~Böling, M.-H. Haghbayan, H.~T. Toivonen, and J.~Plosila, ``Swarms
  of unmanned aerial vehicles — a survey,'' \emph{Journal of Industrial
  Information Integration}, vol.~16, p. 100106, 2019. [Online]. Available:
  \url{https://www.sciencedirect.com/science/article/pii/S2452414X18300086}
\BIBentrySTDinterwordspacing

\bibitem{swarmarchitectures}
\BIBentryALTinterwordspacing
X.~Chen, J.~Tang, and S.~Lao, ``Review of unmanned aerial vehicle swarm
  communication architectures and routing protocols,'' \emph{Applied Sciences},
  vol.~10, no.~10, 2020. [Online]. Available:
  \url{https://www.mdpi.com/2076-3417/10/10/3661}
\BIBentrySTDinterwordspacing

\bibitem{Bekmezci}
\BIBentryALTinterwordspacing
I.~Bekmezci, O.~K. Sahingoz, and S.~Temel, ``Flying ad-hoc networks (fanets): A
  survey,'' \emph{Ad Hoc Networks}, vol.~11, no.~3, pp. 1254 -- 1270, 2013.
  [Online]. Available:
  \url{http://www.sciencedirect.com/science/article/pii/S1570870512002193}
\BIBentrySTDinterwordspacing

\bibitem{swarm5}
A.~Sivakumar and C.-Y. Tan, ``Uav swarm coordination using cooperative control
  for establishing a wireless communications backbone,'' \emph{9th
  International Joint Conference on Autonomous Agents and Multiagent Systems
  2010}, pp. 1157--1164, 2010.

\bibitem{vranics2019electronic}
D.~F. Vr{\'a}nics, M.~Palik, and B.~Zs, ``Electronic administration of unmanned
  aviation with public key infrastructure (pki),'' \emph{Security \& Future},
  vol.~3, no.~4, pp. 152--155, 2019.

\bibitem{swarm6}
\BIBentryALTinterwordspacing
O.~K. Sahingoz, ``Networking models in flying ad-hoc networks ({FANETs}):
  Concepts and challenges,'' \emph{Journal of Intelligent {\&} Robotic
  Systems}, vol.~74, no. 1-2, pp. 513--527, oct 2013. [Online]. Available:
  \url{https://doi.org/10.1007\%2Fs10846-013-9959-7}
\BIBentrySTDinterwordspacing

\bibitem{swarm7}
Y.~{Zhou}, J.~{Li}, L.~{Lamont}, and C.~{Rabbath}, ``Modeling of packet dropout
  for uav wireless communications,'' in \emph{2012 International Conference on
  Computing, Networking and Communications (ICNC)}, 2012, pp. 677--682.

\bibitem{swarm8}
H.~{Shariatmadari}, R.~{Ratasuk}, S.~{Iraji}, A.~{Laya}, T.~{Taleb},
  R.~{Jäntti}, and A.~{Ghosh}, ``Machine-type communications: current status
  and future perspectives toward 5g systems,'' \emph{IEEE Communications
  Magazine}, vol.~53, no.~9, pp. 10--17, 2015.

\bibitem{swarm9}
F.~{Boccardi}, R.~W. {Heath}, A.~{Lozano}, T.~L. {Marzetta}, and P.~{Popovski},
  ``Five disruptive technology directions for 5g,'' \emph{IEEE Communications
  Magazine}, vol.~52, no.~2, pp. 74--80, 2014.

\bibitem{swarm10}
M.~{Agiwal}, A.~{Roy}, and N.~{Saxena}, ``Next generation 5g wireless networks:
  A comprehensive survey,'' \emph{IEEE Communications Surveys Tutorials},
  vol.~18, no.~3, pp. 1617--1655, 2016.

\bibitem{swarm11}
P.~{Demestichas}, A.~{Georgakopoulos}, D.~{Karvounas}, K.~{Tsagkaris},
  V.~{Stavroulaki}, J.~{Lu}, C.~{Xiong}, and J.~{Yao}, ``5g on the horizon: Key
  challenges for the radio-access network,'' \emph{IEEE Vehicular Technology
  Magazine}, vol.~8, no.~3, pp. 47--53, 2013.

\bibitem{swarmAI1}
G.~Beni and J.~Wang, ``Swarm intelligence in cellular robotic systems,'' in
  \emph{Robots and Biological Systems: Towards a New Bionics?}, P.~Dario,
  G.~Sandini, and P.~Aebischer, Eds.\hskip 1em plus 0.5em minus 0.4em\relax
  Berlin, Heidelberg: Springer Berlin Heidelberg, 1993, pp. 703--712.

\bibitem{swarmAI2}
S.~Awasthi, B.~Balusamy, and V.~Porkodi, ``Artificial intelligence supervised
  swarm uavs for reconnaissance,'' in \emph{Data Science and Analytics},
  U.~Batra, N.~R. Roy, and B.~Panda, Eds.\hskip 1em plus 0.5em minus
  0.4em\relax Singapore: Springer Singapore, 2020, pp. 375--388.

\bibitem{swarmAI3}
\BIBentryALTinterwordspacing
J.~Johnson, ``Artificial intelligence, drone swarming and escalation risks in
  future warfare,'' \emph{The RUSI Journal}, vol. 165, no.~2, pp. 26--36, 2020.
  [Online]. Available: \url{https://doi.org/10.1080/03071847.2020.1752026}
\BIBentrySTDinterwordspacing

\bibitem{swarmAI4}
J.~Kusyk, M.~U. Uyar, K.~Ma, J.~J. Wu, W.~Ruan, D.~K. Guha, G.~Bertoli, and
  J.~Boksiner, ``Ai based flight control for autonomous uav swarms,'' in
  \emph{2018 International Conference on Computational Science and
  Computational Intelligence (CSCI)}, 2018, pp. 1155--1160.

\bibitem{swarmAI5}
J.~Kusyk, M.~U. Uyar, K.~Ma, E.~Samoylov, R.~Valdez, J.~Plishka, S.~E. Hoque,
  G.~Bertoli, and J.~Boksiner, ``Artificial intelligence and game theory
  controlled autonomous uav swarms,'' \emph{Evolutionary Intelligence}, pp.
  1--18, 2020.

\bibitem{camp2002survey}
T.~Camp, J.~Boleng, and V.~Davies, ``A survey of mobility models for ad hoc
  network research,'' \emph{Wireless communications and mobile computing},
  vol.~2, no.~5, pp. 483--502, 2002.

\bibitem{broch1998performance}
J.~Broch, D.~A. Maltz, D.~B. Johnson, Y.-C. Hu, and J.~Jetcheva, ``A
  performance comparison of multi-hop wireless ad hoc network routing
  protocols,'' in \emph{Proceedings of the 4th annual ACM/IEEE international
  conference on Mobile computing and networking}.\hskip 1em plus 0.5em minus
  0.4em\relax ACM, 1998, pp. 85--97.

\bibitem{WANG2010399}
\BIBentryALTinterwordspacing
W.~Wang, X.~Guan, B.~Wang, and Y.~Wang, ``A novel mobility model based on
  semi-random circular movement in mobile ad hoc networks,'' \emph{Information
  Sciences}, vol. 180, no.~3, pp. 399 -- 413, 2010. [Online]. Available:
  \url{http://www.sciencedirect.com/science/article/pii/S0020025509004265}
\BIBentrySTDinterwordspacing

\bibitem{yoon2003sound}
J.~Yoon, M.~Liu, and B.~Noble, ``Sound mobility models,'' in \emph{Proceedings
  of the 9th annual international conference on Mobile computing and
  networking}.\hskip 1em plus 0.5em minus 0.4em\relax ACM, 2003, pp. 205--216.

\bibitem{gonzalez2008understanding}
M.~C. Gonzalez, C.~A. Hidalgo, and A.-L. Barabasi, ``Understanding individual
  human mobility patterns,'' \emph{Nature}, vol. 453, no. 7196, pp. 779--782,
  2008.

\bibitem{kumari2015survey}
K.~Kumari, B.~Sah, and S.~Maakar, ``A survey: different mobility model for
  fanet,'' \emph{International Journal of Advanced Research in Computer Science
  and Software Engineering}, vol.~5, no.~6, 2015.

\bibitem{bekmezci2014connected}
I.~Bekmezci, M.~Ermis, and S.~Kaplan, ``Connected multi uav task planning for
  flying ad hoc networks,'' in \emph{Communications and Networking
  (BlackSeaCom), 2014 IEEE International Black Sea Conference on}.\hskip 1em
  plus 0.5em minus 0.4em\relax IEEE, 2014, pp. 28--32.

\bibitem{lee1999demand}
S.-J. Lee, M.~Gerla, and C.-C. Chiang, ``On-demand multicast routing
  protocol,'' in \emph{Wireless Communications and Networking Conference, 1999.
  WCNC. 1999 IEEE}, vol.~3.\hskip 1em plus 0.5em minus 0.4em\relax IEEE, 1999,
  pp. 1298--1302.

\bibitem{6842277}
O.~Bouachir, A.~Abrassart, F.~Garcia, and N.~Larrieu, ``A mobility model for
  uav ad hoc network,'' in \emph{2014 International Conference on Unmanned
  Aircraft Systems (ICUAS)}, May 2014, pp. 383--388.

\bibitem{kuiper2006mobility}
E.~Kuiper and S.~Nadjm-Tehrani, ``Mobility models for uav group reconnaissance
  applications,'' in \emph{Wireless and Mobile Communications, 2006. ICWMC'06.
  International Conference on}.\hskip 1em plus 0.5em minus 0.4em\relax IEEE,
  2006, pp. 33--33.

\bibitem{hybrid_mobility}
E.~{Kieffer}, G.~{Danoy}, P.~{Bouvry}, and A.~{Nagih}, ``Hybrid mobility model
  with pheromones for uav detection task,'' in \emph{2016 IEEE Symposium Series
  on Computational Intelligence (SSCI)}, Dec 2016, pp. 1--8.

\bibitem{TAPASCologne1}
\BIBentryALTinterwordspacing
``Vehicular mobility trace of the city of cologne, germany,'' 2016. [Online].
  Available: \url{http://kolntrace.project.citi-lab.fr/}
\BIBentrySTDinterwordspacing

\bibitem{TAPASCologne2}
S.~Uppoor, O.~Trullols-Cruces, M.~Fiore, and J.~M. Barcelo-Ordinas,
  ``Generation and analysis of a large-scale urban vehicular mobility
  dataset,'' \emph{IEEE Transactions on Mobile Computing}, vol.~13, no.~5, pp.
  1061--1075, 2014.

\bibitem{pedestrian-motion-modeling1999image}
B.~Boghossian and S.~Velastin, ``Image processing system for pedestrian
  monitoring using neural classification of normal motion patterns,''
  \emph{Measurement and Control}, vol.~32, no.~9, pp. 261--264, 1999.

\bibitem{ossama2011extended}
O.~Ossama, H.~M. Mokhtar, and M.~E. El-Sharkawi, ``An extended k-means
  technique for clustering moving objects,'' \emph{Egyptian Informatics
  Journal}, vol.~12, no.~1, pp. 45--51, 2011.

\bibitem{bennewitz2005learning}
M.~Bennewitz, W.~Burgard, G.~Cielniak, and S.~Thrun, ``Learning motion patterns
  of people for compliant robot motion,'' \emph{The International Journal of
  Robotics Research}, vol.~24, no.~1, pp. 31--48, 2005.

\bibitem{vehicle-motion-modeling2009mobility}
J.~Harri, F.~Filali, and C.~Bonnet, ``Mobility models for vehicular ad hoc
  networks: a survey and taxonomy,'' \emph{IEEE Communications Surveys \&
  Tutorials}, vol.~11, no.~4, 2009.

\bibitem{vehiclemotion2006learning}
S.~Atev, O.~Masoud, and N.~Papanikolopoulos, ``Learning traffic patterns at
  intersections by spectral clustering of motion trajectories,'' in
  \emph{Intelligent Robots and Systems, 2006 IEEE/RSJ International Conference
  on}.\hskip 1em plus 0.5em minus 0.4em\relax IEEE, 2006, pp. 4851--4856.

\bibitem{lee2015unifying}
J.-G. Lee, J.~Han, and X.~Li, ``A unifying framework of mining trajectory
  patterns of various temporal tightness,'' \emph{IEEE Transactions on
  Knowledge and Data Engineering}, vol.~27, no.~6, pp. 1478--1490, 2015.

\bibitem{larsen2007route}
C.~Larsen, M.~Zawodniok, and S.~Jagannathan, ``Route aware predictive
  congestion control protocol for wireless sensor networks,'' in
  \emph{Intelligent Control, 2007. ISIC 2007. IEEE 22nd International Symposium
  on}.\hskip 1em plus 0.5em minus 0.4em\relax IEEE, 2007, pp. 13--18.

\bibitem{tan2016short}
H.~Tan, Y.~Wu, B.~Shen, P.~J. Jin, and B.~Ran, ``Short-term traffic prediction
  based on dynamic tensor completion,'' \emph{IEEE Transactions on Intelligent
  Transportation Systems}, vol.~17, no.~8, pp. 2123--2133, 2016.

\bibitem{choi2006learning}
P.~P. Choi and M.~Hebert, ``Learning and predicting moving object trajectory: a
  piecewise trajectory segment approach,'' \emph{Robotics Institute}, p. 337,
  2006.

\bibitem{aoude2011mobile}
G.~Aoude, J.~Joseph, N.~Roy, and J.~How, ``Mobile agent trajectory prediction
  using bayesian nonparametric reachability trees,'' in \emph{Infotech@
  Aerospace 2011}, 2011, p. 1512.

\bibitem{lee2012identification}
G.~Lee, R.~Mallipeddi, and M.~Lee, ``Identification of moving vehicle
  trajectory using manifold learning,'' in \emph{International Conference on
  Neural Information Processing}.\hskip 1em plus 0.5em minus 0.4em\relax
  Springer, 2012, pp. 188--195.

\bibitem{grewal2011kalman}
M.~S. Grewal, \emph{Kalman filtering}.\hskip 1em plus 0.5em minus 0.4em\relax
  Springer, 2011.

\bibitem{lee2009slaw}
K.~Lee, S.~Hong, S.~J. Kim, I.~Rhee, and S.~Chong, ``Slaw: A new mobility model
  for human walks,'' in \emph{INFOCOM 2009, IEEE}.\hskip 1em plus 0.5em minus
  0.4em\relax IEEE, 2009, pp. 855--863.

\bibitem{ming2015interference}
Z.~X. Ming, Z.~Yue, Y.~Fan, and A.~Vasilakos, ``Interference-based topology
  control algorithm for delay-constrained mobile ad hoc networks, in mobile
  computing,'' \emph{IEEE Trans}, vol.~14, no.~4, pp. 742--754, 2015.

\bibitem{lu2005predictiveMobility}
T.-E. Lu, K.-T. Feng \emph{et~al.}, ``Predictive mobility and location-aware
  routing protocol in mobile ad hoc networks,'' in \emph{GLOBECOM'05: IEEE
  Global Telecommunications Conference, Vols 1-6: DISCOVERY PAST AND FUTURE},
  2005, pp. 899--903.

\bibitem{groenevelt2006relaying}
R.~Groenevelt, E.~Altman, and P.~Nain, ``Relaying in mobile ad hoc networks:
  The brownian motion mobility model,'' \emph{Wireless Networks}, vol.~12,
  no.~5, pp. 561--571, 2006.

\bibitem{owen2015implementing}
M.~Owen, R.~W. Beard, and T.~W. McLain, ``Implementing dubins airplane paths on
  fixed-wing uavs,'' in \emph{Handbook of Unmanned Aerial Vehicles}.\hskip 1em
  plus 0.5em minus 0.4em\relax Springer, 2015, pp. 1677--1701.

\bibitem{wiest2012probabilistic}
J.~Wiest, M.~H{\"o}ffken, U.~Kre{\ss}el, and K.~Dietmayer, ``Probabilistic
  trajectory prediction with gaussian mixture models,'' in \emph{Intelligent
  Vehicles Symposium (IV), 2012 IEEE}.\hskip 1em plus 0.5em minus 0.4em\relax
  IEEE, 2012, pp. 141--146.

\bibitem{chandrashekar2004providing}
K.~Chandrashekar, M.~R. Dekhordi, and J.~S. Baras, ``Providing full
  connectivity in large ad-hoc networks by dynamic placement of aerial
  platforms,'' in \emph{IEEE MILCOM 2004. Military Communications Conference,
  2004.}, vol.~3.\hskip 1em plus 0.5em minus 0.4em\relax IEEE, 2004, pp.
  1429--1436.

\bibitem{connectivity0}
W.~T.~L. Teacy, J.~Nie, S.~McClean, and G.~Parr, ``Maintaining connectivity in
  uav swarm sensing,'' in \emph{2010 IEEE Globecom Workshops}, Dec 2010, pp.
  1771--1776.

\bibitem{connectivityAnt}
M.~Rosalie, M.~R. Brust, G.~Danoy, S.~Chaumette, and P.~Bouvry, ``Coverage
  optimization with connectivity preservation for uav swarms applying chaotic
  dynamics,'' in \emph{2017 IEEE International Conference on Autonomic
  Computing (ICAC)}, July 2017, pp. 113--118.

\bibitem{connectECORA}
R.~Costa, D.~RosÃ¡rio, E.~Cerqueira, and A.~Santos, ``Enhanced connectivity
  for robust multimedia transmission in uav networks,'' in \emph{2014 IFIP
  Wireless Days (WD)}, Nov 2014, pp. 1--6.

\bibitem{alsamhi2019convergence}
S.~Alsamhi, O.~Ma, and M.~S. Ansari, ``Convergence of machine learning and
  robotics communication in collaborative assembly: mobility, connectivity and
  future perspectives,'' \emph{Journal of Intelligent \& Robotic Systems}, pp.
  1--26, 2019.

\bibitem{dijkstra}
\BIBentryALTinterwordspacing
E.~W. Dijkstra, ``A note on two problems in connexion with graphs,''
  \emph{Numer. Math.}, vol.~1, no.~1, pp. 269--271, Dec. 1959. [Online].
  Available: \url{http://dx.doi.org/10.1007/BF01386390}
\BIBentrySTDinterwordspacing

\bibitem{backpressure}
L.~Ying, S.~Shakkottai, A.~Reddy, and S.~Liu, ``On combining shortest-path and
  back-pressure routing over multihop wireless networks,'' \emph{IEEE/ACM
  Transactions on Networking}, vol.~19, no.~3, pp. 841--854, June 2011.

\bibitem{routingreview2014}
\BIBentryALTinterwordspacing
O.~K. Sahingoz, ``Networking models in flying ad-hoc networks (fanets):
  Concepts and challenges,'' \emph{Journal of Intelligent {\&} Robotic
  Systems}, vol.~74, no.~1, pp. 513--527, Apr 2014. [Online]. Available:
  \url{https://doi.org/10.1007/s10846-013-9959-7}
\BIBentrySTDinterwordspacing

\bibitem{OLSR}
T.~H. Clausen and P.~Jacquet, ``Optimized link state routing protocol
  (olsrp),'' vol. 3626, 10 2003.

\bibitem{DOLSR}
A.~Alshabtat, L.~Dong, J.~Li, and F.~Yang, ``Low latency routing algorithm for
  unmanned aerial vehicles ad-hoc networks,'' vol.~6, pp. 48--54, 08 2011.

\bibitem{MPEAOLSR}
S.~Y. Dong, ``Optimization of olsr routing protocol in uav ad hoc network,'' in
  \emph{2016 13th International Computer Conference on Wavelet Active Media
  Technology and Information Processing (ICCWAMTIP)}, Dec 2016, pp. 90--94.

\bibitem{CoutinhoMMRSCK:15}
N.~Coutinho, R.~Matos, C.~Marques, A.~Reis, S.~Sargento, J.~Chakareski, and
  A.~Kassler, ``Dynamic dual-reinforcement-learning routing strategies for
  quality of experience-aware wireless mesh networking,'' \emph{Elsevier
  Computer Networks}, vol.~88, pp. 269--285, Sep. 2015.

\bibitem{DSDV}
\BIBentryALTinterwordspacing
C.~E. Perkins and P.~Bhagwat, ``Highly dynamic destination-sequenced
  distance-vector routing (dsdv) for mobile computers,'' in \emph{Proceedings
  of the Conference on Communications Architectures, Protocols and
  Applications}, ser. SIGCOMM '94.\hskip 1em plus 0.5em minus 0.4em\relax New
  York, NY, USA: ACM, 1994, pp. 234--244. [Online]. Available:
  \url{http://doi.acm.org/10.1145/190314.190336}
\BIBentrySTDinterwordspacing

\bibitem{babel}
\BIBentryALTinterwordspacing
J.~Chroboczek, ``The babel routing protocol.''\hskip 1em plus 0.5em minus
  0.4em\relax IETF RFC 6126, 04 2011. [Online]. Available:
  \url{https://tools.ietf.org/html/rfc6126}
\BIBentrySTDinterwordspacing

\bibitem{GCSR}
M.~Al-Ghazal, A.~El-Sayed, and H.~Kelash, ``Routing optimlzation using genetic
  algorithm in ad hoc networks,'' in \emph{2007 IEEE International Symposium on
  Signal Processing and Information Technology}, Dec 2007, pp. 497--503.

\bibitem{WRP}
S.~A.~N. Shaha, V.~Pai, and U.~K. Shenoy, ``Comparison of wireless routing
  protocols over ftp traffic in mobile and non mobile nodes,'' in \emph{2017
  International Conference on Intelligent Computing, Instrumentation and
  Control Technologies (ICICICT)}, July 2017, pp. 349--353.

\bibitem{TBRPF}
B.~Bellur and R.~G. Ogier, ``A reliable, efficient topology broadcast protocol
  for dynamic networks,'' in \emph{INFOCOM '99. Eighteenth Annual Joint
  Conference of the IEEE Computer and Communications Societies. Proceedings.
  IEEE}, vol.~1, Mar 1999, pp. 178--186 vol.1.

\bibitem{batman}
\BIBentryALTinterwordspacing
A.~Neumann, C.~Aichele, M.~Lindner, and S.~Wunderlich, ``Ietf better approach
  to mobile ad hoc networking (b.a.t.m.a.n.).''\hskip 1em plus 0.5em minus
  0.4em\relax IETF Internet Draft draft-wunderlich-openmesh-manet-routing-00,
  04 2008. [Online]. Available:
  \url{https://tools.ietf.org/html/draft-wunderlichopenmesh-manet-routing-00}
\BIBentrySTDinterwordspacing

\bibitem{DSR}
D.~B.~Johnson and D.~Maltz, ``Dynamic source routing in ad hoc wireless
  networks,'' vol. 353, 05 1999.

\bibitem{AODV}
\BIBentryALTinterwordspacing
S.~Murthy and J.~J. Garcia-Luna-Aceves, ``An efficient routing protocol for
  wireless networks,'' \emph{Mobile Networks and Applications}, vol.~1, no.~2,
  pp. 183--197, Jun 1996. [Online]. Available:
  \url{https://doi.org/10.1007/BF01193336}
\BIBentrySTDinterwordspacing

\bibitem{DTAODV}
Y.~Wang and J.~Liu, ``A backup multipath routing protocol for ad hoc networks
  with dynamic topology,'' in \emph{2012 3rd International Conference on System
  Science, Engineering Design and Manufacturing Informatization}, vol.~1, Oct
  2012, pp. 237--241.

\bibitem{ABR}
P.~M. Carthy and D.~Grigoras, ``Multipath associativity based routing,'' in
  \emph{Second Annual Conference on Wireless On-demand Network Systems and
  Services}, Jan 2005, pp. 60--69.

\bibitem{SSA}
R.~Dube, C.~D. Rais, K.-Y. Wang, and S.~K. Tripathi, ``Signal stability-based
  adaptive routing (ssa) for ad hoc mobile networks,'' \emph{IEEE Personal
  Communications}, vol.~4, no.~1, pp. 36--45, Feb 1997.

\bibitem{MPFR}
S.~S. Naik and A.~U. Bapat, ``Message priority based routing protocol in
  manets,'' in \emph{2015 International Conference on Pervasive Computing
  (ICPC)}, Jan 2015, pp. 1--5.

\bibitem{DBR2P}
\BIBentryALTinterwordspacing
Y.-H. Wang and C.-F. Chao, ``Dynamic backup routes routing protocol for mobile
  ad hoc networks,'' \emph{Information Sciences}, vol. 176, no.~2, pp. 161 --
  185, 2006. [Online]. Available:
  \url{http://www.sciencedirect.com/science/article/pii/S0020025504003275}
\BIBentrySTDinterwordspacing

\bibitem{DYMO}
D.~Dimitrova, M.~Brogle, T.~Braun, G.~Heijenk, and N.~Meratnia, \emph{Joint
  ERCIM eMobility and MobiSense Workshop}.\hskip 1em plus 0.5em minus
  0.4em\relax University of Bern, 6 2012.

\bibitem{TSOR}
J.~Hope~Forsmann, R.~Hiromoto, and J.~Svoboda, ``A time-slotted on-demand
  routing protocol for mobile ad hoc unmanned vehicle systems - art. no.
  65611p,'' 05 2007.

\bibitem{ZRP}
\BIBentryALTinterwordspacing
Z.~J. Haas, M.~R. Pearlman, and P.~Samar, ``The zone routing protocol (zrp) for
  ad hoc networks.''\hskip 1em plus 0.5em minus 0.4em\relax IETF Internet Draft
  draft-ietf-manet-zonezrp-04, 07 2002. [Online]. Available:
  \url{https://www.ietf.org/proceedings/55/I-D/draft-ietf-manet-zone-zrp-04.txt}
\BIBentrySTDinterwordspacing

\bibitem{TORA}
\BIBentryALTinterwordspacing
V.~Park and S.~Corson, ``Temporally-ordered routing algorithm (tora).''\hskip
  1em plus 0.5em minus 0.4em\relax IETF Internet Draft
  draft-ietf-manet-tora-spec-04, 07 2001. [Online]. Available:
  \url{http://www.ietf.org/proceedings/52/I-D/draftietf-manet-tora-spec-04.txt}
\BIBentrySTDinterwordspacing

\bibitem{GPSR}
\BIBentryALTinterwordspacing
B.~Karp and H.~T. Kung, ``Gpsr: Greedy perimeter stateless routing for wireless
  networks,'' in \emph{Proceedings of the 6th Annual International Conference
  on Mobile Computing and Networking}, ser. MobiCom '00.\hskip 1em plus 0.5em
  minus 0.4em\relax New York, NY, USA: ACM, 2000, pp. 243--254. [Online].
  Available: \url{http://doi.acm.org/10.1145/345910.345953}
\BIBentrySTDinterwordspacing

\bibitem{GHG}
C.~Liu and J.~Wu, ``Efficient geometric routing in three dimensional ad hoc
  networks,'' in \emph{IEEE INFOCOM 2009}, April 2009, pp. 2751--2755.

\bibitem{GRG}
R.~Flury and R.~Wattenhofer, ``Randomized 3d geographic routing,'' in
  \emph{IEEE INFOCOM 2008 - The 27th Conference on Computer Communications},
  April 2008.

\bibitem{greedyfor}
Y.~Li, S.~Xie, and Y.~Yu, ``Analysis of greedy forwarding in vehicular ad hoc
  networks,'' in \emph{2011 International Conference on System science,
  Engineering design and Manufacturing informatization}, vol.~2, Oct 2011, pp.
  344--347.

\bibitem{EBGR}
S.~Li, H.~Gao, and D.~Wu, ``An energy-balanced routing protocol with greedy
  forwarding for wsns in cropland,'' in \emph{2016 IEEE International
  Conference on Electronic Information and Communication Technology (ICEICT)},
  Aug 2016, pp. 1--7.

\bibitem{GDSTR}
\BIBentryALTinterwordspacing
J.~Zhou, Y.~Chen, B.~Leong, and P.~S. Sundaramoorthy, ``Practical 3d geographic
  routing for wireless sensor networks,'' in \emph{Proceedings of the 8th ACM
  Conference on Embedded Networked Sensor Systems}, ser. SenSys '10.\hskip 1em
  plus 0.5em minus 0.4em\relax New York, NY, USA: ACM, 2010, pp. 337--350.
  [Online]. Available: \url{http://doi.acm.org/10.1145/1869983.1870016}
\BIBentrySTDinterwordspacing

\bibitem{xlingo}
D.~Rosario, Z.~Zhao, T.~Braun, E.~Cerqueira, A.~Santos, and I.~Alyafawi,
  ``Opportunistic routing for multi-flow video dissemination over flying ad-hoc
  networks,'' in \emph{Proceeding of IEEE International Symposium on a World of
  Wireless, Mobile and Multimedia Networks 2014}, June 2014, pp. 1--6.

\bibitem{AFP}
W.~Qingwen, L.~Gang, L.~Zhi, and Q.~Qian, ``An adaptive forwarding protocol for
  three dimensional flying ad hoc networks,'' in \emph{2015 IEEE 5th
  International Conference on Electronics Information and Emergency
  Communication}, May 2015, pp. 142--145.

\bibitem{RGR}
R.~Shirani, M.~St-Hilaire, T.~Kunz, Y.~Zhou, J.~Li, and L.~Lamont, ``Combined
  reactive-geographic routing for unmanned aeronautical ad-hoc networks,'' in
  \emph{2012 8th International Wireless Communications and Mobile Computing
  Conference (IWCMC)}, Aug 2012, pp. 820--826.

\bibitem{imprvRGR}
Y.~Li, M.~St-Hilaire, and T.~Kunz, ``Enhancements to reduce the overhead of the
  reactive-greedy-reactive routing protocol for unmanned aeronautical ad-hoc
  networks,'' in \emph{2012 8th International Conference on Wireless
  Communications, Networking and Mobile Computing}, Sept 2012, pp. 1--4.

\bibitem{Beaconless}
\BIBentryALTinterwordspacing
D.~Rosario, Z.~Zhao, A.~Santos, T.~Braun, and E.~Cerqueira, ``A beaconless
  opportunistic routing based on a cross-layer approach for efficient video
  dissemination in mobile multimedia iot applications,'' \emph{Computer
  Communications}, vol.~45, pp. 21 -- 31, 2014. [Online]. Available:
  \url{http://www.sciencedirect.com/science/article/pii/S0140366414001388}
\BIBentrySTDinterwordspacing

\bibitem{LODMAC}
\BIBentryALTinterwordspacing
S.~Temel and I.~Bekmezci, ``Lodmac: Location oriented directional mac protocol
  for fanets,'' \emph{Computer Networks}, vol.~83, pp. 76 -- 84, 2015.
  [Online]. Available:
  \url{http://www.sciencedirect.com/science/article/pii/S138912861500081X}
\BIBentrySTDinterwordspacing

\bibitem{exor}
S.~Biswas and R.~Morris, ``Exor: Opportunistic routing in multi-hop wireless
  networks[c],'' pp. 133--143, 01 2005.

\bibitem{LAR}
\BIBentryALTinterwordspacing
Y.~Ko and N.~H. Vaidya, ``Location-aided routing (lar) in mobile ad hoc
  networks,'' \emph{Wireless Networks}, vol.~6, no.~4, pp. 307--321, Sep 2000.
  [Online]. Available: \url{https://doi.org/10.1023/A:1019106118419}
\BIBentrySTDinterwordspacing

\bibitem{cbrp}
\BIBentryALTinterwordspacing
J.~Mingliang, L.~Jinyang, and T.~Y. Chiang, ``Cluster based routing
  protocol(cbrp) functional specification,'' Aug 1998. [Online]. Available:
  \url{https://tools.ietf.org/id/draft-ietf-manet-cbrp-spec-00.txt}
\BIBentrySTDinterwordspacing

\bibitem{ecgsr}
K.~Devarajan and V.~Padmathilagam, ``An enhanced cluster gateway switch routing
  protocol (ecgsr) for congestion control using aodv algorithm in manet,''
  \emph{International Journal of Computer Applications}, vol. 123, no.~3, pp.
  37--42, 2015.

\bibitem{FSR1}
\BIBentryALTinterwordspacing
K.~Obraczka, K.~Viswanath, and G.~Tsudik, ``Flooding for reliable multicast in
  multi-hop ad hoc networks,'' \emph{Wireless Networks}, vol.~7, no.~6, pp.
  627--634, Nov 2001. [Online]. Available:
  \url{https://doi.org/10.1023/A:1012323519059}
\BIBentrySTDinterwordspacing

\bibitem{FSR2}
Y.~Sasson, D.~Cavin, and A.~Schiper, ``Probabilistic broadcast for flooding in
  wireless mobile ad hoc networks,'' in \emph{2003 IEEE Wireless Communications
  and Networking, 2003. WCNC 2003.}, vol.~2, March 2003, pp. 1124--1130 vol.2.

\bibitem{randomwalk}
H.~Tian, H.~Shen, and T.~Matsuzawa, ``Randomwalk routing for wireless sensor
  networks,'' in \emph{Sixth International Conference on Parallel and
  Distributed Computing Applications and Technologies (PDCAT'05)}, Dec 2005,
  pp. 196--200.

\bibitem{mimorandom}
X.~Li, X.~Tao, and N.~Li, ``Energy-efficient cooperative mimo-based random walk
  routing for wireless sensor networks,'' \emph{IEEE Communications Letters},
  vol.~20, no.~11, pp. 2280--2283, Nov 2016.

\bibitem{PPMAC}
Z.~Zheng, A.~K. Sangaiah, and T.~Wang, ``Adaptive communication protocols in
  flying ad hoc network,'' \emph{IEEE Communications Magazine}, vol.~56, no.~1,
  pp. 136--142, Jan 2018.

\bibitem{rosati1}
S.~Rosati, K.~Kruzelecki, L.~Traynard, and B.~R. Mobile, ``Speed-aware routing
  for uav ad-hoc networks,'' in \emph{2013 IEEE Globecom Workshops (GC
  Wkshps)}, Dec 2013, pp. 1367--1373.

\bibitem{rosati2}
S.~Rosati, K.~Kruzelecki, G.~Heitz, D.~Floreano, and B.~Rimoldi, ``Dynamic
  routing for flying ad hoc networks,'' \emph{IEEE Transactions on Vehicular
  Technology}, vol.~65, no.~3, pp. 1690--1700, March 2016.

\bibitem{GPMOR}
L.~Lin, Q.~Sun, J.~Li, and F.~Yang, ``A novel geographic position mobility
  oriented routing strategy for uavs,'' vol.~8, pp. 709--716, 02 2012.

\bibitem{MPC}
C.~Zang and S.~Zang, ``Mobility prediction clustering algorithm for uav
  networking,'' in \emph{2011 IEEE GLOBECOM Workshops (GC Wkshps)}, Dec 2011,
  pp. 1158--1161.

\bibitem{RARP}
G.~{Gankhuyag}, A.~P. {Shrestha}, and S.~{Yoo}, ``Robust and reliable
  predictive routing strategy for flying ad-hoc networks,'' \emph{IEEE Access},
  vol.~5, pp. 643--654, 2017.

\bibitem{imprvRGR2}
Y.~Li, M.~St-Hilaire, and T.~Kunz, ``Improving routing in networks of uavs via
  scoped flooding and mobility prediction,'' in \emph{2012 IFIP Wireless Days},
  Nov 2012, pp. 1--6.

\bibitem{QGeo}
W.~{Jung}, J.~{Yim}, and Y.~{Ko}, ``Qgeo: Q-learning-based geographic ad hoc
  routing protocol for unmanned robotic networks,'' \emph{IEEE Communications
  Letters}, vol.~21, no.~10, pp. 2258--2261, 2017.

\bibitem{parrot}
B.~Sliwa, C.~Schüler, M.~Patchou, and C.~Wietfeld, ``Parrot: Predictive ad-hoc
  routing fueled by reinforcement learning and trajectory knowledge,'' 2020.

\bibitem{fuzzy2}
C.~{He}, S.~{Liu}, and S.~{Han}, ``A fuzzy logic reinforcement learning-based
  routing algorithm for flying ad hoc networks,'' in \emph{2020 International
  Conference on Computing, Networking and Communications (ICNC)}, 2020, pp.
  987--991.

\bibitem{adap_q_routing}
Y.~{Shilova}, M.~{Kavalerov}, and I.~{Bezukladnikov}, ``Full echo q-routing
  with adaptive learning rates: A reinforcement learning approach to network
  routing,'' in \emph{IEEE Conf. EIConRusNW}, Feb 2016, pp. 341--344.

\bibitem{aqrerm}
M.~Kavalerov, Y.~Shilova, and Y.~Likhacheva, ``Adaptive q-routing with random
  echo and route memory,'' in \emph{Proceedings of the 20th Conference of Open
  Innovations Association FRUCT}.\hskip 1em plus 0.5em minus 0.4em\relax
  Helsinki, Finland, Finland: FRUCT Oy, 2017, pp. 20:138--20:145.

\bibitem{pbQ-Routing}
D.~Sharma, D.~Kukreja, P.~Aggarwal, M.~Kaur, and A.~Sachan, ``Poisson's
  probability‐based q‐routing techniques for message forwarding in
  opportunistic networks,'' \emph{International Journal of Communication
  Systems}, vol.~31, 05 2018.

\bibitem{q2routing}
T.~{Hendriks}, M.~{Camelo}, and S.~{Latré}, ``Q2-routing : A qos-aware
  q-routing algorithm for wireless ad hoc networks,'' in \emph{2018 14th
  International Conference on Wireless and Mobile Computing, Networking and
  Communications (WiMob)}, Oct 2018, pp. 108--115.

\bibitem{dQ-Routing}
F.~Wang, R.~Feng, and H.~Chen, ``Dynamic routing algorithm with q-learning for
  internet of things with delayed estimator,'' \emph{{IOP} Conference Series:
  Earth and Environmental Science}, vol. 234, p. 012048, mar 2019.

\bibitem{QNGPSR}
N.~{Lyu}, G.~{Song}, B.~{Yang}, and Y.~{Cheng}, ``Qngpsr: A q-network enhanced
  geographic ad-hoc routing protocol based on gpsr,'' in \emph{2018 IEEE 88th
  Vehicular Technology Conference (VTC-Fall)}, 2018, pp. 1--6.

\bibitem{ardeep}
J.~{LIU}, Q.~{WANG}, C.~{HE}, and Y.~{XU}, ``Ardeep: Adaptive and reliable
  routing protocol for mobile robotic networks with deep reinforcement
  learning,'' in \emph{2020 IEEE 45th Conference on Local Computer Networks
  (LCN)}, 2020, pp. 465--468.

\bibitem{TQNGPSR}
\BIBentryALTinterwordspacing
Y.~Chen, N.~Lyu, G.~Song \emph{et~al.}, ``A traffic-aware q-network enhanced
  routing protocol based on gpsr for unmanned aerial vehicle ad-hoc networks,''
  \emph{Front Inform Technol Electron Eng}, vol.~21, pp. 1308--1320, 2020.
  [Online]. Available: \url{https://doi.org/10.1631/FITEE.1900401}
\BIBentrySTDinterwordspacing

\bibitem{QMR}
\BIBentryALTinterwordspacing
J.~Liu, Q.~Wang, C.~He, K.~Jaffrès-Runser, Y.~Xu, Z.~Li, and Y.~Xu,
  ``Qmr:q-learning based multi-objective optimization routing protocol for
  flying ad hoc networks,'' \emph{Computer Communications}, vol. 150, pp.
  304--316, 2020. [Online]. Available:
  \url{https://www.sciencedirect.com/science/article/pii/S0140366419308278}
\BIBentrySTDinterwordspacing

\bibitem{fuzzy1}
\BIBentryALTinterwordspacing
Q.~Yang, S.~Jang, and S.~Yoo, ``Q-learning-based fuzzy logic for
  multi-objective routing algorithm in flying ad hoc networks,'' \emph{Wireless
  Pers Commun}, vol. 113, pp. 115--138, 2020. [Online]. Available:
  \url{https://doi.org/10.1007/s11277-020-07181-w}
\BIBentrySTDinterwordspacing

\bibitem{fuzzy3}
S.~{Jiang}, Z.~{Huang}, and Y.~{Ji}, ``Adaptive uav-assisted geographic routing
  with q-learning in vanet,'' \emph{IEEE Communications Letters}, pp. 1--1,
  2020.

\bibitem{ARPL}
\BIBentryALTinterwordspacing
J.~Wu, M.~Fang, and X.~Li, ``Reinforcement learning based mobility adaptive
  routing for vehicular ad-hoc networks,'' vol. 101, no.~4, p. 2143–2171,
  Aug. 2018. [Online]. Available:
  \url{https://doi.org/10.1007/s11277-018-5809-z}
\BIBentrySTDinterwordspacing

\bibitem{U2RV}
O.~Oubbati, N.~CHAIB, A.~Lakas, S.~Bitam, and P.~Lorenz, ``U2rv: Uav‐assisted
  reactive routing protocol for vanets,'' \emph{International Journal of
  Communication Systems}, vol.~PP, pp. 1--13, 08 2019.

\bibitem{rovirasugranes2021fullyechoed}
A.~Rovira-Sugranes, F.~Afghah, J.~Qu, and A.~Razi, ``Fully-echoed q-routing
  with simulated annealing inference for flying adhoc networks,'' \emph{IEEE
  Transactions on Network Science and Engineering}, 2021.

\bibitem{reinforcementlearning}
R.~S. Sutton and A.~G. Barto, ``Reinforcement learning: An introduction,''
  \emph{IEEE Transactions on Neural Networks}, vol.~16, pp. 285--286, 1988.

\bibitem{q_routing}
J.~A. Boyan and M.~L. Littman, ``Packet routing in dynamically changing
  networks: A reinforcement learning approach,'' in \emph{Int. Conf. on Neural
  Info. Processing Systems}.\hskip 1em plus 0.5em minus 0.4em\relax San
  Francisco, CA, USA: Morgan Kaufmann Publishers Inc., 1993, pp. 671--678.

\bibitem{network_qrouting}
S.~{Khodayari} and M.~J. {Yazdanpanah}, ``Network routing based on
  reinforcement learning in dynamically changing networks,'' in
  \emph{ICTAI'05}, 2005, pp. 5 pp.--366.

\bibitem{pred_q_routing}
S.~P.~M. Choi and D.-Y. Yeung, ``Predictive q-routing: A memory-based
  reinforcement learning approach to adaptive traffic control,'' in \emph{Conf.
  on Neural Info. Processing Systems}.\hskip 1em plus 0.5em minus 0.4em\relax
  Cambridge, MA, USA: MIT Press, 1995, p. 945–951.

\bibitem{drq_routing}
S.~Kumar and R.~Mukkulainen, ``Confidence based dual reinforcement q-routing:
  An adaptive online network routing algorithm,'' in \emph{IJCAI'99}.\hskip 1em
  plus 0.5em minus 0.4em\relax San Francisco, CA, USA: Morgan Kaufmann
  Publishers Inc., 1999, pp. 758--763.

\bibitem{comparison}
F.~Tekiner, Z.~Ghassemlooy, and T.~R. Srikanth, ``Comparison of the q-routing
  and shortest path routing algorithm,'' 2004.

\bibitem{crq_routing}
N.~Gupta, M.~Kumar, A.~Sharma, M.~S. Gaur, V.~Laxmi, M.~Daneshtalab, and
  M.~Ebrahimi, ``Improved route selection approaches using q-learning framework
  for 2d nocs,'' in \emph{Int. Workshop on Many-core Embedded Systems}.\hskip
  1em plus 0.5em minus 0.4em\relax New York, NY, USA: ACM, 2015, pp. 33--40.

\bibitem{xplane}
\BIBentryALTinterwordspacing
``Plane 11 flight simulator: More powerful. made usable.'' Sep 2020. [Online].
  Available: \url{https://www.x-plane.com/}
\BIBentrySTDinterwordspacing

\bibitem{flightgear}
A.~Perry, ``The flightgear flight simulator,'' 01 2004.

\bibitem{gazebo}
\BIBentryALTinterwordspacing
Osrf, ``Why gazebo?'' [Online]. Available: \url{http://gazebosim.org/}
\BIBentrySTDinterwordspacing

\bibitem{jmavsim}
\BIBentryALTinterwordspacing
``jmavsim simulation.'' [Online]. Available:
  \url{https://dev.px4.io/v1.9.0/en/simulation/jmavsim.html}
\BIBentrySTDinterwordspacing

\bibitem{airsim}
\BIBentryALTinterwordspacing
S.~Shah, D.~Dey, C.~Lovett, and A.~Kapoor, ``Airsim: High-fidelity visual and
  physical simulation for autonomous vehicles,'' \emph{CoRR}, vol.
  abs/1705.05065, 2017. [Online]. Available:
  \url{http://arxiv.org/abs/1705.05065}
\BIBentrySTDinterwordspacing

\bibitem{ue4sim}
\BIBentryALTinterwordspacing
M.~M{\"{u}}ller, V.~Casser, J.~Lahoud, N.~Smith, and B.~Ghanem, ``Ue4sim: {A}
  photo-realistic simulator for computer vision applications,'' \emph{CoRR},
  vol. abs/1708.05869, 2017. [Online]. Available:
  \url{http://arxiv.org/abs/1708.05869}
\BIBentrySTDinterwordspacing

\bibitem{qgroundcontrol}
\BIBentryALTinterwordspacing
``Overview - qgroundcontrol user guide.'' [Online]. Available:
  \url{https://docs.qgroundcontrol.com/master/en/index.html}
\BIBentrySTDinterwordspacing

\bibitem{missionplanner}
\BIBentryALTinterwordspacing
``Mission planner ground control station.'' [Online]. Available:
  \url{https://ardupilot.org/planner/docs/mission-planner-ground-control-station.html}
\BIBentrySTDinterwordspacing

\bibitem{mavproxy}
A.-M. Andreescu, M.~Dima, A.~Istrate, C.~Muresan, C.~Visoiu, and P.~Parvu,
  ``Autonomous system for image geo-tagging and target recognition,'' 05 2014.

\bibitem{ugcs}
\BIBentryALTinterwordspacing
S.~E.~. UgCS, ``Leading drone control software.'' [Online]. Available:
  \url{https://www.ugcs.com/}
\BIBentrySTDinterwordspacing

\bibitem{simulation_survey}
A.~I. {Hentati}, L.~{Krichen}, M.~{Fourati}, and L.~C. {Fourati}, ``Simulation
  tools, environments and frameworks for uav systems performance analysis,'' in
  \emph{2018 14th International Wireless Communications Mobile Computing
  Conference (IWCMC)}, 2018, pp. 1495--1500.

\bibitem{SITL}
C.~{Coopmans}, M.~{Podhradský}, and N.~V. {Hoffer}, ``Software- and
  hardware-in-the-loop verification of flight dynamics model and flight control
  simulation of a fixed-wing unmanned aerial vehicle,'' in \emph{2015 Workshop
  on Research, Education and Development of Unmanned Aerial Systems (RED-UAS)},
  2015, pp. 115--122.

\bibitem{MicrosoftAirSim}
\BIBentryALTinterwordspacing
Microsoft, ``Microsoft airsim: Open source simulator based on unreal engine for
  autonomous vehicles from microsoft ai and research,'' Feb 2017. [Online].
  Available: \url{https://github.com/Microsoft/AirSim}
\BIBentrySTDinterwordspacing

\bibitem{airsim2}
\BIBentryALTinterwordspacing
J.~Antunes, ``Airsim: A simulator to help ai research for use in drones,'' Mar
  2017. [Online]. Available:
  \url{https://www.commercialuavnews.com/public-safety/airsim-simulator-help-artificial-intelligence-research-use-drones}
\BIBentrySTDinterwordspacing

\bibitem{ns-3}
\BIBentryALTinterwordspacing
``Ns-3 tutorial,'' Oct 2020. [Online]. Available:
  \url{https://www.nsnam.org/docs/release/3.32/tutorial/html/index.html}
\BIBentrySTDinterwordspacing

\bibitem{opnet}
\BIBentryALTinterwordspacing
``Opnet network simulator,'' Apr 2020. [Online]. Available:
  \url{https://opnetprojects.com/opnet-network-simulator/}
\BIBentrySTDinterwordspacing

\bibitem{ros-netsim}
M.~{Calvo-Fullana}, D.~{Mox}, A.~{Pyattaev}, J.~{Fink}, V.~{Kumar}, and
  A.~{Ribeiro}, ``Ros-netsim: A framework for the integration of robotic and
  network simulators,'' \emph{IEEE Robotics and Automation Letters}, vol.~6,
  no.~2, pp. 1120--1127, 2021.

\bibitem{ub-anc}
\BIBentryALTinterwordspacing
S.~M. Najafabadi, N.~Mastronarde, M.~J. Medley, and J.~D. Matyjas, ``{UB-ANC:}
  an open platform testbed for software-defined airborne networking and
  communications,'' \emph{CoRR}, vol. abs/1509.08346, 2015. [Online].
  Available: \url{http://arxiv.org/abs/1509.08346}
\BIBentrySTDinterwordspacing

\bibitem{AERPAW1}
V.~{Marojevic}, I.~{Guvenc}, R.~{Dutta}, M.~L. {Sichitiu}, and B.~A. {Floyd},
  ``Advanced wireless for unmanned aerial systems: 5g standardization, research
  challenges, and aerpaw architecture,'' \emph{IEEE Vehicular Technology
  Magazine}, vol.~15, no.~2, pp. 22--30, 2020.

\bibitem{AERPAW2}
\BIBentryALTinterwordspacing
M.~L. Sichitiu, I.~Guvenc, R.~Dutta, V.~Marojevic, and B.~Floyd, ``Aerpaw
  emulation overview,'' in \emph{Proceedings of the 14th International Workshop
  on Wireless Network Testbeds, Experimental Evaluation \& Characterization},
  ser. WiNTECH'20.\hskip 1em plus 0.5em minus 0.4em\relax New York, NY, USA:
  Association for Computing Machinery, 2020, p. 1–8. [Online]. Available:
  \url{https://doi.org/10.1145/3411276.3412188}
\BIBentrySTDinterwordspacing

\bibitem{AERPAW3}
\BIBentryALTinterwordspacing
``Nc state named a hot spot for 5g innovation,'' Dec 2019. [Online]. Available:
  \url{https://news.ncsu.edu/2019/12/5g-wireless-network-aerpaw/}
\BIBentrySTDinterwordspacing

\bibitem{POWDER}
J.~Breen, A.~Buffmire, J.~Duerig, K.~Dutt, E.~Eide, M.~Hibler, D.~Johnson,
  S.~K. Kasera, E.~Lewis, D.~Maas, A.~Orange, N.~Patwari, D.~Reading, R.~Ricci,
  D.~Schurig, L.~B. Stoller, J.~Van~der Merwe, K.~Webb, and G.~Wong,
  ``{POWDER}: Platform for open wireless data-driven experimental research,''
  in \emph{Proceedings of the 14th International Workshop on Wireless Network
  Testbeds, Experimental Evaluation and Characterization (WiNTECH)}, Sep. 2020.

\bibitem{ns3gym}
\BIBentryALTinterwordspacing
P.~Gaw{\l}owicz and A.~Zubow, ``{ns-3 meets OpenAI Gym: The Playground for
  Machine Learning in Networking Research},'' in \emph{{ACM International
  Conference on Modeling, Analysis and Simulation of Wireless and Mobile
  Systems (MSWiM)}}, November 2019. [Online]. Available:
  \url{http://www.tkn.tu-berlin.de/fileadmin/fg112/Papers/2019/gawlowicz19_mswim.pdf}
\BIBentrySTDinterwordspacing

\bibitem{ns3ai}
H.~Yin, P.~Liu, L.~Zhang, L.~Cao, Y.~Gao, X.~Hei, and X.-J. Hei, ``Ns3-ai:
  Enable applying artificial intelligence to network simulation in ns-3,'' 06
  2020.

\bibitem{AERPAW4}
\BIBentryALTinterwordspacing
``Aerpaw equipment.'' [Online]. Available:
  \url{https://aerpaw.org/aerpaw-equipment/}
\BIBentrySTDinterwordspacing

\bibitem{blackbird}
A.~Antonini, W.~Guerra, V.~Murali, T.~Sayre-McCord, and S.~Karaman, ``The
  blackbird dataset: A large-scale dataset for uav perception in aggressive
  flight,'' 2018.

\bibitem{KAUST}
\BIBentryALTinterwordspacing
N.~Smith, N.~Moehrle, M.~Goesele, and W.~Heidrich, ``Uav pathplanning dataset
  \& benchmark,'' 2018. [Online]. Available:
  \url{http://hdl.handle.net/10754/630159}
\BIBentrySTDinterwordspacing

\bibitem{midair}
\BIBentryALTinterwordspacing
M.~F. Vandroogenbroeck and Marc, ``Mid-air dataset.'' [Online]. Available:
  \url{https://midair.ulg.ac.be/}
\BIBentrySTDinterwordspacing

\bibitem{uzh}
J.~Delmerico, T.~Cieslewski, H.~Rebecq, M.~Faessler, and D.~Scaramuzza, ``Are
  we ready for autonomous drone racing? the {UZH-FPV} drone racing dataset,''
  in \emph{{IEEE} Int. Conf. Robot. Autom. ({ICRA})}, 2019.

\bibitem{trajectorydeep1}
X.~Pan, P.~Desbarats, and S.~Chaumette, ``A deep learning based framework for
  uav trajectory pattern recognition,'' in \emph{2019 Ninth International
  Conference on Image Processing Theory, Tools and Applications (IPTA)}, 2019,
  pp. 1--6.

\bibitem{trajectoryml}
X.~Liu, Y.~Liu, Y.~Chen, and L.~Hanzo, ``Trajectory design and power control
  for multi-uav assisted wireless networks: A machine learning approach,''
  \emph{IEEE Transactions on Vehicular Technology}, vol.~68, no.~8, pp.
  7957--7969, 2019.

\bibitem{challengescomm}
\BIBentryALTinterwordspacing
H.~Wang, J.~Wang, J.~Chen, Y.~Gong, and G.~Ding, ``Network-connected {UAV}
  communications: Potentials and challenges,'' \emph{CoRR}, vol.
  abs/1806.04583, 2018. [Online]. Available:
  \url{http://arxiv.org/abs/1806.04583}
\BIBentrySTDinterwordspacing

\bibitem{challenges_comm}
\BIBentryALTinterwordspacing
M.~Cooney, ``Drones still face major communications challenges getting onto us
  airspace,'' 2013. [Online]. Available:
  \url{https://www.networkworld.com/article/2224063}
\BIBentrySTDinterwordspacing

\bibitem{challenges_comp}
M.~{Mozaffari}, W.~{Saad}, M.~{Bennis}, Y.~{Nam}, and M.~{Debbah}, ``A tutorial
  on uavs for wireless networks: Applications, challenges, and open problems,''
  \emph{IEEE Communications Surveys Tutorials}, 2019.

\bibitem{challengessurvey}
\BIBentryALTinterwordspacing
A.~Fotouhi, H.~Qiang, M.~Ding, M.~Hassan, L.~G. Giordano,
  A.~Garc{\'{\i}}a{-}Rodr{\'{\i}}guez, and J.~Yuan, ``Survey on {UAV} cellular
  communications: Practical aspects, standardization advancements, regulation,
  and security challenges,'' \emph{CoRR}, vol. abs/1809.01752, 2018. [Online].
  Available: \url{http://arxiv.org/abs/1809.01752}
\BIBentrySTDinterwordspacing

\bibitem{challengessurvey3}
\BIBentryALTinterwordspacing
H.~Shakhatreh, A.~Sawalmeh, A.~I. Al{-}Fuqaha, Z.~Dou, E.~Almaita, I.~Khalil,
  N.~S. Othman, A.~Khreishah, and M.~Guizani, ``Unmanned aerial vehicles: {A}
  survey on civil applications and key research challenges,'' \emph{CoRR}, vol.
  abs/1805.00881, 2018. [Online]. Available:
  \url{http://arxiv.org/abs/1805.00881}
\BIBentrySTDinterwordspacing

\bibitem{futureAI4}
Y.~{Fu}, S.~{Wang}, C.~{Wang}, X.~{Hong}, and S.~{McLaughlin}, ``Artificial
  intelligence to manage network traffic of 5g wireless networks,'' \emph{IEEE
  Network}, vol.~32, no.~6, pp. 58--64, 2018.

\bibitem{futureAI6}
L.~{Xiao}, X.~{Lu}, D.~{Xu}, Y.~{Tang}, L.~{Wang}, and W.~{Zhuang}, ``Uav relay
  in vanets against smart jamming with reinforcement learning,'' \emph{IEEE
  Transactions on Vehicular Technology}, vol.~67, no.~5, pp. 4087--4097, 2018.

\bibitem{futureAI7}
P.~Fraga-Lamas, L.~Ramos, V.~M. Mondéjar-Guerra, and T.~Fernández-Caramés,
  ``A review on iot deep learning uav systems for autonomous obstacle detection
  and collision avoidance,'' \emph{Remote Sensing}, vol.~11, p. 2144, 09 2019.

\bibitem{islam2019fire}
S.~Islam, Q.~Huang, F.~Afghah, P.~Fule, and A.~Razi, ``Fire frontline
  monitoring by enabling uav-based virtual reality with adaptive imaging
  rate,'' in \emph{2019 53rd Asilomar Conference on Signals, Systems, and
  Computers}.\hskip 1em plus 0.5em minus 0.4em\relax IEEE, 2019, pp. 368--372.

\bibitem{erdelj2016uav}
M.~Erdelj and E.~Natalizio, ``Uav-assisted disaster management: Applications
  and open issues,'' in \emph{2016 international conference on computing,
  networking and communications (ICNC)}.\hskip 1em plus 0.5em minus 0.4em\relax
  IEEE, 2016, pp. 1--5.

\bibitem{gonzalez2016unmanned}
L.~F. Gonzalez, G.~A. Montes, E.~Puig, S.~Johnson, K.~Mengersen, and K.~J.
  Gaston, ``Unmanned aerial vehicles (uavs) and artificial intelligence
  revolutionizing wildlife monitoring and conservation,'' \emph{Sensors},
  vol.~16, no.~1, p.~97, 2016.

\bibitem{futurerouting1}
M.~Y. {Arafat} and S.~{Moh}, ``Location-aided delay tolerant routing protocol
  in uav networks for post-disaster operation,'' \emph{IEEE Access}, vol.~6,
  pp. 59\,891--59\,906, 2018.

\bibitem{futureenergy1}
B.~{Uragun}, ``Energy efficiency for unmanned aerial vehicles,'' in \emph{2011
  10th International Conference on Machine Learning and Applications and
  Workshops}, vol.~2, 2011, pp. 316--320.

\bibitem{futureenergy2}
\BIBentryALTinterwordspacing
H.~Peng, A.~Razi, F.~Afghah, and J.~D. Ashdown, ``A unified framework for joint
  mobility prediction and object profiling of drones in {UAV} networks,''
  \emph{CoRR}, vol. abs/1808.00058, 2018. [Online]. Available:
  \url{http://arxiv.org/abs/1808.00058}
\BIBentrySTDinterwordspacing

\bibitem{futureenergy3}
J.~{Chen}, U.~{Mitra}, and D.~{Gesbert}, ``Optimal uav relay placement for
  single user capacity maximization over terrain with obstacles,'' in
  \emph{2019 IEEE 20th International Workshop on Signal Processing Advances in
  Wireless Communications (SPAWC)}, 2019, pp. 1--5.

\bibitem{futureenergy5}
Z.~Pan, L.~An, and C.~Wen, ``Recent advances in fuel cells based propulsion
  systems for unmanned aerial vehicles,'' \emph{Applied Energy}, vol. 240, pp.
  473--485, 04 2019.

\bibitem{6futureenergy}
E.~F. Costa, D.~A. Souza, V.~P. Pinto, M.~S. Ara{\'u}jo, A.~Peixoto, and
  E.~D.~C. J{\'u}nior, ``Prediction of lithium-ion battery capacity in uavs,''
  \emph{2019 6th International Conference on Control, Decision and Information
  Technologies (CoDIT)}, pp. 1865--1869, 2019.

\bibitem{7futureenergy}
Q.~{Wu}, G.~{Zhang}, D.~W.~K. {Ng}, W.~{Chen}, and R.~{Schober}, ``Generalized
  wireless-powered communications: When to activate wireless power transfer?''
  \emph{IEEE Transactions on Vehicular Technology}, vol.~68, no.~8, pp.
  8243--8248, 2019.

\bibitem{8futureenergy}
A.~Kurs, A.~Karalis, R.~Moffatt, J.~Joannopoulos, P.~Fisher, and M.~Soljacic,
  ``Wireless power transfer via strongly coupled magnetic resonances,''
  \emph{Science (New York, N.Y.)}, vol. 317, pp. 83--6, 08 2007.

\bibitem{9futureenergy}
M.~Khonji, M.~Alshehhi, C.-M. Tseng, and C.-K. Chau, ``Autonomous inductive
  charging system for battery-operated electric drones,'' 05 2017, pp.
  322--327.

\bibitem{10futureenergy}
M.~{Nguyen}, L.~D. {Nguyen}, T.~Q. {Duong}, and H.~D. {Tuan}, ``Real-time
  optimal resource allocation for embedded uav communication systems,''
  \emph{IEEE Wireless Communications Letters}, vol.~8, no.~1, pp. 225--228,
  2019.

\bibitem{11futureenergy}
S.~{Bi} and R.~{Zhang}, ``Placement optimization of energy and information
  access points in wireless powered communication networks,'' \emph{IEEE
  Transactions on Wireless Communications}, vol.~15, no.~3, pp. 2351--2364,
  2016.

\bibitem{12futureenergy}
A.~M.~R. {Tolba}, ``Trust-based distributed authentication method for collision
  attack avoidance in vanets,'' \emph{IEEE Access}, vol.~6, pp.
  62\,747--62\,755, 2018.

\bibitem{futureenergy_neural}
T.-J. Yang, Y.-H. Chen, and V.~Sze, ``Designing energy-efficient convolutional
  neural networks using energy-aware pruning,'' in \emph{Proceedings of the
  IEEE Conference on Computer Vision and Pattern Recognition}, 2017, pp.
  5687--5695.

\bibitem{energymetrics}
T.~Johann, M.~Dick, S.~Naumann, and E.~Kern, ``How to measure energy-efficiency
  of software: Metrics and measurement results,'' in \emph{2012 First
  International Workshop on Green and Sustainable Software (GREENS)}.\hskip 1em
  plus 0.5em minus 0.4em\relax IEEE, 2012, pp. 51--54.

\bibitem{energyml}
E.~Garc{\'\i}a~Mart{\'\i}n, ``Energy efficiency in machine learning: A position
  paper,'' in \emph{30th Annual Workshop of the Swedish Artificial Intelligence
  Society SAIS, Karlskrona}, vol. 137.\hskip 1em plus 0.5em minus 0.4em\relax
  Link{\"o}ping University Electronic Press, 2017, pp. 68--72.

\bibitem{spectrum1}
C.~{Zhang} and W.~{Zhang}, ``Spectrum sharing for drone networks,'' \emph{IEEE
  Journal on Selected Areas in Communications}, vol.~35, no.~1, pp. 136--144,
  Jan 2017.

\bibitem{spectrum2}
\BIBentryALTinterwordspacing
J.~Kakar and V.~Marojevic, ``Waveform and spectrum management for unmanned
  aerial systems beyond 2025,'' \emph{CoRR}, vol. abs/1708.01664, 2017.
  [Online]. Available: \url{http://arxiv.org/abs/1708.01664}
\BIBentrySTDinterwordspacing

\bibitem{fiberoptic}
F.~Delgado, J.~Carvalho, T.~Coelho, and A.~D. Santos, ``An optical fiber sensor
  and its application in uavs for current measurements,'' \emph{Sensors},
  vol.~16, no.~11, p. 1800, 2016.

\bibitem{laser}
W.~Griethe, M.~Gregory, F.~Heine, and H.~Kampfner, ``High speed laser
  communications in uav scenarios,'' 05 2011.

\bibitem{lifi}
\BIBentryALTinterwordspacing
B.~O. Sadiq, ``A feasibility study of using light fidelity with multiple
  unmanned aerial vehicles for indoor collaborative and cooperative
  networking,'' \emph{CoRR}, vol. abs/1707.08627, 2017. [Online]. Available:
  \url{http://arxiv.org/abs/1707.08627}
\BIBentrySTDinterwordspacing

\bibitem{certificate}
{Jun Luo}, J.~. {Hubaux}, and P.~T. {Eugster}, ``Dictate: Distributed
  certification authority with probabilistic freshness for ad hoc networks,''
  \emph{IEEE Transactions on Dependable and Secure Computing}, vol.~2, no.~4,
  pp. 311--323, 2005.

\bibitem{gligor}
\BIBentryALTinterwordspacing
L.~Eschenauer and V.~D. Gligor, ``A key-management scheme for distributed
  sensor networks,'' in \emph{Proceedings of the 9th ACM Conference on Computer
  and Communications Security}, ser. CCS '02.\hskip 1em plus 0.5em minus
  0.4em\relax New York, NY, USA: Association for Computing Machinery, 2002, p.
  41–47. [Online]. Available: \url{https://doi.org/10.1145/586110.586117}
\BIBentrySTDinterwordspacing

\bibitem{pakp}
M.~{Gharib}, E.~{Emamjomeh-Zadeh}, A.~{Norouzi-Fard}, and A.~{Movaghar}, ``A
  novel probabilistic key management algorithm for large-scale manets,'' in
  \emph{2013 27th International Conference on Advanced Information Networking
  and Applications Workshops}, 2013, pp. 349--356.

\bibitem{jensen2019blockchain}
I.~J. Jensen, D.~F. Selvaraj, and P.~Ranganathan, ``Blockchain technology for
  networked swarms of unmanned aerial vehicles (uavs),'' in \emph{2019 IEEE
  20th International Symposium on" A World of Wireless, Mobile and Multimedia
  Networks"(WoWMoM)}.\hskip 1em plus 0.5em minus 0.4em\relax IEEE, 2019, pp.
  1--7.

\bibitem{alladi2020secauthuav}
T.~Alladi, N.~Naren, G.~Bansal, V.~Chamola, and M.~Guizani, ``Secauthuav: A
  novel authentication scheme for uav-base station scenario,'' \emph{IEEE
  Transactions on Vehicular Technology}, 2020.

\bibitem{uavlocalization}
A.~Couturier and M.~Akhloufi, ``A review on absolute visual localization for
  uav,'' \emph{Robotics and Autonomous Systems}, vol. 135, p. 103666, 01 2021.

\bibitem{jamming2}
T.~{Erpek}, Y.~E. {Sagduyu}, and Y.~{Shi}, ``Deep learning for launching and
  mitigating wireless jamming attacks,'' \emph{IEEE Transactions on Cognitive
  Communications and Networking}, vol.~5, no.~1, pp. 2--14, March 2019.

\bibitem{futurerouting2}
L.~{Abusalah}, A.~{Khokhar}, and M.~{Guizani}, ``A survey of secure mobile ad
  hoc routing protocols,'' \emph{IEEE Communications Surveys Tutorials},
  vol.~10, no.~4, pp. 78--93, 2008.

\end{thebibliography}

\end{document}